\documentclass[aps,prb,twocolumn,showpacs,preprintnumbers]{revtex4}

\usepackage{bm,epsf}
\usepackage{graphicx}
\usepackage{amsmath}

\newcommand{\bk}{{\bf k}}

\newcommand{\bv}{{\bf v}}

\newcommand{\bl}{{\bf l}}
\newcommand{\bq}{{\bf q}}
\newcommand{\br}{{\bf r}}
\newcommand{\bR}{{\bf R}}
\newcommand{\bB}{{\bf B}}
\newcommand{\bA}{{\bf A}}

\newcommand{\eps}{{\epsilon}}

\newcommand{\half}{\frac{1}{2}}
\newcommand{\hx}{\hat{x}}
\newcommand{\hy}{\hat{y}}
\newcommand{\htau}{\hat{\tau}}
\newcommand{\ba}{{\bf a}}
\newcommand{\bdelta}{{\boldsymbol{\delta}}}

\newcommand{\brho}{{\boldsymbol\rho}}
\newcommand{\phidot}{{\dot{\phi}}}
\newcommand{\lnabla}{\overline{\nabla}}
\newcommand{\fbar}{{\overline{f}}}
\newcommand{\Jbar}{{\overline{J}}}
\newcommand{\bg}{{\mathbf g}}

\newcommand{\dJ}{{\delta J}}

\newcommand{\Qbar}{{\overline{Q}}}

\newcommand{\mB}{{ b}}
\newcommand{\bmB}{{\mathbf  b}}
\newcommand{\be}{{\mathbf e}}
\newcommand{\bu}{{\mathbf u}}
\newcommand{\bw}{{\mathbf w}}
\newcommand{\bLambda}{{\mathbf \Lambda}}
\newcommand{\spLambda}{{\check{\Lambda}}}
\newcommand{\spl}{{\check{l}}}
\newcommand{\spb}{{\check{b}}}
\newcommand{\spZ}{{\check{Z}}}
\newcommand{\cA}{{\cal A}}
\newcommand{\Dbar}{{\bar{D}}}
\newcommand{\clD}{\hat{{\mathcal{D}}}}
\newcommand{\clK}{\hat{{\mathcal{K}}}}

\begin{document}

\title{A model of phase fluctuations in a lattice
$d$-wave superconductor:
application to the Cooper pair charge-density-wave in underdoped cuprates}%
\author{Ashot Melikyan and Zlatko Te\v sanovi\'c}
\address{Department of Physics and Astronomy, Johns Hopkins
University, Baltimore, MD 21218, USA 
\\ {\rm(\today)}
}
\begin{abstract}
\medskip
We introduce and study an XY-type model of thermal and quantum phase 
fluctuations in a two-dimensional correlated lattice $d$-wave superconductor
based on the QED$_3$ effective theory of high temperature
superconductors.
General features of and selected results obtained 
within this model were reported earlier
in an abbreviated format (Z. Te\v sanovi\' c, cond-mat/0405235).
The model is geared toward describing not only the long
distance but also the {\em intermediate} lengthscale
physics of underdoped cuprates. In particular, we elucidate
the dynamical origin and investigate specific features
of the charge-density-wave of Cooper pairs, which we argue
is the state behind the periodic charge density
modulation discovered in recent
scanning-tunneling-microscopy experiments. 
We illustrate how Mott-Hubbard correlations near half-filling suppress
superfluid density and favor an incompressible state which
breaks translational symmetry of the underlying atomic lattice.
We show how the formation of the Cooper pair charge-density-wave in
such a strongly quantum fluctuating superconductor
can naturally be understood as an Abrikosov-Hofstadter problem in a
type-II {\em dual} superconductor, with the role of the dual magnetic field
played by the electron density. The resulting Abrikosov lattice of
dual vortices translates into the periodic modulation of the 
Bogoliubov-deGennes gap function and the electronic density.
We numerically study the energetics of various 
Abrikosov-Hofstadter dual vortex arrays and compute their detailed
signatures in the single-particle local tunneling density of states. 
A 4$\times$4 checkerboard-type modulation pattern naturally arises as an
energetically favored ground state at and near the $x=1/8$ doping 
and produces the local density of states
in good agreement with experimental observations. 

\end{abstract}

\maketitle
\section{Introduction}
Several recent experiments \cite{corson,ong,campuzano}
support the proposal that the
pseudogap state in underdoped cuprates should be viewed as 
a phase-disordered superconductor \cite{emerykivelson}.
The effective theory based on this viewpoint 
was derived in Ref. \onlinecite{qed}:
One starts with the observation \cite{zt} that
in a phase fluctuating cuprate superconductor 
the Cooper pairing amplitude is large and
robust, resulting in a short coherence length $\xi\sim k_F^{-1}$
and small, tight cores for {\em singly quantized}
(anti)vortices. As a result, the phase fluctuations are greatly enhanced,
with $hc/2e$ vortex and antivortex excitations, their cores containing hardly
any electrons, quantum tunneling from place to place with the greatest of
ease, scrambling off-diagonal order in the process -- incidentally, 
this is the obvious interpretation of  
the Nernst effect experiments \cite{ong}.
Simultaneously, the largely inert pairing amplitude takes on
a dual responsibility of suppressing multiply quantized
(anti)vortices, which require larger cores and cost more kinetic
energy, while continuously maintaining the pseudogap effect in the 
single electron excitation spectrum.
The theory of Ref. \onlinecite{qed} uses this 
large $d$-wave pairing pseudogap $\Delta$ 
to set the stage upon which the low-energy degrees 
of freedom, identified as electrons organized into
Cooper pairs and BdG nodal fermions, and fluctuating 
$hc/2e$ vortex-antivortex pairs, mutually interact via
two emergent non-compact U(1) gauge fields, $v_\mu$ and $a_\mu$. 
$v$ and $a$ couple to electronic
charge and spin degrees of freedom, respectively, 
and mediate interactions which are responsible for
the three major phases of the theory \cite{qed,herbut}:
A $d$-wave superconductor, an insulating
spin-density wave  (SDW, which at half-filling
turns into a Mott-Hubbard-Neel antiferromagnet),
and an intermediate ``algebraic Fermi liquid'', a 
non-Fermi liquid phase characterized by critical, power-law 
correlations of nodal fermions. 

In the context of the above physical picture, the recent discovery
in scanning tunneling 
microscopy (STM) experiments \cite{yazdani,davis,kapitulnik} of 
the ``electron crystal'', manifested by
a periodic modulation of the local density of states (DOS),
and the subsequent insightful theoretical analysis \cite{zhang0}
of this modulation in terms of the {\em pair} density-wave,
comes not entirely unexpected. Such modulation originates from
the charge Berry phase term involving $v_0$ \cite{qed,preprint}, 
the time-like component of $v_\mu$, and the long-distance physics behind 
it bears some resemblance to that
of the elementary bosons, like $^4$He \ (Ref.~\onlinecite{fisherlee}).
As the quantum phase fluctuations become very strong, they
occasion a suppression of the compressibility
of the underlying electron system, via the phase-particle number uncertainty
relation $\Delta\varphi\Delta N\agt 1$, whose effective theory 
manifestation is precisely the above charge Berry phase.
Once the off-diagonal order disappears, the system
inevitably turns incompressible and the diagonal positional order sets in,
leading to a Mott insulating state. The resulting
charge-density-wave of Cooper pairs (CPCDW) \cite{preprint}
causes a periodic modulation
of the electron density and the size of the pseudogap $\Delta$ and
induces a similarly modulated local tunneling DOS. In this context, the
observed ``electron crystal'' state \cite{davis}
should be identified as a CPCDW.

While the above CPCDW scenario is almost certainly qualitatively correct,
the ultimate test of the theory is whether it can explain and predict some of 
the {\em specific details} of the modulation patterns
as they are actually observed in cuprates.
This brings us to the main theme of this paper. Typically, when
constructing an effective low energy theory of a condensed matter
system, we are solely concerned with the long distance, low energy
behavior. In the present case, however, this will not suffice. The
modulation in question is associated with length and energy scales
which are {\em intermediate}, between the short distance scale physics
of a single lattice spacing and the ultimate long distance behavior. 
Our aim should thus be to construct a description which will be valid not
only over very long lengthscales but also on the scale of several
lattice spacings, which are the 
periodicities observed in experiments \cite{yazdani,davis,kapitulnik}.
Furthermore, we ideally should be aiming for a ``bosonized'' version
of the theory, within which a mean-field type approximation for the
CPCDW state could be gainfully formulated. A natural
question that needs to be answered first is what should be the
objects that play the role of these bosons? 

One choice is to designate {\em real-space} pairs of electrons (or holes) as 
such ``elementary'' bosons and to endow them with some
``manageable'', i.e. pairwise and short-ranged, effective
interactions. Such a model indeed generically leads 
to a phase diagram with compressible superfluid and incompressible 
Wigner crystal states just as our long-distance argument has suggested.
This picture of real-space pairs arises within the SO(5)-based
theory \cite{zhang,auerbach}, where the low
energy sector of the theory assumes the form of hard-core
plaquette bosons with nearest and next-nearest neighbor interactions.
When one is dealing with extremely strongly bound {\em real
space} $s$-wave or $d$-wave pairs this is undoubtedly the natural choice.
In our view, however, in cuprates one 
is faced not with the real-space pairs
but with the {\em momentum space} Cooper pairs. 
Apparently, one encounters here an echo of
the great historical debate on Blatt-Schafroth versus BCS pairs -- while
certain long distance features are the same
in both limits, many crucial properties
are quite different \cite{footothers}. 
To be sure, the Cooper pairs in cuprates are
not far from the real-space boundary; the coupling is strong,
the BCS coherence length is short, and the fluctuations are 
greatly  enhanced. Still, there is a simple litmus test
that places cuprates squarely on the BCS side: they 
are $d$-wave superconductors {\em with nodal fermions}.

This being the case, constructing a theory with Cooper pairs as
``elementary'' bosons turns into a daunting enterprise. Cooper
pairs in nodal $d$-wave superconductors are highly non-local
objects in the real space, and the effective 
theory in terms of their center-of-mass
coordinates will reflect this non-locality in an essential way,
with complicated intrinsically multi-body, extended-range interactions.
The basic idea behind the QED$_3$ theory \cite{qed} is 
that in these circumstances the role of ``elementary'' 
bosons should be accorded to {\em vortices}
instead of Cooper pairs. Vortices in cuprates, with their small cores,
are simple real space objects and 
the effective theory of quantum fluctuating
vortex-antivortex pairs can be written in the form that is 
local and far simpler to analyze. In this dual language
the formation of the CDW of Cooper pairs translates -- via the
charge Berry phase discussed above -- to the 
familiar Abrikosov-Hofstadter problem in a dual
superconductor \cite{preprint, sachdev}. The solution of 
this problem intimately reflects the non-local character of 
Cooper pairs and their interactions, and the specific CPCDW modulation
patterns that arise in such theory are generally {\em different}
from those of a real-space pair density wave.
These two limits, the Cooper versus the real-space pairs, correspond to
two different regimes of a dual superconductor, 
reminiscent of the strongly type-II versus strongly type-I
regimes in ordinary superconductors. 
This difference is fundamental and, while both descriptions are
legitimate, only one has a chance of being relevant for cuprates.

A well-informed reader will immediately protest that the key tenet of 
the QED$_3$ theory is that we {\em cannot} write down a useful
``bosonized'' version of the theory at all -- the 
nodal BdG fermions in a fluctuating $d$-wave
superconductor must be kept as an integral part of the quantum
dynamics in underdoped cuprates. This, while true, mostly
reflects the central role of nodal fermions
in the {\em spin} channel. In contrast,
the formation of CPCDW is predominantly 
a {\em charge} sector affair and there, provided the theory is reexpressed
in terms of vortex-antivortex fluctuations -- i.e. properly ``dualized'' -- 
the effect of nodal fermions is less singular and definitely treatable. 
This gives one hope that a suitably ``bosonized'' dual 
version of the charge sector might be devised which will 
provide us with a faithful representation of underdoped cuprates. 
This is the main task we undertake in this paper. 

To this end, following a brief review of the QED$_3$ theory in Section II,
we propose in Section III a simple but realistic XY-type model of
a thermally phase-fluctuating $d$-wave superconductor. Starting
from this model we derive its effective Coulomb gas representation 
in terms of vortex-antivortex pairs. This representation
is employed in Section IV to construct an effective
action for quantum fluctuations of vortex-antivortex pairs and
derive its field theory representation in terms of a dual type-II
superconductor, incorporating the effect
of nodal fermions. In Section V, we discuss some general features 
of this dual type-II superconductor and relate our results to
the recent STM experiments \cite{yazdani,davis,kapitulnik}.
This section is written in the style that seeks to elucidate basic
concepts at the expense of overbearing mathematics; we hope the
presentation can be followed by a general reader.
After this we plunge in Section VI into a detailed study of 
the dual Abrikosov-Hofstadter-like problem which arises in the
mean-field approximation applied to a dual superconductor
and which regulates various properties
of the CPCDW state. This particular variant of the
Abrikosov-Hofstadter problem arises from the said charge Berry phase effect:
In the dual representation quantum bosons representing fluctuating
vortex-antivortex pairs experience an overall {\em dual} magnetic
field, generated by Cooper pairs, whose flux per plaquette of the (dual)
CuO$_2$ lattice is set by doping: $f=p/q=(1-x)/2$.
Finally, a brief summary of our results and conclusions are presented
in Section VII.

\section{Brief review of the QED$_3$ theory}

The purpose of this section is mainly pedagogical: 
Before we move on to our main topic, we provide some background on
the QED$_3$ effective theory of the pairing pseudogap in underdoped cuprates.
This will serve to motivate our interest in constructing an XY-type
model of fluctuating $d$-wave superconductors and to make a
casual reader aware of the connection between the theoretical notions
discussed in the rest of this paper and the actual physics of real electrons
in CuO$_2$ planes. The readers well versed in the art of construction 
of effective field theories or those already familiar with the 
approach of Ref. \onlinecite{qed} can safely jump directly to Section III. 

The effective theory \cite{qed} of a strongly
fluctuating $d_{x^2-y^2}$-wave superconductor
represents the interactions of 
fermions with $hc/2e$ vortex-antivortex
excitations in terms of two non-compact emergent gauge fields,
$v_\mu$ and $a_\mu^i$. The Lagrangian ${\cal L}={\cal L}_f + {\cal
L}_0$ is
\setlength{\multlinegap}{0pt}
\begin{multline}
\bar\Psi[D_0 +iv_0 -ieA_0+
\frac{({\bf D} +i{\bf v}-ie{\bf A})^2}{2m}-\mu]\Psi \\-
i\Delta\Psi^T\sigma_2 \hat\eta\Psi + {\rm c.c.}+  {\cal L}_0[v,a]~~,
\label{lagrangiani} 
\end{multline}  
where $\bar\Psi = (\bar\psi_\uparrow,\bar\psi_\downarrow)$, $\sigma_i$'s
are the Pauli matrices, $\Delta$ is the amplitude of the
$d_{x^2-y^2}$ pairing pseudogap, $A_\mu$ is the external electromagnetic
gauge field, $D_\mu =\partial_\mu +i2a_\mu^i(\sigma_i/2)$ is
SU(2) covariant derivative and $\hat\eta\equiv D_x^2 - D_y^2$.
${\cal L}_0[v,a]$ is generated by the 
Jacobian of the Franz-Te\v sanovi\' c (FT) singular gauge transformation:
\begin{multline}
\exp\bigl ({-\int_0^\beta d\tau\int d^2r {\cal L}_0[v_\mu,a_{\mu}^i]}\bigr )=
\int d\Omega_{\bf\hat n}\sum_{A,B}\int {\cal D}\varphi ({\bf
r},\tau) \\
\times  2^{-N_l}\delta[(w/\pi) - (n_A+n_B)]
\delta[(\tilde F^i/\pi) - G^i_{A,B}]~~,
\label{jacobian}  
\end{multline}
where
$2\pi n_{A,B}=\partial\times\partial\varphi_{A,B}$, $w=\partial\times v$,
$\tilde F^i_\lambda = \frac{1}{2}\epsilon_{\lambda\mu\nu}F^i_{\mu\nu}$,
$F^i_{\mu\nu}= \partial_\mu a^i_\nu - \partial_\nu a^i_\mu + 2\epsilon^{ijk}a^j_\mu a^k_\nu$,
$G^i_{A,B} = \Omega ({\bf\hat n})^{ij}m_j$,
$m_j=(0,0,(n_A-n_B))$,
$\Omega ({\bf\hat n})$ rotates the
spin quantization axis from ${\bf\hat z}$ to ${\bf\hat n}$, and
$\int d\Omega_{\bf\hat n}$ is the integral over
such rotations.

Its menacing appearance notwithstanding, the physics behind 
(\ref{lagrangiani}) is actually quite clear.
In ${\cal L}_f$, which is just the effective $d$-wave pairing
Lagrangian, the original electrons
$c_\sigma (x)$, with the spin quantized along an {\em arbitrary} direction 
${\bf\hat n}$, have been turned into topological fermions
$\bar\Psi = (\bar\psi_\uparrow,\bar\psi_\downarrow)$ through the
application of the FT transformation:
$(\bar c_\uparrow,\bar c_\downarrow)\to (\exp(i\varphi_A)\bar\psi_\uparrow,\exp(i\varphi_B)\bar\psi_\downarrow)$, where
$\exp(i\varphi_A (x)+i\varphi_B(x))$ 
equals $\exp(i\varphi (x))$, the center-of-mass
fluctuating superconducting phase (see Ref. \onlinecite{qed} for details). 
The main purpose of this transformation is to strip the awkward
phase factor $\exp(i\varphi (x))$ from the center-of-mass gap function,
leaving behind a $d$-wave pseudogap amplitude $\Delta$ and two gauge fields,
$v$ and $a^i$, which mimic the effects of phase fluctuations.
The Jacobian given by ${\cal L}_0[v,a^i]$ (\ref{jacobian}) insures that
the fluctuations in {\em continuous} fields $v$ and $a^i$ faithfully
represent configurations of {\em discrete} (anti)vortices.
Note that topological fermions do not carry a definite charge and 
are neutral on the average -- they
do, however, carry a definite spin $S=\half$, 
reflecting the fact that the spin
SU(2) symmetry remains intact in a spin-singlet superconductor.
As a consequence, the spin density operator 
$S_z= \bar c_\uparrow c_\uparrow -\bar c_\downarrow c_\downarrow =\bar\psi_\uparrow\psi_\uparrow -\bar\psi_\downarrow\psi_\downarrow$
is an invariant of the FT transformation. 

The two gauge fields, $v$ and $a^i$, are the main dynamical agents
of the theory. They describe the interactions
of fermionic BdG quasiparticles with vortex-antivortex pair excitations
in the fluctuating superconducting phase $\varphi(x)$. A
U(1) gauge field $v_\mu$ is the quantum fluctuating superflow 
and enters the non-conserved
charge channel. In the presence of the external electromagnetic
field $A_\mu$, one has $v_\mu\to v_\mu-eA_\mu$ everywhere in
${\cal L}_f$ (\ref{lagrangiani}) (but {\em not} in ${\cal L}_0$), to maintain
the local gauge symmetry of Maxwell electrodynamics. The Jacobian
${\cal L}_0$ (\ref{jacobian}) is the
exception since it is a purely mathematical object,
generated by the change of variables from discrete (anti)vortex coordinates
to continuous fields $v$ (and $a^i$). $v_\mu$ appearing in ${\cal L}_0$
contains only the vortex part of the superflow and is intrinsically
gauge invariant. This point is of much importance since it is
the state of the fluctuating (anti)vortices, manifested through
${\cal L}_0[v,a^i]$, which determines the state of our system as a whole, as
we will see momentarily. 

By comparison to $v$,
an SU(2) gauge field $a_\mu^i$ is of a more intricate origin. It encodes
topological frustration experienced by BdG quasiparticles as they encircle
$hc/2e$ vortices; the frustration arises from the
$\pm 1$ phase factors picked up by fermions
moving through the spacetime filled with fluctuating
vortex-antivortex pairs. These phase factors, unlike those
produced by superflow and emulated by $v$, are insensitive to
\begin{figure}[tbh]
\includegraphics[width=\columnwidth]{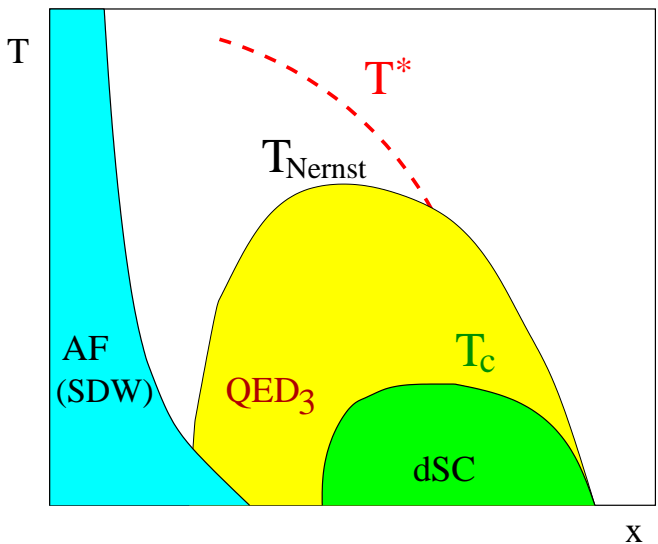}
\caption{\label{figqed}
The phase diagram of cuprates following from the QED$_3$ theory. Under the 
$T_{\rm Nernst}$ dome, the Nernst effect experiments \cite{ong}
indicate strong vortex-antivortex fluctuations; this is where the conditions
for the applicability of the theory are met. As the doping is
reduced, the superconducting ground state is followed by an
``algebraic'' Fermi liquid, a non-Fermi liquid state described by
the symmetric phase of
QED$_3$ and characterized by critical, power-law correlations
\cite{thermalmetal}. 
At yet lower doping, the BdG chiral symmetry is broken, resulting
in an incommensurate antiferromagnet (SDW), which eventually morphs
into the Neel antiferromagnet at half-filling. Depending on material
parameters (anisotropy, strength of residual interactions, degree
of interlayer coupling, etc.) of a 
particular HTS system, the transition between a superconductor and
a SDW could be a direct one, without the intervening chiral
symmetric ground state. The way CPCDW, the main subject of this paper,
fits into this phase diagram is detailed later in the text.
}
\end{figure}
vorticity and depend only on the spacetime configuration of
vortex loops, not on their internal orientation: if we picture a fixed
set of closed loops in 2+1-dimensional spacetime we can change
the orientation of any of the loops any which way without affecting
the topological phase factors, although the ones associated with
superflow will change dramatically. This crucial
symmetry of the original problem dictates the manner in which
$a^i$ couples to fermions in the above
effective theory \cite{qed}. Its topological origin is betrayed by its
coupling to a {\em conserved} quantum number -- in spin-singlet
superconductors this is a quasiparticle spin -- and the absence of
direct coupling to the non-conserved charge channel. 
By strict rules of the effective theory, such minimal coupling 
to BdG quasiparticles is what endows $a_\mu^i$ 
with its ultimate punch at the lowest energies.
While in this limit $v_\mu$ -- which couples to fermions in the
non-minimal, non-gauge invariant way -- turns massive 
and ultimately irrelevant, $a_\mu^i$ ends up generating
the interacting infrared (IR) critical point of QED$_3$ theory
which regulates the low energy fermiology of the pseudogap state.

${\cal L}$ (\ref{lagrangiani}) manifestly displays the
U(1) charge and SU(2) spin global symmetries,
and is useful for general considerations. 
To perform explicit calculations in the above theory
one must face up to a bit of algebra that goes into computing
${\cal L}_0 [v,a^i]$. The main technical obstacle is
provided by two constraints: 
{\em i)} the sources of $a_\mu^i$ should be 
an SU(2) gauge equivalent of half-flux Dirac strings, 
and {\em ii)} these sources are
permanently confined to the sources of $v_\mu$, which
themselves are also half-flux Dirac strings.  The confined objects 
formed by these two half-string species are 
nothing but the physical fluctuating
$hc/2e$ vortex excitations. These constraints are
the main source of technical hurdles encountered in trying to reduce
(\ref{lagrangiani}) to a more manageable form -- nevertheless,
it is imperative they be carefully enforced lest we lose
the essential physics of the original
problem. The constraints can be solved by introducing 
two dual complex fields $\Phi (x)= |\Phi |e^{i\phi}$ 
and $\Phi_s(x)= |\Phi_s|e^{i\phi_s}$ to provide coherent functional
integral representation of the half-strings corresponding to  
$v_\mu$ and $a_\mu^i$, respectively. In this dual language the half-strings
are simply the worldlines of dual bosons $\Phi$ and $\Phi_s$ propagating
through $(2+1)$-dimensional spacetime. It is important
to stress that this is just a mathematical tool used to properly
couple vortex-antivortex pair fluctuations to the electronic
degrees of freedom -- readers less intimate
with the technology of dualization will find it explained
in detail in the Appendix. The confinement of the two species 
of half-strings into 
$hc/2e$ (anti)vortices is accomplished by demanding
$|\Phi(x)|=|\Phi_s(x)|$. This finally
results in ${\cal L}_0$ expressed as
\begin{multline}
\exp\bigl ({-\int_0^\beta d\tau\int d^2r {\cal L}_0[v_\mu,a_{\mu}^i]}\bigr )\to
\int d\Omega_{\bf\hat n}
\int {\cal D}[\Phi,\phi_s,A_d,\kappa^i]
\\ 
\times
{\cal C}^{-1}[|\Phi |]
\exp\{\int d^3x(i2A_d\cdot w +
i2\kappa^i\cdot\tilde F^i-{\cal L}_d)\}~~,
\label{unwieldyjacobian}
\end{multline}
where ${\cal L}_d[\Phi,A_d,\kappa^i]$ is
a dual Lagrangian given by
\begin{multline}
m_d^2|\Phi |^2 + |(\partial_\mu 
- i2\pi A_{d\mu})\Phi |^2 + \frac{g}{2}|\Phi|^4\\
+ |\Phi|^2(\partial_\mu\phi_s-2\pi\Omega^{i3}\kappa^i_\mu)^2~~,
\label{duallagrangiani}
\end{multline}
and ${\cal C}[|\Phi |]$ is a normalization factor determined by
\begin{equation}
\int d\Omega_{\bf\hat n}{\cal D}[a^i,\phi_s,\kappa^i]\exp\{i2\kappa^i\cdot\tilde F^i
+|\Phi|^2(\partial_\mu\phi_s-2\pi\Omega^{i3}\kappa^i_\mu)^2\}~.
\label{normalization}
\end{equation}
Here $A_d$ and $\kappa^i$ are the charge and 
spin dual gauge fields, respectively, having been introduced to enforce the
two $\delta$-functions in (\ref{jacobian}).
Relative to our notation here,  in Ref. \onlinecite{preprint} both gauge fields are 
rescaled as $A_d(\kappa^i)\to 2\pi A_d(\kappa^i)$.
Notice that $A_d$ and $\kappa^i$ couple to $v$ and $a^i$ as
$\exp\bigl(i2A_d\cdot w +i2\kappa^i\cdot\tilde F^i\bigr)=\exp\bigl(i2A_d\cdot(\partial\times v)+i\kappa^i_\lambda\epsilon_{\lambda\mu\nu}(\partial_\mu a^i_\nu - \partial_\nu a^i_\mu + 2\epsilon^{ijk}a^j_\mu a^k_\nu)\bigr)$ (\ref{unwieldyjacobian}).

At this point, one should take note of the following convenient
property of ${\cal L}$ (\ref{lagrangiani},\ref{unwieldyjacobian}),
one that will prove handy repeatedly as we go along:
$A_d$ and $\kappa^i$ provide -- via the above coupling to
$v$ and $a^i$ -- the only link of communication
between the fermionic matter Lagrangian ${\cal L}_f$ and
the vortex-antivortex state Jacobian ${\cal L}_0$. This
way of formulating the theory is more than a mere mathematical nicety:
It is particularly
well-suited for the strongly fluctuating superconductor which
nonetheless does {\em not} belong to the 
extreme strongly bound limit of real-space
pairs, when they can be viewed as ``elementary'' bosons. Such
extreme limit of ``preformed'' pairs or pair ``molecules''
is frequently invoked, inappropriately in our view, to describe the
cuprates. Were the cuprates in this extreme limit, there would be
{\em no} gapless nodal quasiparticle excitations, the situation
manifestly at odds with the available experimental information
\cite{taillefer}. In addition, such demise of nodal fermions would rob
the effective low-energy theory of any {\em spin} degrees of freedom,
a most undesirable theoretical feature in the context of all we know
about the cuprates.
 
In a superconductor, the vortex-antivortex pairs are bound
and the dual bosons representing vortex loop worldlines
are in the ``normal'' (i.e., non-superfluid state):
$\langle\Phi\rangle=\langle\Phi_s\rangle =0$ (see the Appendix). Consequently,
both dual gauge fields are $A_d$ and $\kappa^i$ are massless; this
is as it should be in the dual normal state. This implies that upon
integration over $A_d$ and $\kappa^i$ both
$v_\mu$ and $a_\mu^i$ will be {\em massive}:
${\cal L}_0 \to M^2_0v_0^2 +M^2_\perp{\bf v}^2 + (v\to a)$,
where $M^2_0\propto M^2_\perp\sim \xi_d$, the
dual correlation length. As advertised above, the massive character of 
$v$ immediately makes the system
a physical superconductor: If we introduce a static external source vector
potential ${\bf A}^{ex}(\bq)$, the response to 
this perturbation is determined
by the fermionic ``mean-field'' stiffness originating from  
${\cal L}_f$ and the ``intrinsic'' stiffness set by ${\cal L}_0$,
whichever is {\em smaller}. In a strongly fluctuating superconductor
this in practice means ${\cal L}_0$. The mass $M^2_\perp (x,T)$ 
of the Doppler gauge field
${\bf v}$ in ${\cal L}_0$ is effectively the helicity modulus of such a 
superconductor and determines its superfluid density $\rho_s$,
reduced far bellow what would follow from ${\cal L}_f$
alone. In a similar vein, the mass $M^2_0 (x,T)$ of $v_0$ 
sets the compressibility of the system $\chi_c$, again far smaller than
the value for the non-interacting system of electrons at the
same density:
\begin{equation}
\begin{aligned}
\rho_s\sim \lim_{\bq\to 0}
\frac{\delta^2E_0}{\delta{\bf A}^{ex}(\bq)\delta{\bf A}^{ex}(-\bq)}=
\frac{\Pi^{jj}_f({\bq})M^2_\perp}{\Pi^{jj}_f({\bq})+ M^2_\perp}\sim M^2_\perp\\
\chi_c\sim \lim_{\bq\to 0}\frac{\delta^2E_0}{\delta A_0^{ex}(\bq)\delta A_0^{ex}(-\bq)}
=\frac{\Pi^{\rho\rho}_f({\bq})M^2_0}{\Pi^{\rho\rho}_f({\bq})+ M^2_0}\sim M^2_0
\label{superfluidstiffness}
\end{aligned}~,
\end{equation}
where $E_0$ is the ground state energy
and $\Pi^{jj}_f(\bq;x,T)$ and $\Pi^{\rho\rho}_f(\bq;x,T)$ are the 
current-current and density-density responses of
the fermionic Lagrangian ${\cal L}_f$, respectively.

Evidently, the ``intrinsic'' stiffness of 
$v$ as it appears in ${\cal L}_0$ encodes at the level of
the effective theory the
strong Mott-Hubbard short-range correlations which are the
root cause of reduced compressibility and superfluid density
and enhanced phase fluctuations in underdoped cuprates. 
This is entirely consistent with the spirit of the effective theory
and should be contrasted with the application of a Gutzwiller-type projector
to a mean-field BCS state, to which it is clearly superior:
In the latter approach one is suppressing on-site double occupancy
by a local projector mimicking correlations that are
{\em already} included 
at the level of the renormalized ``mean-field'' fermionic
Lagrangian ${\cal L}_f$ in (\ref{lagrangiani}), i.e. they are built-in
into $\Pi^{jj}_f$ and $\Pi^{\rho\rho}_f$ (\ref{superfluidstiffness}).
In this approach the superconductivity is always present at $T=0$ as long as 
the doping is finite, albeit with a reduced superfluid density. 
In contrast, within the effective 
theory (\ref{lagrangiani}), the microscopic Mott-Hubbard
correlations are {\em additionally} echoed by the appearance 
of well-defined vortex-antivortex excitations, 
whose core size is kept small by a large pairing
pseudogap $\Delta$. Such vortex-antivortex excitations further suppress
superfluid density from its renormalized ``mean-field'' value
and ultimately destroy the off-diagonal
long range order (ODLRO) altogether -- even at $T=0$ and finite doping --
all the while remaining sharply defined with no 
perceptible reduction in $\Delta$.

As the vortex-antivortex pairs unbind, 
the superconductivity is replaced by the dual superfluid order: 
$\langle\Phi\rangle\not= 0$, $\langle\Phi_s\rangle \not= 0$,
and $v_\mu$ and $a_\mu^i$ turn massless:
${\cal L}_0\to c_0\xi (\nabla v_0)^2 +c\xi (\nabla\times {\bf v})^2 +(v\to a)$,
where $\xi (x,T\to 0)$ is the superconducting correlation length,
$c_0$ and $c$ are numerical constants, and we have used radiation
gauge $\nabla\cdot {\bf v}=0$.
Physically, the massless character of $v_\mu$ and $a_\mu^i$ describes
the admixing of {\em free} quantum vortex-antivortex
excitations into the ground state of the system which started as
a $d$-wave superconductor.
Now, it is clear from Eq. (\ref{superfluidstiffness}) that
the response to ${\bf A}^{ex}(\bq)$ is {\em entirely}
determined by ${\cal L}_0$. The massless character of ${\bf v}$ in
${\cal L}_0$ implies the vanishing of the helicity modulus and
superfluid density, despite the fact that the contribution from 
${\cal L}_f$ to both remains {\em finite} and hardly changes 
through the transition. Similarly, the response to $A_0^{ex}(\bq)$
vanishes as well, since $v_0$ is now massless. Thus, the system
simultaneously loses superfluidity ($\rho_s\to 0$) and turns
incompressible ($\chi_c\sim c_0\xi |\bq|^2\to 0$) for doping
smaller than some critical $x_c$.

Note that the theory \cite{qed} predicts a universal relation between
the superfluid density $\rho_s$ and the fluctuation
diamagnetic susceptibility $\chi_{\rm dia}$ in underdoped cuprates,
as $T\to 0$. Right before the superconductivity disappears, for $x>x_c$,
the discussion surrounding Eq. (\ref{superfluidstiffness}) implies
$\rho_s\sim M^2_\perp\sim 1/\xi_d$. Similarly, in the region of strong
superconducting fluctuations for $x<x_c$, $M^2_\perp$ in 
(\ref{superfluidstiffness}) is replaced by $c\xi{\bq}^2$ as is
clear from the previous paragraph. The prefactor $c\xi$ is nothing
but the diamagnetic susceptibility $\chi_{\rm dia}\sim c\xi$. Assuming
that the superconducting correlation length and its dual are proportional
to each other, $\xi\sim\xi_d$, one finally
obtains $\chi_{\rm dia}\sim \rho_s^{-1}$.
In the case where strong phase fluctuations deviate from ``relativistic''
behavior, the dynamical critical exponent $z\not =1$ needs to be introduced
and the above expressions generalized to
$\rho_s\sim M^2_\perp \sim 1/\xi_d^z$,
$\chi_{\rm dia}\sim \xi^{2-z}$, and $\chi_{\rm dia}\sim \rho_s^{(z-2)/z}$.
Experimental observation of such a universal relation between
$\rho_s$ and $\chi_{\rm dia}$ would provide a powerful evidence
for the dominance of {\em quantum} phase fluctuations in the pseudogap state
of underdoped cuprates.

In the ``dual mean-field''
approximation $\langle\Phi\rangle =\langle\Phi_s\rangle = \Phi$
one finds:
\begin{multline}
{\cal L}_0 =\frac{1}{4\pi ^2|\Phi |^2}(\partial\times v)^2
+ {\cal L}_0^a[a^i]~;\\
e^{{-\int d^3x {\cal L}_0^a}}
={\cal C}^{-1}\int {\cal D}\kappa^i \exp\{\int d^3x\bigl (
i2\kappa^i\cdot\tilde F^i+{\cal R} [\kappa^i]\bigr )\}~~,
\label{jacobian2}
\end{multline}
where $\int d^3x{\cal R}\equiv \ln\int d\Omega_{\bf\hat n}\exp(-4\pi^2\int d^3x |\Phi|^2(\kappa_\mu ^i\hat n_i)^2)$.

${\cal L}_0^a[a^i]$ appears somewhat unwieldy, chiefly through
its non-local character. The non-locality is the penalty
we pay for staying faithful to the underlying physics:
While the topological origin of $a^i$ demands
coupling to the conserved SU(2)
spin three-currents, its sources are permanently confined to
Dirac half-strings of the superflow Doppler field $v$, which
by its very definition is a non-compact U(1) gauge field (see
below). The manifestly SU(2) invariant form (\ref{lagrangiani})
subjected to these constraints is therefore even more elegant than it is
useful. 

The situation, however, can be remedied entirely by a judicious choice of
gauge: $a_\mu^1=a_\mu^2=0$, $a_\mu^3=a_\mu$.
In this ``spin-axial'' gauge -- which amounts to selecting
a fixed (but arbitrary) spin quantization 
axis --  the integration over $\kappa_i$ can be
performed and ${\cal L}_0^a$ (\ref{jacobian2}) reduces to a 
simple {\em local} Maxwellian:
\begin{equation}
{\cal L}_0 [v,a]\to
\frac{1}{4\pi ^2|\Phi |^2}(\partial\times v)^2 +
\frac{1}{4\pi ^2|\Phi |^2}(\partial\times a)^2~~,
\label{spin-axial}
\end{equation}
where $\Phi$ is the dual
order parameter of the pseudogap state, i.e. the 
condensate of loops formed by 
vortex-antivortex creation and annihilation 
processes in (2+1)-dimensional Euclidean spacetime. This is
the quantum version of the Kosterlitz-Thouless unbound vortex-antivortex
pairs and is discussed in detail in the Appendix.  
The effective Lagrangian ${\cal L}={\cal L}_f + {\cal L}_0[v,a]$ 
of the quantum fluctuating $d$-wave superconductor finally takes the form:
\begin{multline}
\bar\Psi[D_0 +iv_0 -ieA_0+
\frac{({\bf D} +i{\bf v}-ie{\bf A})^2}{2m}-\mu]\Psi \\-
i\Delta\Psi^T\sigma_2 \hat\eta\Psi + {\rm c.c.}+ {\cal L}_0[v,a]~~,
\label{lagrangianqed3} 
\end{multline}  
where $D_\mu \to\partial_\mu +ia_\mu\sigma_3$, other quantities remain
as defined below Eq. (\ref{lagrangiani}), and  
${\cal L}_0[v,a]$ of the pseudogap state
is given by Eq. (\ref{spin-axial}). More generally, the 
full spin-axial gauge expression for ${\cal L}_0[v,a]$ in 
(\ref{lagrangianqed3}), valid
both in the pseudogap ($\langle\Phi\rangle\not =0$) and
superconducting ($\langle\Phi\rangle =0$) states, is:
\begin{multline}
\exp\bigl ({-\int_0^\beta d\tau\int d^2r {\cal L}_0[v_\mu,a_{\mu}]}\bigr )\to
\int {\cal D}[\Phi,\phi_s,A_d,\kappa]
\\ 
\times
{\cal C}^{-1}[|\Phi |]
\exp\{\int d^3x(i2A_d\cdot w +
i2\kappa\cdot\tilde F-{\cal L}_d)\}~~,
\label{wieldyjacobian}
\end{multline}
where dual Lagrangian ${\cal L}_d[\Phi,A_d,\kappa]$ is
\begin{multline}
m_d^2|\Phi |^2 + |(\partial_\mu 
- i2\pi A_{d\mu})\Phi |^2 + \frac{g}{2}|\Phi|^4\\
+ |\Phi|^2(\partial_\mu\phi_s-2\pi\kappa_\mu)^2~~,
\label{duallagrangianiv}
\end{multline}
and normalization factor ${\cal C}[|\Phi |]$ equals
\begin{equation}
\int {\cal D}[a,\phi_s,\kappa]\exp\{i2\kappa\cdot\tilde F
+|\Phi|^2(\partial_\mu\phi_s-2\pi\kappa_\mu)^2\}~.
\label{normalizationiv}
\end{equation}
Again, $A_d$ and $\kappa$ are the charge and 
spin dual gauge fields, respectively, which, in the spin-axial gauge,
couple to $v$ and $a$ as
$\exp\bigl(i2A_d\cdot w +i2\kappa\cdot\tilde F\bigr)=\exp\bigl(i2A_d\cdot(\partial\times v)+i\kappa_\lambda\epsilon_{\lambda\mu\nu}(\partial_\mu a_\nu - \partial_\nu a_\mu)\bigr)$ (\ref{wieldyjacobian}).

The above is just the standard form of the
QED$_3$ theory discussed earlier \cite{qed}. It resurfaces here as a 
particular gauge edition of a more symmetric, 
but also far more cumbersome
description -- fortunately, in contrast to its high symmetry parent,
the Lagrangian (\ref{lagrangianqed3}) itself is
eminently treatable \cite{footqed}. 

A committed reader should note that the
ultimate non-compact U(1) gauge theory 
form of ${\cal L}$ (\ref{lagrangianqed3})
arises as a natural consequence of the constraints described
above. Once the sources of $v$ and $a$ are confined into
physical $hc/2e$ vortices, the non-compact U(1) character is the
shared fate for both. While we do have the choice of selecting
the singular gauge in which to represent the Berry gauge field
$a$, there is no similar choice for $v$, which must be a non-compact
U(1) gauge field.  It is then only natural to 
use the FT gauge and represent $a$ as a 
non-compact U(1) field as well, in order to straightforwardly enforce the 
confinement of their respective sources \cite{footupdown}. The non-compactness 
in this context reflects nothing but an elementary property of
a phase-fluctuating superconductor:
The conservation of a topological vortex charge 
$\partial\cdot n =0 =\partial\cdot n_{A,B}$ \cite{footcompactqed}.

We end this section with the following remarks: In explicit calculations
with (\ref{lagrangianqed3}) or its lattice equivalent, it is often
useful to separate the low energy
nodal BdG quasiparticle excitations from the rest
of the electronic degrees of freedom by linearizing
${\cal L}_f$ near the nodes. The nodal fermions
$\psi_{\sigma,\alpha}$, where $\alpha=$ 1, $\bar 1$, 2 and $\bar 2$
is a node index, can then be arranged into $N$ 
four-component BdG-Dirac spinors
following the conventions of Ref. \onlinecite{qed},
where $N=2$ for a single CuO$_2$ layer. These massless Dirac-like
objects carry no overall charge and are at zero chemical
potential -- reflecting the fact that pairing in the
particle-particle channel always pins the $d$-wave nodes 
to the true  chemical potential -- and can be thought of as the 
particle-hole excitations of the BCS ``vacuum''. They, however,
can be polarized and their polarization will 
renormalize the fluctuations of the
Doppler gauge field $v$, the point which will be emphasized later.
Furthermore, the nodal fermions carry spin $S=\half$ and
interact strongly with the Berry gauge field $a$ to which they
are minimally coupled. In contrast, the rest of
the electronic degrees of freedom, which we label
``anti-nodal'' fermions, $\psi_{\sigma,\langle\alpha\beta\rangle}$,
where $\langle\alpha\beta\rangle =$ $\langle 12\rangle$, 
$\langle 2\bar1\rangle$, $\langle \bar 1\bar 2\rangle$, 
and $\langle \bar 2 1\rangle$,
are combined into spin-singlet Cooper pairs and do not contribute
significantly to the spin channel. On the other hand, these anti-nodal
fermions have finite density and carry the overall electric charge.
Their coupling to $v$ is the dynamical driving force behind the
formation of the CPCDW. Meanwhile, the nodal fermions are not
affected by the CPCDW at the leading order -- their low energy
effective theory is still the symmetric QED$_3$ even though
the translational symmetry is broken by the CPCDW \cite{preprint}.

The presence of these massless Dirac-like excitations in 
${\cal L}_f$ is at the heart of the QED$_3$ theory of cuprates \cite{qed}.
QED$_3$ theory is an effective low energy description of a
fluctuating $d_{x^2-y^2}$-wave superconductor, gradually losing
phase coherence by progressive admixing of quantum vortex-antivortex
fluctuations into its ground state. All the while, even as the ODLRO
is lost, the amplitude of the BdG gap function remains finite and
largely undisturbed as the doping is reduced toward half-filling as depicted
in Fig. \ref{figqed}.
How is this possible? A nodal $d_{x^2-y^2}$-wave superconductor 
 -- in contrast to its fully gapped cousin or a conventional
s-wave superconductor -- possesses {\em two} fundamental symmetries in
its ground state; the familiar one is just the presence of the ODLRO,
shared by all superconductors. In addition, there is a more
subtle symmetry of its low energy fermionic spectrum which is exclusively
tied to the presence of nodes. This symmetry is
{\em emergent}, in the sense that it is not a symmetry of the full
microscopic Hamiltonian but only of its low energy nodal sector, and
is little more than the freedom intrinsic to arranging two-component
nodal BdG spinors into four-component massless Dirac fermions \cite{qed} -- by
analogy to the field theory we call this symmetry {\em chiral}.
The QED$_3$ theory first formulates and then answers 
in precise mathematical terms
the following question: can a nodal $d_{x^2-y^2}$-wave superconductor 
lose ODLRO but nonetheless retain the BdG chiral symmetry of its low energy
fermionic spectrum \cite{qed}? In conventional BCS theory the 
answer is a straightforward ``no'':
as the gap goes to zero all vestiges of the superconducting state are
erased and one recovers a Fermi liquid normal state. 
However, in a strongly quantum fluctuating superconductor considered here,
which loses ODLRO via vortex-antivortex unbinding, the answer is
a remarkable ``yes''. The chirally {\em symmetric}, IR
{\em critical} phase of QED$_3$ is the explicit
realization of this new, non-Fermi liquid state of quantum matter, with the
phase order of a $d_{x^2-y^2}$-wave superconductor gone but
the chiral symmetry of fermionic excitations left in its wake.
While these excitations are incoherent, being strongly scattered by the
massless Berry gauge field, the BdG chiral symmetry 
of the low energy sector remains intact \cite{thermalmetal}. Importantly,
as interactions get too strong -- the case in point being
the cuprates at very low dopings  -- the
BdG chiral symmetry will be eventually 
spontaneously broken, leading to antiferromagnetism
and possibly other fully gapped states \cite{qed,herbut}. Nonetheless,
the BdG chiral symmetry breaking is not fundamentally tied to the loss
of ODLRO: different HTS materials are likely to have different $T=0$ 
phase diagrams with larger or smaller portions of a stable critical
``normal'' state -- described by the symmetric QED$_3$ -- sandwiched
between a superconductor and an antiferromagnet (SDW) (Fig. \ref{figqed}).
In all cases, provided our starting assumption of the predominantly 
pairing nature of the pseudogap
is correct, the symmetric QED$_3$ emerges as the underlying effective
theory of underdoped cuprates, echoing the role
played by Landau Fermi liquid theory in conventional metals.

\section{Thermal (``classical'') phase fluctuations}

\begin{figure}[tbh]
\includegraphics[width=\columnwidth]{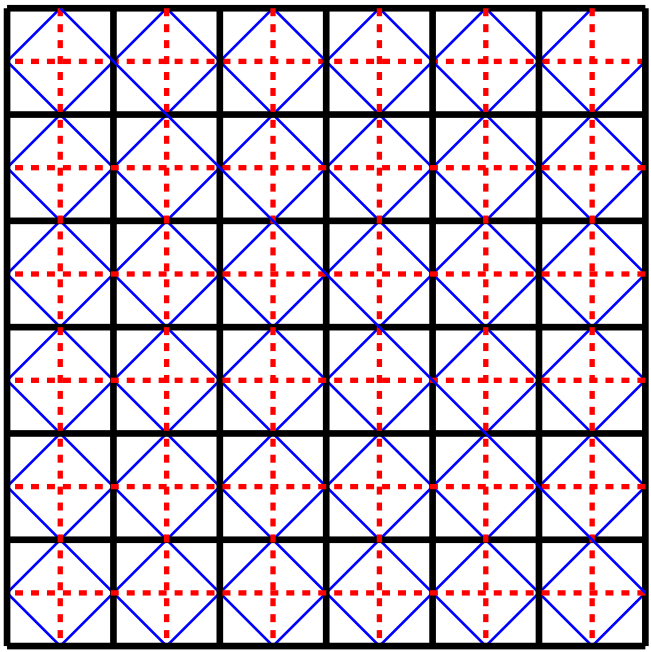}
\caption{\label{fig1}
CuO$_2$ lattice represented by thick black lines (vertices are copper
atoms and oxygens are in the middle of each bond).
The nodes of the blue lattice (thin solid lines) represent sites of
our ``effective'' XY model (\ref{hxy}),
with the phase factor $\exp(i\varphi_i)$
located at each blue vertex. The dual of the copper lattice is shown in red dashed lines.
The nearest neighbor coupling on the
blue lattice is denoted by $J$ (blue link)
and the two next-nearest neighbor couplings are
$J_1$ (red link) and $J_2$ (black link). Generically, $J_1\not =J_2$
and the translational symmetry of the blue lattice is broken
by the ``checkerboard'' array of ``red'' and ``black'' plaquettes, as
is evident from the figure. This doubles up the unit cell which
then coincides with that of the
original (black) CuO$_2$ lattice, as it should be.
}
\end{figure}

Our first goal is to introduce an XY model type representation
of thermal (or ``classical'') phase fluctuations. 
To this end we first observe that
in cuprates, we are dealing with a $d$-wave superconductor on a
tightly bound two-dimensional CuO$_2$ lattice (the black
lattice in Fig. \ref{fig1}). The simplest starting point is the lattice $d$-wave 
superconductor (LdSC) model discussed 
in detail by Vafek {\em et al.} \cite{vafek}.
The model describes fermions hopping between nearest
neighbor sites $\langle ij\rangle$ on a square lattice with a
renormalized matrix element $t^*$ and contains a 
nearest neighbor spin-singlet pairing term with an effective coupling constant
$\lambda_{\rm eff}$ adjusted to stabilize the $d_{x^2-y^2}$ state
with the maximum pairing gap $\Delta$:
\begin{multline}
H_{\rm LdSC}= -t^*\sum_{\langle ij\rangle,\sigma}c^\dag_{i\sigma}c_{j\sigma}
+ \sum_{\langle ij\rangle}\Delta_{ij}[c^\dag_{i\uparrow}c^\dag_{j\downarrow}
- c^\dag_{i\downarrow}c^\dag_{j\uparrow}] 
+ ({\rm c.c.})\\ +(1/\lambda_{\rm eff})\sum_{\langle ij\rangle}|\Delta_{ij}|^2
+ (\cdots)~~,
\label{ldsc}
\end{multline}
where $(\cdots)$ denotes various residual interaction terms. 
$c^\dag_{i\sigma}$ and $c_{i\sigma}$ are creation and annihilation
operators of some {\em effective} electron states, appropriate for energy
scales below and around $\Delta$, which {\em already include} renormalizations
generated by integration of higher energy configurations, particularly
those associated with strong Mott-Hubbard correlations.
This effective  LdSC model is phenomenological but can be justified within
a more microscopic approach, an example being the one based on a $t-J$-style
effective Hamiltonian. The complex gap function is defined on 
the bonds of the CuO$_2$ (black) lattice:  
$\Delta_{ij} = A_{ij}\exp(i\theta_{ij})$. The amplitude $A_{ij}$
is frozen below the pseudogap energy scale $\Delta$
and equals $A_{ij}=\pm\Delta$ along horizontal (vertical)
bonds. Thus, the $d$-wave character of the 
pairing has been incorporated directly into $A_{ij}$ from the start. 

What remains are the fluctuations of the bond phase $\theta_{ij}$. 
It is advantageous to
represent these {\em bond} phase fluctuations in terms of {\em site}
fluctuations by identifying $\exp(i\theta_{ij})\to\exp(i\varphi_k)$,
where $\varphi_k$ is a site phase variable associated with the
middle of the bond $\langle ij\rangle$. Consequently, we are now
dealing with the set of {\em site} phases $\varphi_k$ located at
the vertices of the blue lattice in Fig. \ref{fig1}.
Note that the blue
lattice has twice as many sites as the original CuO$_2$ lattice or
its dual. This is an important point and will be discussed shortly.

We can now integrate over the fermions in the LdSC model and 
generate various couplings
among phase factors  $\exp(i\varphi_k)$ residing on different ``blue'' sites.
The result is the minimal site XY-type model Hamiltonian
representing a fluctuating classical lattice $d$-wave superconductor:
\begin{multline}
{\cal H}^d_{XY} = -J\sum_{nn}\cos(\varphi_i - \varphi_j)
-J_1\sum_{rnnn}\cos(\varphi_i - \varphi_j)\\
-J_2\sum_{bnnn}\cos(\varphi_i - \varphi_j)~~,
\label{hxy} 
\end{multline}  
where $\sum_{nn}$ runs over nearest neighbors on the blue lattice
while $\sum_{rnnn}$ and $\sum_{bnnn}$ run over red and black next-nearest
neighbor bonds depicted in Fig. \ref{fig1}, respectively. The 
couplings beyond next-nearest neighbors are neglected\cite{footcompact} in
(\ref{hxy}). Such
more distant couplings are not important for the basic physics which is
of interest to us here, and their effect can be 
simulated by a judicious choice of
$J_1$ and $J_2$ -- the reader should be warned that
the situation changes when
$T\to 0$ as will be discussed in the next section. 
Keeping $J_1$ and $J_2$, however, is essential. If only
$J$ is kept in (\ref{hxy}), the translational symmetry of the blue
lattice would be left intact and we would find ourselves in a situation 
more symmetric than warranted by the microscopic physics of a
$d$-wave superconductor. In this sense $J_1$ and $J_2$ or, more precisely,
their difference $J_1-J_2\not = 0$, are dangerously irrelevant couplings
and should be included in any approach which aims to describe
the physics at intermediate lengthscales.

In general, near half-filling, 
$J$, $J_1$ and $J_2$ tend to be all positive and mutually 
different. 
This ensures that the ground state is indeed a $d$-wave superconductor
(the reader should recall that $d$-wave signs had already been absorbed into
$A_{ij}$'s so the ground state is an XY ferromagnet instead of an
antiferromagnet -- the antiferromagnet would be an extended $s$-wave state.).
The explicit values of $J$, $J_1$ and $J_2$ can be
computed in a particular model. In this paper we treat 
them as adjustable parameters.

\subsection{Vortex-antivortex Coulomb gas representation of ${\cal H}^d_{XY}$}

The fact that $J_1\not = J_2$ breaks translational symmetry of the
blue lattice. This is as it should be since the blue lattice has
twice as many sites as the original CuO$_2$ lattice. This is
an important point to which we will return shortly. Now, however,
let us assume for the moment that $J_1 = J_2$. This assumption is
used strictly for pedagogical purposes. Then, the
blue lattice has a full translational symmetry and it is 
straightforward to derive the effective Coulomb gas representation
for fluctuating vortex-antivortex pairs. Here we follow 
the derivation of Ref. \onlinecite{nelson}.
The cosine functions in (\ref{hxy}) are expanded to second order
and one obtains the effective continuum theory:
\begin{equation}
{\cal H}_{cont}
= \frac{1}{2} \tilde J\int d^2r |\nabla\varphi (\br) |^2 +
(\cdots)~~,
\label{hcont}
\end{equation}
where $\tilde J= J+J_1+J_2$ and $\varphi (\br)$ is the continuum version
of the site phase $\varphi_i$ on the blue lattice. ($\cdots$) denotes
higher order terms in the expansion of the cosine function. 
As usual, the superfluid velocity
part $v(\br)=\nabla\varphi (\br)$ is separated into a regular (XY spin-wave)
part and a singular (vortex) part $(\nabla\varphi )_v$:
\begin{equation}
v(\br) = \nabla\chi(\br) + 2\pi (\hat z\times\nabla )
\int d^2r' n(\br ')G(r,r')~~,
\label{velocity}
\end{equation}
where $\chi$ is a regular free field and $n(\br)$ is the density
of topological vortex charge: 
$n(\br) =\sum_i\delta(\br -\br_i^v)-\sum_j\delta(\br -\br_j^a)$, 
with $\br _i^v$ and $\br _j^a$ being
the positions of vortex and antivortex defects, respectively. We
are limiting ourselves to $\pm 1$ vortex charges but higher charges
are easily included. $G(r,r')$ is the 2D electrostatic Green's function
which satisfies:
\begin{equation}
\nabla ^2 G(r,r') = \delta (\br - \br')~~.
\label{green}
\end{equation}

Far from systems boundaries, the solution of (\ref{green}) is:
\begin{equation}
G(r,r') = \frac{1}{2\pi} \ln (|\br -\br'|/a) + C~~,
\label{solution}
\end{equation}
where $a$ is the UV cutoff, of the order of the (blue) lattice
spacing and $C$ is an integration constant, to be 
associated with the core energy.

This ultimately gives the vortex part of the Hamiltonian as:
\begin{equation}
{\cal H}_{v}
= 2\pi^2 \tilde J\!\int \!\! d^2r d^2r' d^2r''
n(\br ')n(\br'')\nabla G(r,r')\cdot\nabla G(r,r'')~,
\label{hvortex}
\end{equation}
After integration by parts this results in the desired
Coulomb gas representation of ${\cal H}_{v} $:
\begin{equation}
-\pi \tilde J\int d^2r d^2r'
n(\br)n(\br')\ln (|\br -\br'|/a) 
+ E_c\int d^2r n^2(\br)~~,
\label{hcoulomb}
\end{equation}
where $E_c \sim C$ is the core energy. Note that the
integrals in the first part of (\ref{hcoulomb}) include only the
regions outside vortex cores (the size of which is $\sim a^2$).

How is the above derivation affected if we now restore
$J_1\not = J_2$, as is the case in real cuprates? The long
distance part remains the same since $J_1 - J_2$ enters 
only at the $O(k^4)$ order of the gradient expansion of
cosine functions in (\ref{hxy}), i.e. $J_1 - J_2\not =0$ affects
only the terms denoted by $(\cdots)$ in ${\cal H}_{cont}$ (\ref{hcont}). 
So, the strength of the Coulomb interaction between 
(anti)vortices in ${\cal H}_v$
(\ref{hcoulomb}) is still given by $\tilde J=J+J_1+J_2$.
The core energy changes, however. This change can be
traced back to (\ref{green}) and (\ref{solution}); depending
on whether the vortex core is placed in a red (i.e., containing
the red cross in Fig. \ref{fig1}) or a black plaquette of the blue lattice, the
constant of integration $C$ will generally be different. This is just
the reflection of the fact that for $J_1\not = J_2$ the
original translational symmetry of the blue lattice is broken
down to the {\em checkerboard} pattern as is obvious from Fig. \ref{fig1}.
So, Eq. (\ref{solution}) must be replaced by:

\begin{equation}
G(r,r') = \frac{1}{2\pi} \ln (|\br -\br'|/a) + C_{r(b)}~~.
\label{solution_alt}
\end{equation}

$G(r,r')$ is just the electrostatic potential at point $\br$ produced
by a vortex charge at $\br'$. Since not all locations for
$\br '$ are equivalent, there are
two constants of integration $C_r\not = C_b~(C_r-C_b\sim J_1-J_2)$,
corresponding to whether $\br'$ is in the red or black plaquettes of
the blue lattice (Fig. \ref{fig1}). Retracing the steps leading to the
Coulomb gas representation we finally obtain:
\begin{multline}
{\cal H}^d_{v} 
= -\pi \tilde J\int d^2r\int d^2r'
n(\br)n(\br')\ln (|\br -\br'|/a)\\
+ E_c^r\int d^2r \,n^2(\br) + E_c^b\int d^2r \,n^2(\br)~~,
\label{hcoulomb1}
\end{multline}
where $E_c^{r(b)}$ are the core energies of (anti)vortices located
in red (black) plaquettes and $E_c^r - E_c^b\sim J_1-J_2$. 
$E_c^{r(b)}$ are treated as adjustable parameters, chosen to best
reproduce the energetics of the original Hamiltonian (\ref{hxy}).
The explicit lattice version of (\ref{hcoulomb}) follows from
Ref. \onlinecite{jose}, where a duality transformation
and a Migdal-style renormalization procedure have been applied to the XY
model:
\begin{multline}
{\cal H}^d_{v} 
= -\pi \tilde J\sum_{\br\not =\br'}
s(\br)s(\br')\ln (|\br -\br'|/a) 
\\+ E_c^r\sum_{\br\in{\cal R}} s^2(\br) +
E_c^b\sum_{\br\in{\cal B}}s^2(\br)~~.
\label{latticecoulomb}
\end{multline}
In (\ref{latticecoulomb}) $s(\br) = 0,\pm 1 (\pm 2, \pm 3,\dots)$ and
$\br$ are summed over the red $({\cal R})$
and black $({\cal B})$ sites of the lattice
dual to the blue lattice in Fig. \ref{fig1}. Eq. (\ref{latticecoulomb}) 
can be viewed as a convenient lattice regularization of (\ref{hcoulomb1}).
It is rederived in the Appendix with the help of Villain approximation.

We can also recast  ${\cal H}_{v}$ in a form which interpolates between
the continuum (\ref{hcoulomb1}) and lattice (\ref{latticecoulomb}) Coulomb
gases. For convenience, we now
limit our attention to only $\pm 1$ vortex charges (vortices and antivortices)
since these are the relevant excitations in the pseudogap regime.
The underlying lattice in
(\ref{latticecoulomb}) is ``smoothed out'' into a periodic potential
$V(\br)$ whose minima are located in the plaquettes of the blue lattice
(at vertices of the dual blue lattice). The precise form of $V(\br)$ can
be computed in a continuum model of realistic cuprates involving
local atomic orbitals, self-consistent computations of the pseudogap
$\Delta$, etc. and is not important for our purposes; only 
its overall symmetry matters and the
fact that it is sufficiently ``smooth''. This finally produces:
\begin{multline}
{\cal H}^d_{v} 
= -\pi \tilde J\int d^2r \int d^2r'
n(\br)n(\br')\ln (|\br -\br'|/a)
\\+ \int d^2r \,V(\br)\rho (\br)~~,
\label{hpotential}
\end{multline}
where $n (\br)$ is the vortex charge density as before and
$\rho (\br) =\sum_i\delta(\br -\br_i^v)+\sum_j\delta(\br -\br_j^a)$ is
the vortex particle density {\em irrespective} of vortex charge.
If $E_c^r =E_c^b$, the periodicity of
$V(\br)$ coincides with that of the blue lattice -- the value at the
minimum is basically $E_c^r =E_c^b$. On the other hand, if
$E_c^r \not =E_c^b$ as is the case in real cuprates,
$V(\br)$ has a {\em checkerboard} symmetry on the
blue lattice, with {\em two} different kinds of local
extrema; the red one, at energy set by $E_c^r$ and a black one, 
at energy $E_c^b$ (this is all depicted in Fig. \ref{fig1}).

The most important consequence of these two different local
extrema of $V(\br)$ is that there are two special locations for the
position of vortices; on the original CuO$_2$ lattice of Fig. \ref{fig1}
these correspond to vortices either residing in its plaquettes
or at its vertices. On general grounds, we expect one of these
positions to be the absolute minimum while the other assumes the role of
either a locally stable minimum or
a saddle point with an unstable direction. 
Whether $E_c^r < E_c^b$ or the other
way around and whether the higher energy is a local minimum or
a saddle point, however, can be answered with certainty
only within a model more macroscopic
than the one used here. In particular, a specific analysis of the 
electronic structure of a 
strongly correlated vortex core is necessary which goes 
far beyond the XY-type model used in this
paper. For our purposes $E_c^r$ and $E_c^b$ can be treated as
adjustable parameters and it suffices only to know that the microscopic
physics selects either the red or the black plaquettes in Fig. \ref{fig1}
as the favored vortex core sites and relegates the other to either the
metastable or a weakly unstable status. 

\section{Quantum phase fluctuations}

The previous discussion pertains to 
the thermal (or classical) 2D XY model. The key difference
from the usual Coulomb gas representation for vortex-antivortex
pairs turned out to be the inequivalence of vortex core positions on the blue
lattice (Fig. \ref{fig1}), which served to recover the translational
symmetry of the original CuO$_2$ lattice. This effect was represented by
the blue lattice potential $V(\br)$ in the vortex Hamiltonian
(\ref{hpotential}).

To obtain the {\em quantum} version of the fluctuation problem
or an effective (2+1)D XY-type model we need to include
the imaginary time dependence in $\varphi_i\to\varphi_i(\tau)$. In general,
this will result in vortex-antivortex pairs propagating through 
imaginary (Euclidean) time, as vortex cores move from one plaquette
to another through a sequence of quantum tunneling processes. There are two
aspects of such motion: The first is the tunneling of the vortex
core. This is a ``microscopic'' process, in the  sense that a
detailed continuum model, with self-consistently computed core structure,
is needed to describe it quantitatively. 
For the purposes of this paper, all we need
to know is that one of the final components characterizing
such a tunneling process is the ``mass'' $M$ of the vortex
core as it moves through imaginary time \cite{mizel}.  This implies a term 
\begin{equation}
\half \int_0^\beta d\tau \sum_i M\Bigl (\frac{d\br _i^{v(a)}}{d\tau}\Bigr )^2
\label{vortexmass}
\end{equation}
in the quantum mechanical vortex action, 
where $\br _i^{v(a)}(\tau)$ is the position of the $i$th (anti)vortex in 
the CuO$_2$ plane at time $\tau$. Note that $M$ is here
just another parameter in the theory which, like $J$, $J_1$, and
$J_2$, is more microscopic than our model \cite{mizel}.
However, the microscopic physics of the cuprates clearly
points to $M$ being far smaller than in conventional superconductors. 
Cuprates are strongly coupled systems,
with a short coherence length $\xi\sim k_F^{-1}$ and the
vortex core of only several lattice spacings in size. Consequently,
there are only a few electrons ``inside a core'' at any given
time, making (anti)vortices ``light'' and 
highly quantum objects, with strong zero-point fluctuations.
Furthermore, the core excitations appear to be gapped \cite{core}
by the combination of
strong coupling and local 
Mott-Hubbard correlations \cite{mottvortex,ogata,dhlee,palee,millis}.
This has an important implication for the second component
of the core tunneling process, the familiar
Bardeen-Stephen form of dissipation, which is such a ubiquitous and
dominating effect in conventional superconductors. In conventional systems
the core is hundreds of nanometers in diameter and contains thousands
of electrons. $M$ is large in such a superconductor and, as a 
vortex attempts to tunnel to a different site, its motion is
damped by these thousands of effectively normal electrons, resulting
in significant Bardeen-Stephen dissipation and high viscosity. The motion
of such a huge, strongly damped object is effectively classical.
Vortices in underdoped cuprates are precisely the opposite, with small $M$,
the Bardeen-Stephen dissipation nearly absent and a very low viscosity.
In this sense, the quantum motion of a vortex core in 
underdoped cuprates is closer to superfluid $^4$He
than to other, conventional superconductors \cite{galilean}. We are therefore
justified in assuming that it is adequately described 
by (\ref{vortexmass}) and ignoring small vortex core viscosity
for the rest of this paper \cite{footviscosity}.
The special feature of the cuprates, however, is the presence
of gapless nodal quasiparticles away from the cores
which generate their own peculiar brand of (weak) dissipation -- such
effect is discussed and included later in this section. 

The second aspect of the vortex propagation through imaginary time 
involves the motion of the superflow velocity field
surrounding the vortex {\em outside} its core, i.e., in the region
of space where the magnitude of $\Delta$ is approximately uniform.
This is a long range effect and, unlike the vortex core energies,
mass, or Bardeen-Stephen dissipation,
exhibits certain universal features, shared by all superconductors
and superfluids. For example, in superfluid $^4$He this 
effect would result in a Magnus force
acting on a vortex. In a superconductor, the effect arises from the
time derivative of the phase in the regions of uniform $\Delta$.
The origin of such time derivative and its physical
consequences are most easily appreciated 
by considering an ordinary (fictitious) strongly fluctuating
$s$-wave superconductor with a
phase factor $\exp(i\phi_i(\tau))$ at each {\em site} of the CuO$_2$
lattice. For reader's benefit we discuss this case first as a 
pedagogical warm-up for what lies ahead.

\subsection{Pedagogical exercise: quantum fluctuating $s$-wave superconductor}

After performing the FT transformation and forgetting the
double valuedness problem (ignoring the Berry gauge field $a$)
since we are concentrating on the charge channel,
the quantum action will contain a purely imaginary term (see Section II):
\begin{equation}
\frac{i}{2}\int_0^\beta d\tau\sum_i (n_{i\uparrow} (\tau)+n_{i\downarrow}(\tau))
\partial_\tau\phi_i
~~,
\label{berry}
\end{equation}
where $n_{i\sigma}\equiv \bar\psi_{i\sigma}\psi_{i\sigma}$.
It is useful to split the electron density into its average and
the fluctuating parts: $n=n_\uparrow +n_\downarrow\to\bar n +\delta n$.
The average part is unimportant for the spin-wave phase due to the
periodicity of $\phi_i(\tau)$ in the interval $[0,\beta]$. In the
vortex part of the phase, however, this 
average density acts as a {\em magnetic flux} 
$\half\bar n$ seen by (anti)vortices \cite{fisherlee}.
This first time derivative is just the charge
Berry phase and it couples to (half of) the total 
electronic charge density. After the fermions are integrated out,
the fluctuating part of the density will simply generate
a $\dot\phi_i^2$ term whose stiffness is set by the fermionic
compressibility. 

For simplicity, we will first ignore the charge Berry phase
and set $\bar n = 0~({\rm mod}~2)$. This results in a (2+1)D XY model:
\begin{equation}
{\cal L}_{XY} = \frac{1}{2}K_0\sum_i\dot\phi_i^2 -
J\sum_{nn}\cos(\phi_i - \phi_j)~~,
\label{lxy} 
\end{equation}
where $K_0$ is effectively the fermionic charge 
susceptibility (compressibility):
$K_0\sim \chi_c\sim\langle\delta n_i^2\rangle$. 
Again, we expand the cosine to the leading order (the same results are
obtained in the Villain approximation, see Appendix) 
and separate the regular and singular
(vortex) contributions to 
$\partial_\mu \phi\to \partial_\mu\chi +(\partial_\mu\varphi)_v$
As a result,  ${\cal L}_{XY}$ is transformed into
\begin{equation}
\frac{1}{2}K_\mu (\partial_\mu\phi)^2\to 
iW_\mu\partial_\mu\chi + iW_\mu(\partial_\mu\phi )_v + \frac{1}{2}K_\mu^{-1}W_\mu^2~~,
\label{lagrangian} 
\end{equation}
where $K_\mu=(K_0,K,K)$ and 
$W$ is a real  Hubbard-Stratonovich three-vector field.
The integral over the free field $\chi$ can be carried out, producing the
constraint $\partial\cdot W=0$. The constraint is solved by demanding
that $W=\partial\times A_d$, where $A_d$ will soon assume the meaning of
the dual gauge field. The vortex part is now manipulated into:
\begin{multline}
{\cal L}_{XY} \to i(\partial\times A_d)\cdot (\partial\phi )_v 
+ \frac{1}{2}K_\mu^{-1}(\partial\times A_d)_\mu^2 \\
\to
-2\pi iA_d\cdot n + \frac{1}{2}K_\mu^{-1}(\partial\times A_d)_\mu^2~~,
\label{lagrangianv} 
\end{multline}
where the integration by parts and $\partial\times(\partial\phi)_v=2\pi n$ have
been used and $n_\mu$ is the vorticity in the (2+1) dimensional spacetime
(vorticity, by its very nature, is conserved: $\partial\cdot n=0$).
Now, we use the standard transition from Feynman path integrals to
coherent functional integrals \cite{kleinert}. 
A relativistic vortex boson complex field
$\Phi (x)$ is introduced, whose worldlines in (2+1) dimensional spacetime
coincide with fluctuating vortex loops (see the Appendix for a more
detailed account). The first term in (\ref{lagrangianv}) is
just the current of these relativistic vortex particles coupled
to a vector potential $A_d$. Furthermore, vortices have some intrinsic line
action $\int ds {\cal S}_0$, coming from core terms (and/or 
lattice regularization) which supply the (2+1)
kinetic term. In the end, one obtains a dual Lagrangian:
\begin{equation}
{\cal L}_d = |(\partial +i2\pi A_d)\Phi |^2 + m^2(\br)|\Phi |^2 +
\frac{g}{2}|\Phi |^4 
+ \frac{K_\mu^{-1}}{2}(\partial\times A_d)_\mu^2,
\label{duallagrangian} 
\end{equation}
where $g>0$ is a short range repulsion describing
the penalty for vortex core overlap. 
The mass term will in general
be spatially modulated, reflecting the underlying lattice
potential, as in (\ref{hpotential}). A detailed derivation of
the above dual Lagrangian can be found in the Appendix.

It is useful to underscore the following relation between
the steps that led to (\ref{duallagrangian}) and
the formalism discussed in Section II, which is coached
in the language closer to the original electrons: 
Only the familiar long range Biot-Savart
(anti)vortex interactions, mediated by dual gauge field
$A_d$, arise from the fermionic action in Section II and 
Refs. \onlinecite{preprint, footswave}. This corresponds to large regions of space
where the pseudogap $\Delta$ is large and approximately
uniform. In contrast, small (anti)vortex core regions, where
$\Delta$ might exhibit significant non-uniformity, supply the
the core kinetic part (\ref{vortexmass}), vortex
core energy and the short range core
repulsion term ($g$), which are all stored in the ``Jacobian'' for $v$ and
$a$ gauge fields. This is significant since it enables us to deal
rather straightforwardly with the charge Berry phase which
now must be restored. 

First, notice that the self-action of $A_d$,
which was introduced in this section through a Hubbard-Stratonovich
decoupling in Eq. (\ref{lagrangian}), actually follows ``microscopically''
from the integration over the fermions and the gauge field
$v$ in Section II and \cite{preprint}. There $A_d$ was introduced as a field
enforcing a $\delta$-function constraint
$\delta [(\partial\times v)/\pi-n]$ but the physics is
precisely the same, the formal difference being just the order
in which we integrate the fermionic matter.
Once we restore the Berry phase term 
$(i/2)\int d^3x\bar n(\partial_\tau\phi)_v$ back to the action,
we note that in the formalism of Section II and \cite{preprint} none of the
fields in the dual Lagrangian (\ref{duallagrangian}) couples
to it directly. Instead, it is $v_0$ that enters the Berry phase
via the  $\delta$-function constraint $\partial_\tau\phi\to v_0$.
The Berry phase term will then affect (\ref{duallagrangian})
via the coupling of $v_0$ to the spatial part of $A_d$, i.e., the
dual magnetic induction ${\bf B}_d=\nabla\times {\bf A}_d$.

The above observations suggest that
it useful to separate out the saddle point part of
$v$ as: $v\to -i\bar v +\delta v$. The saddle point part $\bar v$
is determined by minimizing the total action (the ``$-$'' sign
is chosen so that $\bar v_0$ couples to the electron density as a
chemical potential). Of course, if there were no Berry phase
term we would have $\bar v=0$. With the Berry phase term
included ${\bf\bar v}=0$ but $\bar v_0$ is now finite. 
Observe from Eqs. (1,2) of Ref. \onlinecite{preprint}
that the saddle point equation for $\bar v_0$ simply
reduces to: $\half\bar n= {\bf B}_d=\nabla\times {\bf A}_d$ 
(note that, relative to our notation, dual gauge field in
Ref. \onlinecite{preprint} is rescaled as $A_d\to 2\pi A_d$). So,
with the Berry phase included, the final form of the dual theory
still remains given by Eq. (\ref{duallagrangian}) but must
be appended by the constraint 
$\half\langle n(x)\rangle= \langle {\bf B}_d (x)\rangle$.

If we now apply the dual mean-field approximation to (\ref{lagrangian})
one obtains ${\cal F}_{mf}$ given by:
\begin{eqnarray}
|(\nabla +i2\pi{\bf A}_d)\Phi |^2 + m^2(\br)|\Phi |^2 +
\frac{g}{2}|\Phi |^4 + \frac{K^{-1}_0}{2}{\bf B}_d^2~~,
\label{dualmeanfield} 
\end{eqnarray}
with the constraint taking the form 
${\bf B}_d(\br)=\half\langle n(\br)\rangle$.
One immediately recognizes (\ref{dualmeanfield}) as
the Abrikosov-Hofstadter problem for a dual type-II superconductor --
we are in type-II regime \cite{abrikosov} since small compressibility implies
large dual Ginzburg parameter 
$\kappa_d\sim 1/\sqrt{K_0}\sim 1/\sqrt{\chi_c}$ --
subjected to the overall (dual) magnetic field 
flux per plaquette of the dual lattice, $f$, given by
$f=p/q=(1-x)/2$, where $x$ is doping.
Note that at half-filling $\bar n = 1 \Rightarrow f=1/2$. 
The solution that minimizes
(\ref{dualmeanfield}) at half-filling is a checkerboard array
of vortices and antivortices in $\Phi$, with the associated
modulation in ${\bf B}_d(\br)=\half\langle n(\br)\rangle$, as discussed
in Refs. \onlinecite{preprint, footswave}. This is nothing but the
dual equivalent of a CPCDW in an $s$-wave superconductor at half-filling,
with a two-fold degenerate array of alternating enhanced and suppressed 
charge densities on sites of the original lattice. An overwhelming
analytic and numerical support exists for this state being the actual
ground state of the negative $U$ Hubbard model \cite{singer}, 
the prototypical theoretical toy model in this category.
This gives us confidence that the mean-filed approximation captures
basic features of the problem.

\subsection{Quantum fluctuating $d$-wave superconductor}

The above pedagogical exercise applies to an idealized strongly fluctuating
$s$-wave superconductor and its associated ordinary (2+1)D XY model. 
What about our $d$-wave
lattice superconductor (LdSC) and its effective XY-type model
(\ref{hxy}), with the bond
phases mapped into site phases on the blue lattice? 
Again, we go from $\exp(i\theta_{ij})$
to $\exp(i\varphi_k)$ as before. The cosine part of (\ref{lxy}) is
replaced by ${\cal H}_{XY}^d$ (\ref{hxy}) and is handled in the same way as 
for the classical case. This will ultimately
result in a modulated potential $V(\br)$ of Eq. (\ref{hpotential}).
In the context of the dual theory (\ref{duallagrangian}) this
will translate into a position-dependent mass term $m^2(\br)|\Phi (x)|^2$
in the dual Lagrangian (\ref{duallagrangian}) except now
this mass term has the checkerboard symmetry on the blue
lattice, with $m_r^2$ in the red plaquettes different from
$m_b^2$ in the black plaquettes ($m_r^2-m_b^2\sim E_c^r-E_c^b$). 
Including the constraint on the overall dual induction ${\bf B}_d$
seems to complete the process.

Alas, the situation is a bit more involved. First,
the time derivative part and the
Berry phase are more complicated. The difficulty arises
since the {\em bond} superconducting phase $\theta_{ij}$ of a LdSC
couples in a more complicated way to the {\em site} fermionic variables. 
However, we can still deal with this
by enlisting the help of 
the FT gauge transformation which screens the long-range
part of $\theta_{ij}$ by $\half\phi_i +\half\phi_j$, where
$\phi_i$ are the phases in the fermionic fields. After this
transformation (we are again ignoring $a_\mu$ which we can put back in
at the end) one obtains 
$i\half n_i\partial_\tau \phi_i$ at each site of the CuO$_2$
lattice. This translates into 
$\frac{i}{8}n_i\partial_\tau \phi_i+\frac{i}{8}n_j\partial_\tau \phi_j$ for
each bond $\langle ij\rangle$ of the CuO$_2$ lattice. 
This bond expression can be rewritten as:
\begin{multline}
\frac{i}{8}[n_i+n_j]\half(\partial_\tau \phi_i+\partial_\tau \phi_j)+
\frac{i}{8}[n_i-n_j]\half(\partial_\tau \phi_i-\partial_\tau \phi_j)\\
\cong \frac{i}{8}[n_i+n_j]\partial_\tau\theta_{ij} + (\cdots)~~,
\label{dberry}
\end{multline}
where $(\cdots)$ contains higher order derivatives 
and is typically not important
in the discussion of critical phenomena (but see below).
In the end, following our replacement of bond phases
on CuO$_2$ lattice with site phases on the blue lattice
$\theta_{ij}\to\varphi_k$ we finally obtain the Berry phase term of
our lattice $d$-wave superconductor: 
\begin{equation}
\frac{i}{8}[n_i+n_j]\partial_\tau\varphi_k \to
\frac{i}{8}[\bar n_i+\bar n_j]\partial_\tau\varphi_k + \frac{i}{8}\delta\Delta_{ij}\partial_\tau\varphi_k~~.
\label{dberryi}
\end{equation}

The expression (\ref{dberry}) seems too good to be true and it is --
the replacement $\theta_{ij} \to \half\phi_i +\half\phi_j$ holds only
far away from vortices, when the phases change slowly between nearby
bonds (sites). In general, the bond phase of the LdSC will not couple
to the electron densities in a simple way suggested by (\ref{dberry}). Still,
we have gained an important insight; its coupling to the {\em overall}
electron density represented by the leading term in (\ref{dberryi})
is effectively exact. This follows directly from the electrodynamic
gauge invariance which mandates that both regular 
and singular configurations of the
phase must have the same first time derivative in the action. For
arbitrarily smoothly varying phase, with no vortices present, the 
replacement $\theta_{ij} \to \half\phi_i +\half\phi_j$ is accurate
to any desired degree and the Berry phase given by the leading term
in (\ref{dberryi}) follows. Incidentally, this result is not spoiled
by the higher order terms, represented by $(\cdots)$ in (\ref{dberry}),
since they do not contribute to the {\em overall} 
dual magnetic field, by virtue of $\bar n_i -\bar n_j=0$. 
We, however, cannot claim with the same degree of certainty that
$\delta\Delta_{ij}$ is in a simple relationship to variations in
the electronic densities on sites $i$ and $j$ as (\ref{dberry}) would
have it \cite{footdeltadelta}. Instead, we should
view $\delta\Delta_{ij}$ in (\ref{dberryi})
as a Hubbard-Stratonovich field used to decouple 
the $\dot\theta_{ij}^2$ term in the quantum action. 
This makes $\delta\Delta_{ij}$ inherently 
a {\em bond} field whose spatial modulation
translates into variation in the pseudogap $\Delta$ and 
will be naturally related to the dual 
induction ${\bf B}_d$ below \cite{footdeltabdeltadelta}.
This is different from our pedagogical $s$-wave exercise where
the modulation in ${\bf B}_d$ was directly related to the variation
in the electronic density.

The second source of complication is more intricate and
relates to the fact that, as $T\to 0$,
we must be concerned about nodal fermions. In the $s$-wave case
and the finite temperature $d$-wave XY-type model (\ref{hxy}) we know 
that coupling constants $J$, $J_1$, $J_2$, and so on, are finite or at
least can be made so in simple, reasonably realistic models.
In general, we need to keep only several of the nearby neighbor
couplings to capture the basic physics. There
is a tacit assumption that such expansion is in fact well-behaved. In the
quantum $d$-wave case, however, the expansion in near neighbor
couplings is not possible -- at $T=0$ its presumed analytic structure
is obliterated by gapless nodal fermions. 
Naturally, after the electrons are integrated out we will
still be left with some effective action for the phase degrees
of freedom $\theta_{ij}(\tau)$ but this action will be both
non-local and non-analytic in terms of differences of phases on
various bonds. The only option appears to be to keep the gapless
fermionic excitations in the theory. This in itself of course is perfectly
fine but it does not advance our stated goal of ``bosonizing'' the 
CPCDW problem a least bit. Note that the issue here is not
the Berry gauge field $a$ and its coupling to nodal fermions -- even when we
ignore coupling of vortices to spin by setting $a\to 0$ 
and concentrate solely on the charge sector as we have done so far, the 
gapless BdG quasiparticles still produce
non-analytic contributions to the phase-only action.
In short, what one has here is more than a problem; it is a calamity. 
The implication is nothing less but that there is 
{\em no} usable XY-type model for the charge sector of a quantum $d$-wave
superconductor, the optimistic title of the present paper notwithstanding.

Fortunately, there is a way out of this conundrum. 
While indeed we cannot write down a simple XY-like model
for the fluctuating phases it turns out that the dual
version of the theory can be modified in a relatively 
straightforward way to incorporate the non-analyticity caused
by nodal fermions. To see how this is accomplished we go
back to Section II and use our division of electronic degrees of freedom into
low-energy nodal ($\psi_{\sigma,\alpha}$) and ``high-energy''
anti-nodal fermions, tightly bound into spin-singlet
Cooper pairs ($\psi_{\sigma,\langle\alpha\beta\rangle}$).
Imagine for a moment that we ignore nodal fermions 
altogether, by, for example, setting
the number of Dirac nodal flavors $N\to 0$ (in a single CuO$_2$ plane
$N=2$). Exact integration over anti-nodal electrons
$c_{\sigma,\langle\alpha\beta\rangle}$
leads to an XY-type model precisely of the type discussed at the
beginning of this subsection. Since anti-nodal electrons are fully
gapped the effective Hamiltonian has the conjectured form (\ref{hxy})
with {\em finite} couplings $J$, $J_1$ and $J_2$, and with higher
order terms which decay sufficiently rapidly. The explicit
values of $J$, $J_1$ and $J_2$ at $T=0$ might be difficult to determine
but not more so than for an $s$-wave superconductor.
Furthermore, since anti-nodal fermions carry all the charge density,
the Berry phase term remains as determined earlier and so does the
timelike stiffness of $\varphi_i$. Derivation of the dual theory proceeds
as envisioned, with all the short range effects stored in vortex
core terms and ultimately in $m^2(\br)$,
and with only remaining long range interactions mediated by
the dual gauge field $A_d$. We now restore 
nodal fermions $(N\not =0)$; as already
emphasized, the only coupling of nodal fermions to the dual theory is
via the gauge fields $v$ and $a$. Since we are ignoring the spin channel
we can neglect $a$. As far as $v$ is concerned, including nodal fermions
leads to a non-local, non-analytic correction to its stiffness, expressed as 
$$
v_\mu K_\mu v_\mu\to 
v_\mu(K_\mu\delta_{\mu\nu} + NQ_{\mu\nu})v_\nu~~,
$$
where $K_\mu$ is determined by anti-nodal electrons.
$Q_{\mu\nu}(q)$ is the contribution from nodal fermions,
linear in $q$ and in general quite complicated.
We give here few simpler limits: $Q_{00}(0,\omega_n)=c|\omega_n|$,
$Q_{0i}=Q_{i0}=0$, $Q_{xx}(\bq,0)=Q_{yy}(\bq,0)=-cq_>$,
$Q_{xy}(\bq,0)=Q_{yx}(\bq,0)= -cq_<{\rm sgn}(q_x,q_y)$, where $q_>={\rm max}(|q_x|,|q_y|)$,
$q_<={\rm min}(|q_x|,|q_y|)$ and $c\sim \pi^2/16\sqrt{2}$ \cite{franzaffleck}.

In the language of dual theory, this translates into a modified
self-action for the dual gauge field:
\begin{multline}
\half(\partial\times A_d)_\mu K^{-1}_\mu(\partial\times A_d)_\mu \\
\to
\half(\partial\times A_d)_\mu[K_\mu\delta_{\mu\nu} + NQ_{\mu\nu}]^{-1}(\partial\times A_d)_\nu
~~,
\label{nodalselfaction}
\end{multline}
where $[K_\mu\delta_{\mu\nu} + NQ_{\mu\nu}]^{-1}$ is the tensor inverse
of $K_\mu\delta_{\mu\nu} + NQ_{\mu\nu}$. 
This action is clearly more complicated 
than its Maxwellian $s$-wave counterpart
but nevertheless decidedly manageable -- most importantly, 
the part of the dual action involving dual boson field $\Phi$ remains
{\em unaffected}.
Note that nodal fermions induce
{\em subleading} but still long-ranged interactions between vortices,
in addition to the familiar Biot-Savart interactions. These interactions
have square symmetry on the blue lattice, reflecting their nodal origin.
Furthermore, when $A_d$ is integrated over,
they will produce a peculiar dissipative term 
$\sim |\omega_n|v_0^2$ which describes the damping of 
collective quantum vorticity fluctuations by gapless 
nodal quasiparticles. The importance of this effect is 
secondary relative to the mass term for $v_0$ which is 
always generated by anti-nodal
fermions, basically because the density of states $N(E)$ 
for nodal fermions vanishes as $E\to 0$. However, given the smallness
of traditional Bardeen-Stephen core viscosity this ``nodal'' mechanism 
is an important source of dissipation of vortex motion
in underdoped cuprates. Since our focus in the present paper is 
dual mean-field theory, this dissipative term will play no direct role.

Armed with the above analysis we can finally write down the Lagrangian
of the quantum XY-type model describing a LdSC:
\begin{multline}
{\cal L}_{XY}^d = i\sum_if_i\dot\varphi_i+ \frac{K_0}{2}\sum_i\dot\varphi_i^2 
+{\cal H}_{XY}^d + {\cal L}_{\rm nodal} + {\cal L}_{\rm core}\\=
i\sum_if_i\dot\varphi_i+ \frac{K_0}{2}\sum_i\dot\varphi_i^2 
-J\sum_{nn}\cos(\varphi_i - \varphi_j)\\ 
-J_1\sum_{rnnn}\cos(\varphi_i - \varphi_j)
-J_2\sum_{bnnn}\cos(\varphi_i - \varphi_j)\\ 
+ {\cal L}_{\rm nodal}[\cos(\varphi_i - \varphi_j)] +{\cal L}_{\rm core}~~,
\label{lxyd} 
\end{multline}
where the sums run over sites $i$ of the blue lattice and
the notation is the same as below Eq. (\ref{hxy}),
$f_i=\frac{1}{8}(\bar n_k+\bar n_j)$ with $k,j$ being the end sites
of bond $i$, the time-like phase stiffness $K_0$ 
results from the Hubbard-Stratonovich
integration over $\delta\Delta_{ij}$ as 
detailed above, and ${\cal L}_{\rm core}$ contains core terms
coming from a small region around the (anti)vortex where $\Delta$ itself
is significantly suppressed, an example being (\ref{vortexmass}). 
The explicit form of ${\cal L}_{\rm nodal}[\cos(\varphi_i - \varphi_j)]$ 
in the XY model language is unknown but it is a non-analytic,
non-local functional of $\cos(\varphi_i - \varphi_j)$; 
its effect will be 
incorporated once we arrive at the dual description, following
the above arguments.
Note that the short range parts of cosine functions and time derivative in
(\ref{lxyd}) will be subsumed into ${\cal L}_{\rm core}$ once we go
to continuum or Villain representations of the problem, as described
elsewhere in the paper. Furthermore, observe that at half-filling
the dual flux per plaquette of the blue lattice has an {\em intrinsic}
checkerboard pattern: 
$f=1/2$ for black plaquettes and $f=0$ for red plaquettes.
This is a direct consequence of $\varphi(\tau)$ being
a bond phase, residing on the blue lattice sites in Fig. \ref{fig1},
its Berry phase given by Eq. (\ref{dberryi}).
Thus, in the quantum problem, the Berry phase {\em combines} with the
next-nearest neighbor bonds (for $J_1\not =J_2$) in breaking translational
symmetry of the blue lattice down to the checkerboard pattern of 
Fig. \ref{fig1}. The average flux through the unit cell 
on the blue lattice is $f=1/2$ which
is just as it should be since there are two plaquettes
of the blue lattice per single plaquette of the original
black lattice and the dual flux per plaquette
of the CuO$_2$ lattice is $f=1/2$ at half-filling. 

It is now straightforward to derive a dual representation of
(\ref{lxyd}) by retracing the path that led from (\ref{lxy}) to
(\ref{duallagrangian}). In this fashion we obtain the
dual Lagrangian of the (2+1)D XY model appropriate for a
strongly fluctuating $d$-wave
superconductor near half-filling \cite{footnospin}:
\begin{multline}
{\cal L}_d = |(\partial +i2\pi A_d)\Phi |^2 + m^2(\br)|\Phi |^2 +
\frac{g}{2}|\Phi |^4 \\+
\half(\partial\times A_d)_\mu[K_\mu\delta_{\mu\nu} + NQ_{\mu\nu}]^{-1}(\partial\times A_d)_\nu
+ (\cdots)~~,
\label{dduallagrangian} 
\end{multline}
where $\partial\times A_d $ satisfies the constraint 
arising from the charge Berry phase.
The dual Lagrangian for a $d$-wave
superconductor (\ref{dduallagrangian}) differs from its
$s$-wave counterpart (\ref{duallagrangian})
in the following important respects:
{\em i}) the periodicity of the mass potential is different and reflects
the checkerboard periodicity on the (dual) blue
lattice set by our vortex core potential $V(\br)$, {\em ii})
the self-action for the dual gauge field contains non-local,
non-analytic contribution of nodal fermions, {\em iii}) 
the modulation of the dual flux ${\bf B}_d$ relates to the variation
in the pseudogap $\delta\Delta_{ij}$ (as opposed to the electronic
density $\delta n_i$) via the constraint
$\delta{\bf B}_d(\br)\sim (1/16)\langle \delta\Delta_{ij}(\br)\rangle $, 
and {\em iv}) at doping $x$,
the overall dual flux through the unit cell
of the blue lattice imposed by the Berry phase constraint
is $f=p/q=(1-x)/2$, comprised of $f=p/q=(1-x)/2$ at
a black plaquette and $f=0$ at a red one. 
We therefore expect the results for stable CPCDW states to be different
from those of a hypothetical strongly
fluctuating  $s$-wave superconductor. Also, note that $K_0$ is now set by
$K_0\sim \langle |\delta\Delta_{ij}|^2\rangle$, which is
still related to fermion compressibility
\cite{footdeltadelta} and still relatively small since the basic aspect
of our approach is that amplitude fluctuations in $\Delta_{ij}$ are
suppressed. This implies our dual superconductor remains in the
type-II regime ($\kappa_d\sim 1/\sqrt{K_0}\gg 1$). 
Finally, $(\cdots)$ denotes higher order terms which have been neglected.
For those readers who find the above road to (\ref{dduallagrangian})
perhaps a bit too slick, we give a detailed step-by-step derivation
in the Appendix.

An important point should be noted here: 
The neglected higher other terms $(\cdots )$ involve 
higher order kinetic terms, additional
intermediate range interactions, short range current-current interactions
and numerous other contributions. All are irrelevant in the sense of
critical behavior. In our problem, however, we are not interested
only in the critical behavior. In particular, we would like to
determine the charge modulation of the CPCDW on lengthscales which are not
extremely long compared to the CuO$_2 $ lattice constant. These
additional terms could be important for this purpose. 
Particularly significant in this respect are the higher 
order corrections to the Berry
phase (\ref{dberry}) which, while not affecting the value of overall $f$,
do influence the form of the effective dual Abrikosov-Hofstadter problem
that ensues.

The strategy of keeping a large number of otherwise irrelevant
terms is not a practical one. We will therefore introduce a simplification
here which is actually quite natural and allows us to retain
the essential realistic features of the original model and maintain
the particle-hole symmetry as well. Consider a situation
where $m^2(\br)$ in (\ref{dduallagrangian}) is very
strongly modulated. This is a ``tight-binding'' limit, universally
considered appropriate for cuprates, and
we can simply view quantum (anti)vortices as being able
to hop only between the nearby plaquettes of the
blue lattice, as is clearly implicit in (\ref{lxyd}). Furthermore, we 
assume that the two extrema in $m^2(\br)$ are 
separated by an energy scale larger than such hopping. This
is a perfectly reasonable assumption since it illustrates
an already important characteristic of our effective model,
the fact that red and black plaquettes of the blue lattice
are intrinsically not equivalent (by the form of the Berry phase and
$J_1\not =J_2$). Under these
circumstances we can rewrite the dual Lagrangian in a tight-binding form:
\begin{multline}
{\cal L}_d =
\sum_r|(\partial_\tau +i2\pi A_{d0})\Phi_r |^2 + 
\sum_b|(\partial_\tau +i2\pi A_{d0})\Phi_b |^2 \\
-\sum_{\langle rr'\rangle}t_{rr}\exp(i2\pi\int_r^{r'} d{\bf s}\cdot{\bf A}_d)\Phi_r^*\Phi_{r'} \\
-\sum_{\langle rb\rangle}t_{rb}\exp(i2\pi\int_r^b d{\bf s}\cdot{\bf A}_d)\Phi_r^*\Phi_b - ({\rm c.c.}) \\
+\sum_r(m^2_r|\Phi_r |^2+\frac{g}{2}|\Phi_r |^4)
 +\sum_b(m^2_b|\Phi_b|^2 + \frac{g}{2}|\Phi_b |^4)  \\
+\half(\partial\times A_d)_\mu[K_\mu\delta_{\mu\nu} + NQ_{\mu\nu}]^{-1}(\partial\times A_d)_\nu
+ (\cdots)~~,
\label{ltightbinding} 
\end{multline}
where vortex boson 
fields $\Phi_{r(b)}$ reside on red (black) plaquettes of the
blue lattice, $t_{rr(b)}$ is the vortex hopping between 
the nearest red-red(black) neighbors, 
$\exp(i2\pi\int_r^{r'(b)} d{\bf s}\cdot{\bf A}_d)$ are the appropriate Peierls
factors, $|m^2_r-m^2_b|\gg t_{rb}$ (core energies on red and
black plaquettes are significantly different), and $\nabla\times {\bf A}_d$
in the last term is the lattice flux of the dual magnetic field,
equal $f=p/q=(1-x)/2$ per unit cell of the original CuO$_2$ lattice.
We have also assumed that it is the red plaquettes that are favored by
vortex cores, making it unnecessary to include  $t_{bb}$ explicitly.
The  assumption $E_r\ll E_b$ appears naturally warranted by 
the overall symmetry of the problem
but, should the details of microscopic physics intervene and 
reverse the situation in favor
of the black plaquettes, all one needs to do is exchange labels
$r\leftrightarrow b$ (note that $m^2_r<0$, $m^2_b>0$ in a dual superconductor).

The resulting tight-binding dual Hamiltonian:
\begin{eqnarray}
{\cal H}_d &=&
-\sum_{\langle rr'\rangle}t_{rr}\exp(i2\pi\int_r^{r'} d{\bf s}\cdot{\bf A}_d)\Phi_r^*\Phi_{r'} \nonumber \\
&-&\sum_{\langle rb\rangle}t_{rb}\exp(i2\pi\int_r^b d{\bf s}\cdot{\bf A}_d)\Phi_r^*\Phi_b - ({\rm c.c.})
+ \ \ \ \ \nonumber \\
&+&\sum_r(m^2_r|\Phi_r |^2+\frac{g}{2}|\Phi_r |^4) \nonumber \\
&+& \sum_b(m^2_b|\Phi_b |^2+\frac{g}{2}|\Phi_b |^4)+(\cdots)
\label{hofstadter}
\end{eqnarray}
will be analyzed in two limits: 
for $t_{rr}\gg t_{rb}^2/ |m^2_r-m^2_b|$, in which case we can simply
set $t_{rb}\to 0$, the
effective Hofstadter problem assumes the form equivalent to the standard
$s$-wave case defined on the sites of the red lattice (Fig. \ref{fig1}). Importantly,
however, the relation between the modulation 
in ${\bf B}_d$ and $\delta\Delta_{ij}$ remains
{\em different} from the $s$-wave case and peculiar
to a $d$-wave, as explained above. We 
will call this limit an H1 model.
Similarly, in the opposite case  $t_{rr}\ll t_{rb}^2/ |m^2_r-m^2_b|$
we can set $t_{rr}\to 0$ and obtain an effective Hofstadter problem
with hoppings between red and black sites only. Note that this 
situation is {\em not} equivalent to the standard
$s$-wave case: In order to hop from one red plaquette to another a vortex
must go through a black site, picking up a Peierls phase factor
different from the one for a 
conventional direct red-to-red hop. In considering these Peierls factors
we must exercise caution since the ``assisted'' hops between red
plaquettes pass directly through the dual fluxes $f=p/q=(1-x)/2$
located on black sites.
By infinitesimally  displacing the said flux, one is back
to the situation where all closed paths of hops are composed of
fluxes $f=p/q=(1-x)/2$ through black and $f=0$ through red plaquettes
of the blue lattice.
The resulting Hamiltonian has an exact particle-hole symmetry around
half-filling ($x\to -x$), as it should. We call this the H2 model.
Finally, still in the limit $t_{rr}\ll t_{rb}^2/ |m^2_r-m^2_b|$ with
 $t_{rr}\to 0$, we can ``spread'' the dual flux so it is uniformly
distributed throughout each blue lattice plaquette and 
equal $f=p/q=(1-x)/4$.
In this situation, dubbed an H3 model, the $x\to -x$ symmetry is obeyed only 
approximately, for small $x$ (near
half-filling), but this is all we are interested in. We have
established by explicit computations that the H3 model
satisfies the particle-hole symmetry at some
Hofstadter fractions $f$ while it does not appear to
do so at others; it is for this reason that we tend stay away
from the H3 model in this paper. The reader should note that
the issue of how to deal with dual fluxes when hopping through black
plaquettes arises only in tight-binding models H2 and H3 (but not H1!) --
the original dual Lagrangian (\ref{dduallagrangian}) is free of such
issues and has manifest $x\to -x$ symmetry. The down side, of course,
is that such a continuum theory is far more difficult to solve, both
analytically and numerically.

The Hamiltonian
(\ref{hofstadter}), in its three editions H1, H2, and (to a limited
extent!) H3,
defines probably the simplest dual version of the Hofstadter problem
which appropriately builds in the $d$-wave character of the fluctuating
lattice superconductor and the essential
phenomenology of the pseudogap state in
underdoped cuprates. The extra terms $(\cdots)$ can still be
used for fine-tuning (for example, $g\to g_r, g_b$, additional
$|\Phi _r|^2|\Phi_b|^2$ repulsions, etc.)
but the important particle-hole symmetry around half-filling
is already present without them. All the detailed numerical
calculations reported in the latter
pages of this paper and described in Ref. \onlinecite{preprint}
use the mean-field version of (\ref{hofstadter}) and the three models
based on it, H1 ($t_{rr}\gg t_{rb}^2/ |m^2_r-m^2_b|$) 
and H2  and H3 ($t_{rr}\ll t_{rb}^2/ |m^2_r-m^2_b|$), as the point
of departure.

\section{Dual superconductor and Cooper pair CDW in underdoped cuprates}

\begin{figure}[tbh]
\includegraphics[width=\columnwidth]{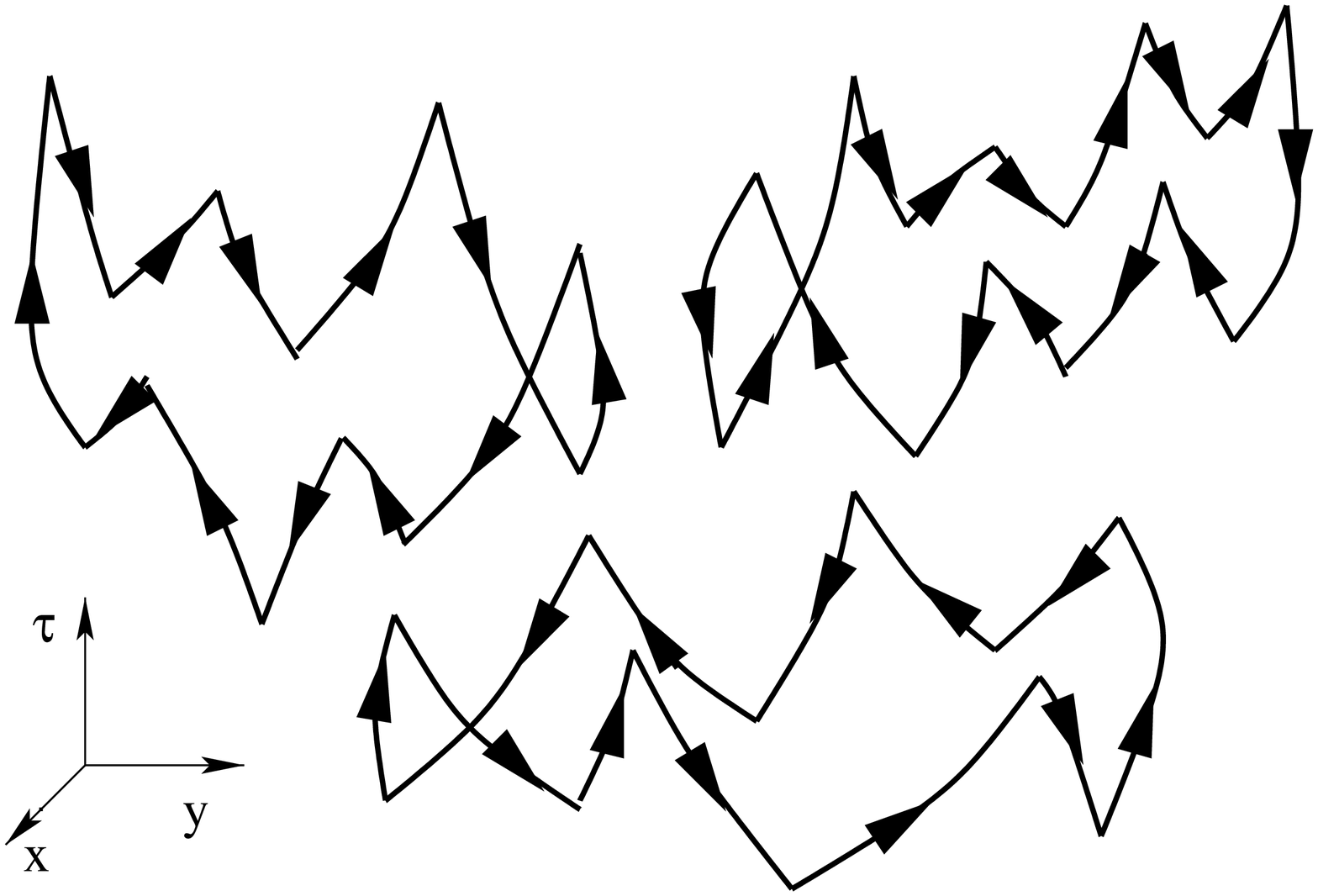}
\caption{\label{vortexloops}
Quantum fluctuations of vortex-antivortex pairs. $xy$-plane is the
CuO$_2$ layer and $\tau$-axis shows direction of imaginary time. Vortices
(arrows pointing upwards) and antivortices (arrows pointing downwards)
are always created and annihilated in pairs. Note that the structures
arising from linked creation-annihilation events form oriented loops
carrying  $\pm 1$ vorticity. These loops are just the virtual
particle-antiparticle creation and annihilation processes in the 
quantum vacuum of the relativistic bosonic field $\Phi(x)$, as described
in the text and the Appendix. $\Phi$ is our {\em dual} order parameter: In
a physical superconductor such vorticity loops 
are finite and  $\langle\Phi\rangle =0$. Note that the intersections of
such finite loops with the  $xy$-plane at any given time $\tau$ define a
set of bound vortex-antivortex dipoles in the CuO$_2$ layer.
The superconducting order is lost when vortex-antivortex pairs
unbind and the average size of the above 
vorticity loops diverges -- some of the loops become as large as
the system size. This is the pseudogap
state, with $\langle\Phi\rangle\not =0$.
The reader should note the following amusing aside: The above figure
can easily be adapted to depict the low energy {\em fermionic} excitations
of theory (\ref{lagrangiani}). The creation and annihilation processes
now describe spin up (arrows pointing upwards) and spin down 
(arrows pointing downwards) quasiparticle excitations from the
BCS-type spin-singlet
vacuum. The loops carry a well-defined spin and can be though
of as relativistic BdG Dirac particles/antiparticles -- their massless
character in a nodal $d$-wave superconductor
implies the presence of loops of arbitrarily large size. These fermionic
loops move in the background of fluctuating vortex loops discussed
above, their mutual interactions encoded in gauge fields $v$ and $a$.
This is a pictorial representation of the theory (\ref{lagrangiani}).
}
\end{figure}

The previous sections concentrated on the derivation of the
effective quantum XY-type model for phase fluctuations in
underdoped cuprates and its dual counterpart (\ref{dduallagrangian}) 
and (\ref{ltightbinding}).  In this section we pause to take stock of 
where we are with respect to the real world
and to consider some general features of the physical picture
that emerges from Eqs. (\ref{dduallagrangian},\ref{ltightbinding}).
First, the dual superconductor (\ref{dduallagrangian}) describes
the physics of strongly fluctuating superconductors in terms
of a complex bosonic field $\Phi$, which creates and 
annihilates quantum vortex-antivortex pairs viewed as ``charged''
relativistic particles (vortices) and antiparticles (antivortices) -- this
is depicted in Fig. \ref{vortexloops}.
The conserved ``charge'' is just the vorticity associated with these
topological defects, +1 for vortices and -1 for antivortices, and
the gauge field coupled to it is nothing but $A_d$. 
Physically, $A_d$ describes the familiar logarithmic interactions
between (anti)vortices. We stress that the dual description is just
a convenient mathematical tool:  Its main advantage 
is that it allows a theorist to access a strongly quantum
fluctuating regime of a superconductor, where the superfluid density
$\rho_s$ at $T=0$ can be very small or even vanish, while the pairing pseudogap
$\Delta$ remains large. According to the theory of Ref. \onlinecite{qed}
this is the regime that governs the properties of the
pseudogap state in underdoped cuprates. 
This regime is entirely inaccessible by more 
conventional theoretical approaches which use
the mean-field BCS state as the starting point around which to compute
fluctuation corrections. 

\begin{figure}[tbh]
\includegraphics[width=\columnwidth]{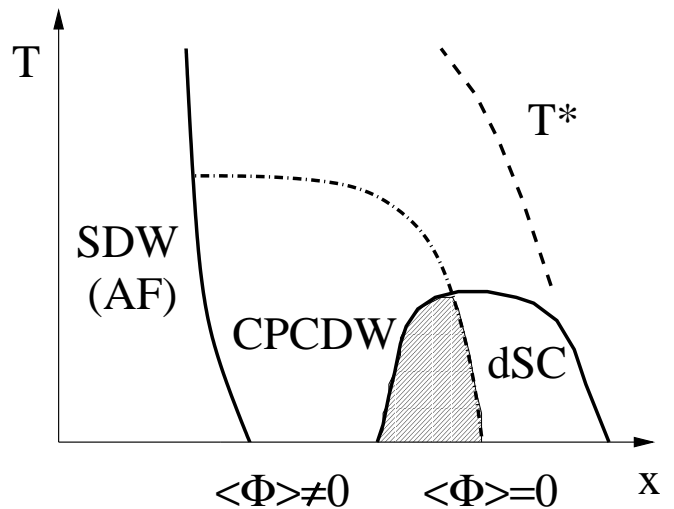}
\caption{\label{dphasediagram}
The schematic phase diagram of underdoped cuprates in the theory of Refs. 
\onlinecite{qed, preprint}. $T^*$ denotes the pseudogap temperature
($T^*\sim \Delta$). Dual superconducting order (finite $\langle\Phi\rangle $)
implies the absence of the true, physical superconductivity. Conversely,
the absence of dual order ($\langle\Phi\rangle =0$) corresponds to
the physical superconductor.
The shaded area represents the region of
coexistence between a strongly fluctuating superconductor and
a CPCDW state, which will generally occur in the dual theory. 
This region of ``supersolid'' behavior is
characterized by $\langle\Phi\rangle = 0$ but finite
$\langle|\Phi|^2\rangle$, which is modulated within the CuO$_2$ layer
according to our theory of a dual superconductor. The precise size
of a coexistence region, however, is difficult to estimate from
the mean-field theory and a more elaborate approach, including
fluctuations in $\Phi$ and $A_d$, needs to be employed. Finally,
the finite temperature phase boundary of the CPCDW (dashed-dotted line)
should not be taken quantitatively apart from the fact that it is located
below $T^*$.
}
\end{figure}

The dual superconductor description predicts
two basic phases of cuprates: $\langle\Phi\rangle =0$ is just 
the familiar superconducting state. Quantum and thermal vortex-antivortex
pair fluctuations are present (and thus $\langle|\Phi|^2\rangle\not =0$)
but these pairs are always bound, resulting in a finite, but considerably
suppressed superfluid density. As vortex-antivortex 
pairs unbind  at some doping $x=x_c$,
the superfluid density goes to zero and
the superconducting state is replaced by its dual counterpart,
$\langle\Phi\rangle \not =0$. The meaning of dual ODLRO is actually
quite simple: Finite $\langle\Phi\rangle$ means that vortex
loops (loops of vortex-antivortex pairs being created and annihilated
in (2+1) dimensional spacetime) can now extended over the 
whole sample, i.e. the worldline of a dual relativistic boson
can make its way from any point to infinity (see
Fig. \ref{vortexloops}). The presence of such
unbound vortex-antivortex excitations directly implies vanishing helicity
modulus and thus  the absence of the Meissner effect and $\rho_s = 0$
(see Section II for details). The phase diagram of a dual
superconductor as it applies to cuprates is shown in
Fig. \ref{dphasediagram}.

In the dual mean-field approximation, just as in a conventional
one, we ignore fluctuations in $\Phi$ and minimize the action
specified by (\ref{dduallagrangian})
with respect to a complex function $\Phi (\br)$. In a physical
superconductor ($x>x_c$) $\Phi (\br)=0$. For $x<x_c$, the minimum
action corresponds to finite  $\Phi (\br)$. However, since the
charge Berry phase translates into a dual magnetic field ${\bf B}_d$
this finite $\Phi (\br)$ must contain $N_d$ vortices, where
$N_d$ is the number of the dual flux quanta piercing the system.
Note that this number is nothing but half of the total number
of electrons $N_d=N/2$, the factor of one half
being due to the fact that we are considering Cooper pairs.
In a tight binding representation (\ref{ltightbinding}) this
implies a dual flux $f=p/q= (1-x)/2$ per each CuO$_2$ unit cell.
The presence of such vortex array in $\Phi (\br)$ will be accompanied
by spatial variation in $B_d(\br)$, which translates into the
modulation of the pseudogap $\delta\Delta_{ij}$ and thus into a
CDW of Cooper pairs (CPCDW). Consequently, in 
the vocabulary of the dual mean-field 
approximation, the question of
formation and the structure of the CPCDW is equivalent
to finding the Abrikosov vortex array on a tight binding
lattice, i.e. the Abrikosov-Hofstadter problem defined by
(\ref{ltightbinding},\ref{hofstadter}).
The formation of CPCDW results in a modulation
of  the local tunneling DOS and one is led to identify the ``electron crystal''
state observed in STM experiments \cite{yazdani,davis,kapitulnik}
 as the CPCDW. 

Our task is to determine the specific pattern in which
the vortices in $\Phi (\br)$ arrange themselves to minimize
the expectation value of the dual Hamiltonian (\ref{hofstadter}).  
Once this is known we can determine the modulation in the 
dual induction ${\bf B}_d$ and the structure of the CPCDW
follows from the ${\bf B}_d\leftrightarrow \delta\Delta_{ij}$ correspondence.
In this section we are interested in general qualitative results
and we therefore focus on the H1 model where such results are
more transparent. The H2 and H3 models turn out to be more opaque
and have to be studied numerically as soon
as one is away from half-filling. 

The solution of the Abrikosov-Hofstadter problem
is obtained as follows: one first sets $g=0$ in
(\ref{hofstadter}) and finds the ground state of the resulting quadratic
Hofstadter Hamiltonian ${\cal H}_d (g=0)$ given by
\begin{multline}
-\sum_{\langle rr'\rangle}t_{rr}
e^{2\pi i \int_r^{r'} d{\bf s}\cdot{\bf A}_d}
\Phi_r^*\Phi_{r'}
-\sum_{\langle rb\rangle}t_{rb}
e^{2\pi i \int_r^b d{\bf s}\cdot{\bf A}_d}
\Phi_r^*\Phi_b \\- ({\rm c.c.})
+\sum_rm^2_r|\Phi_r |^2 + \sum_bm^2_b|\Phi_b |^2~~.
\label{hofstadteri}
\end{multline}
At flux $f=p/q$ the ground state is $q$-fold degenerate
and we denote it by $\Phi^{(q)}(\br)$.
One then turns on finite $g$, forms a linear combination
of these degenerate states $\sum_q\alpha_q\Phi^{(q)}(\br)$, and
determines variationally the set of coefficients $\{\alpha_q\}$ 
which minimizes the full Hamiltonian (\ref{hofstadter}).
With $\{\alpha_q\}$ fixed in this fashion
the only remaining degeneracy in the ground state
consists of lattice translations and rotations. Once the ground state
$\Phi^{(0)}$ is found, the magnetic induction ${\bf B}_d$ follows from
Maxwell equation:
\begin{equation}
\frac{\delta{\cal L}_d}{\delta {\bf A}_d} = 
\frac{1}{\tilde K_0}\Delta\times \delta {\bf B}_d(\br) - {\bf
j}^\Phi=0~~,
\label{maxwell}
\end{equation}
where ${\cal L}_d$ is given by Eq. (\ref{ltightbinding}),  
$\delta {\bf B}_d = \Delta\times{\bf A}$, 
and
${\bf j}^\Phi$ is the current in the ground state of dual Hamiltonian
(\ref{hofstadteri}) with the {\em uniform} dual flux $f$.
All quantities in (\ref{maxwell}) are defined on the blue lattice of 
Fig. \ref{fig1}: $\Delta$ is a lattice derivative, ${\bf j}^\Phi$ 
and ${\bf A}_d$ are link variables, and ${\bf B}_d=\Delta\times{\bf A}_d$
is a site variable. The detailed definitions of all these objects
are given in the next section. The non-local, non-analytic 
self-energy for the dual gauge field
in (\ref{ltightbinding}) was replaced by an effective 
Maxwellian ($K_\mu\to \tilde K_\mu$) -- this approximation is
valid for weak modulation $\delta {\bf B}_d$. 
We should stress that this way of determining
the ground state $\Phi^{(0)}$ and dual induction ${\bf B}_d(\br)$ is valid only
if $\tilde K_0$ is sufficiently small so that the dual Ginzburg parameter
$\kappa_d\sim 1/\sqrt{\tilde K_0}$ is 
sufficiently large -- in underdoped cuprates
this is a justified assumption since strong Mott-Hubbard
correlations strongly suppress all charge density fluctuations.
At higher modulation, i.e. for intermediate values of $\kappa_d$,
the nodal contribution becomes more significant and its intrinsic
square symmetry (see discussion around Eq. (\ref{nodalselfaction}))
will act to orient $\delta {\bf B}_d$ relative to the blue lattice.
If this is the case,
the interplay between this ``nodal'' effect and the one arising from the
Abrikosov-Hofstadter problem itself can lead to interesting new patterns
for the CPCDW state; a detailed study of such an interplay is left for the
future. Finally, from $\delta{\bf B}_d (\br)\equiv {\bf B}_d (\br) -f$
we obtain $\delta\Delta_{ij}$ which can be fed back into the
electronic structure via the expressions given in Section II.

Notice that the above method of solving the problem
corresponds to the strongly type-II regime of a dual superconductor
($\kappa_d\gg 1$). In this limit, the CPCDW pattern is 
{\em primarily} given
by the dual supercurrent ${\bf j}^\Phi$ in the Maxwell equation
(\ref{maxwell}), itself determined by the solution to
the Abrikosov-Hofstadter problem (\ref{hofstadteri}). 
The modulation in $\delta {\bf B}_d$
only reflects this pattern of vortices
in ${\bf j}^\Phi$ (\ref{maxwell}), and the dual magnetic energy is
only a small fraction of the Abrikosov-Hofstadter condensation 
energy. Imagine now
that we ask the following question: What are the interactions
among vortices in $\Phi$ that have resulted in $\Phi^{(0)}$ being
the ground state of the Abrikosov-Hofstadter problem? This question
is analogous to the one inquiring about the interactions that have
led to the triangular lattice of vortices in the continuum version
of the problem, i.e. the interactions inherent in the famed
Abrikosov participation ratio $\beta_A$. In both cases, these
are far from pairwise and short-ranged -- they are in
fact intrinsically multi-body interactions, involving 
two-, three-, and multiple-body terms, all of comparable sizes
and all long-ranged \cite{read}.  They can be thought of
as the interactions among the center-of-mass coordinates of
Cooper pairs. This should be contrasted with the picture of
real-space pairs, interacting with some simple, pairwise and short-ranged
interactions. The pair density-wave patterns in this case are
determined not by the charge Berry phase and the
Abrikosov-Hofstadter problem but by the
Wigner-style crystallization. This is precisely the opposite limit
of the Maxwell equation (\ref{maxwell}), in which
it is the dual magnetic energy that determines the pattern rather than
simply reflecting the one set by the Abrikosov-Hofstadter
condensation energy, encoded in $\Phi^{(0)}$, and communicated 
by ${\bf j}^\Phi$. This is clearly seen in the Wigner crystal limit: consider
an array of real-space pairs fixed in their positions. Such particles
carry a unit dual flux and are completely {\em invisible} to 
vortices. As a result ${\bf j}^\Phi=0$ and (\ref{maxwell}) turns
into the minimization of the dual magnetic energy. In the
real-space pair limit this dual magnetic energy is nothing but
the original assumed interactions between the pair bosons:
$\half\sum_{ij}V(\br_i-\br_j)=\half\int d^2rd^2r' {\bf B}_d(\br)V(\br-\br'){\bf B}_d(\br')$, since ${\bf B}_d (\br)=\sum_i\delta(\br - \br_i)$, 
where $\{\br_i\}$ are the pairs' positions -- this 
is just the minimization of the
potential energy in the Wigner problem. Consequently,
the two descriptions, that of the Cooper pairs versus the
real-space pairs, correspond to the two {\em opposite} limits of
(\ref{maxwell}, \ref{hofstadteri}), the first to the strongly type-II,
the second to the strongly type-I limit of a dual superconductor.
The density-wave patterns associated with these two limits
are generally different and can be distinguished in experiments.

\begin{figure*}[tbh]
\includegraphics[width=0.8\textwidth]{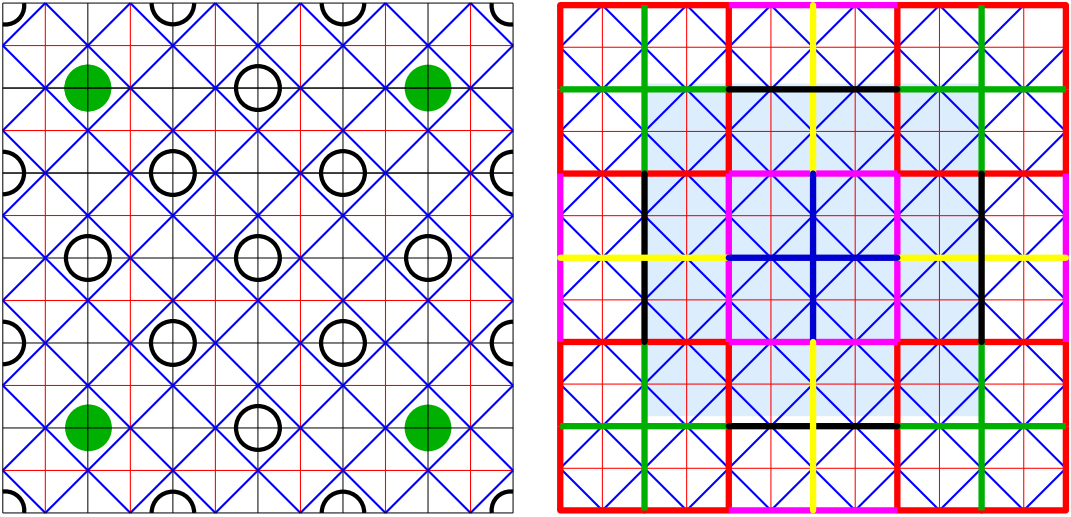}
\caption{\label{dvortexarray} 
Left panel: the circles depict the checkerboard array of vortices in 
$\Phi$ forming the ground state of Hamiltonian H1 at half-filling
($f=1/2$ per plaquette of the atomic lattice). The empty black plaquettes are 
actually occupied by dual antivortices, brought into
existence by a simple geometric constraint on the 
phase of $\Phi$ in such a checkerboard array. The full green
circles denote ``dual vortex holes'' \cite{qed}, i.e. the
dual vortices that are missing relative to the half-filling
checkerboard pattern once the system is doped to $x=1/8$ ($f=7/16$). 
The centers of these green circles form a square which defines
the $4\times 4$ elementary block of 7 dual vortices (black circles) per
16 sites of the CuO$_2$ lattice discussed in text.
Right panel: the most general distribution of modulations $\delta \Delta_{ij}$ 
consistent with the vortex pattern just  described. The green ``crosses'' correspond to 
``dual vortex holes''.  In general, there are six \cite{fiveindependent} phenomenological parameters
$\delta \Delta_{ij}$ shown in different colors. 
The number of inequivalent sites of the black lattice 
is also six (see Fig. \ref{ldos}).
}
\end{figure*}

\begin{figure*}[tbh]
\includegraphics[width=0.8\textwidth]{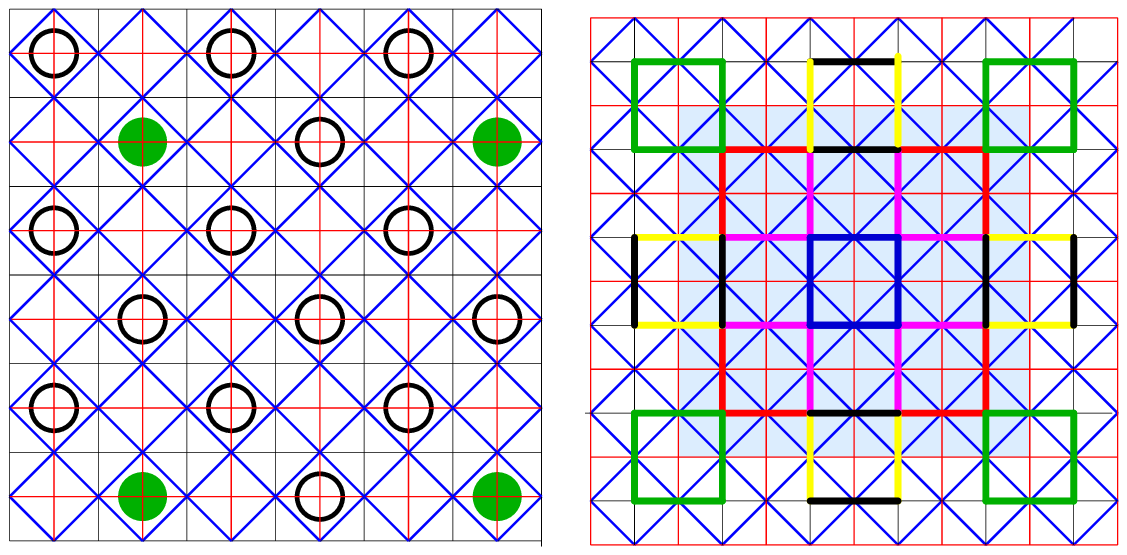}
\caption{\label{dvortexarray1}
The same as Fig. \ref{dvortexarray} except now the role of red and
black plaquettes is reversed. This corresponds to either
$E_c^b < E_c^r$ or the (anti)vortex core location at the
black plaquette being a {\em local} minimum of the vortex lattice
potential $V(\br)$. In the latter case, the 
pictured array would be a metastable
configuration of dual vortices at $x=0$ and $x=1/8$, 
ultimately unstable to the true ground 
state at those dopings depicted in Fig. \ref{dvortexarray}. 
The most general pattern of $\delta\Delta_{ij}$ is shown on the right. In this case, 
the ``dual vortex holes'' (solid green circles) correspond to the centers of $\delta\Delta_{ij}$
bonds shown in green. Note that in general there are six \cite{fiveindependent} distinct parameters controlling  
modulations of the pairing pseudogap $\Delta_{ij}$. The number of non-equivalent
 (black) sites, however, is three (see Fig. \ref{ldos1}).
} 
\end{figure*}

We now resume our discussion of the Abrikosov-Hofstadter problem.
The simplest case, for all models, is $f=1/2$ or $x=0$. For the H1 model, 
the (anti)vortices prefer red plaquette locations and 
$\Phi_r$ will be large compared to $\Phi_b$, which can be safely ignored. 
The resulting dual vortex array at half-filling is
depicted in Fig. \ref{dvortexarray}. The structure is a checkerboard
with vortices in $\Phi_r$ located on alternating
black plaquettes. There is a single dual vortex per two unit
cells of the CuO$_2$ lattice, as expected for $f=1/2$. Such
close packing of dual vortices results in ``empty'' black plaquettes
actually being occupied by dual antivortices (manifestation of the
fact that our $f=1/2$ Hofstadter problem does not break the dual
version of time reversal invariance). 
There is, however, {\em no} modulation in bond 
variables $\delta\Delta_{ij}$ located on vertices of the blue lattice
due to its peculiar relation to dual
magnetic induction ${\bf B}_d$ -- all blue sites are in a 
perfectly symmetric relation
to the dual vortex-antivortex checkerboard pattern on black plaquettes,
as is obvious from Fig. \ref{dvortexarray}.
This implies  $\delta\Delta_{ij} = 0$ and the pseudogap $\Delta$ remains
uniform despite $\Phi (\br)\not =0$. From this one would
tend to conclude that there will be no CPCDW and
no modulation in the local DOS. Still, there is a clear lattice
translational symmetry breaking in the dual sector as evident
from Figs. \ref{dvortexarray} and \ref{dvortexarray1}. Consequently,
if we go beyond the leading derivatives, for example by including
corrections to the Berry phase discussed around
(\ref{berry}), we expect that a weak
checkerboard modulation will develop in quantities like the electron
density $\delta n_i$. The above is a special feature of the fluctuating 
LdSC which, however, is altered when we include the spin channel in 
our consideration. The Berry gauge field $a_\mu$ must then be 
restored -- its coupling
to nodal fermions induces antiferromagnetic order at half-filling
and thus breaks the above symmetry in the leading
order by a commensurate spin-density wave (SDW)
\cite{herbut},\cite{qed}.

As the system is doped $f$ decreases away form $1/2$
according to $f=p/q=(1-x)/2$.
The ground state energy of (\ref{hofstadter}) is particularly
low for dopings such that $q$ is
a small integer, (integer)$^2$ or a multiple
of 2, reflecting the square symmetry of the CuO$_2$ planes.
Such dopings are thus identified as ``major fractions'' in the
sense of the Abrikosov-Hofstadter problem.
In the window of doping which is of interest in cuprates, these
fractions are $7/16$, $4/9$, $3/7$, $6/13$, $11/24$,
$15/32$, $13/32$, $29/64$, $27/64$, $\dots$, 
corresponding to dopings $x=0.125~(1/8)$, $0.111~(1/9)$,
$0.143~(1/7)$, $0.077~(1/13)$, $0.083~(1/12)$, 
$0.0625~(1/16)$, $0.1875~(3/16)$,
$0.09375~(3/32)$, $0.15625~(5/32)$, etc. 
Other potentially prominent fractions, like $1/4$, $1/3$, $2/5$, or $3/8$, 
are associated with dopings outside the underdoped regime
of strong vortex-antivortex fluctuations. 
In general, we expect that particularly low energy states
correspond to fractions such that the pattern of dual vortices in $\Phi$
can be easily accommodated by the underlying CuO$_2$ lattice. Furthermore,
we expect that the quartic repulsion in (\ref{hofstadter}) will
favor the most uniform array of dual vortices that can be constructed
from the $q$-dimensional degenerate Hofstadter manifold. In the
window of dopings one deals with in cuprates, these 
conditions single out doping $x=0.125$ ($f=7/16$) 
as a particularly prominent fraction. 
At $x=0.125$ ($q=16$), the dual vortex pattern and the accompanying
modulation in ${\bf B}_d$ can take advantage of a $4\times 4$ elementary block 
which, when oriented along the $x(y)$ direction,  
fits neatly into plaquettes of the dual lattice, as depicted in
Fig. \ref{dvortexarray}. This $4\times 4$ elementary block embedded into
the original CuO$_2$ lattice and containing 7 dual vortices ($f=7/16$)
is clearly the most prominent geometrical 
structure among all the ones we have found in our study, both in its
intrinsic simplicity and its favorable 
commensuration with the underlying atomic
 lattice. It is bound to be among the highly energetically preferred
states in underdoped cuprates, as it is indeed found in the next section.

\begin{figure*}[tbh]
\includegraphics{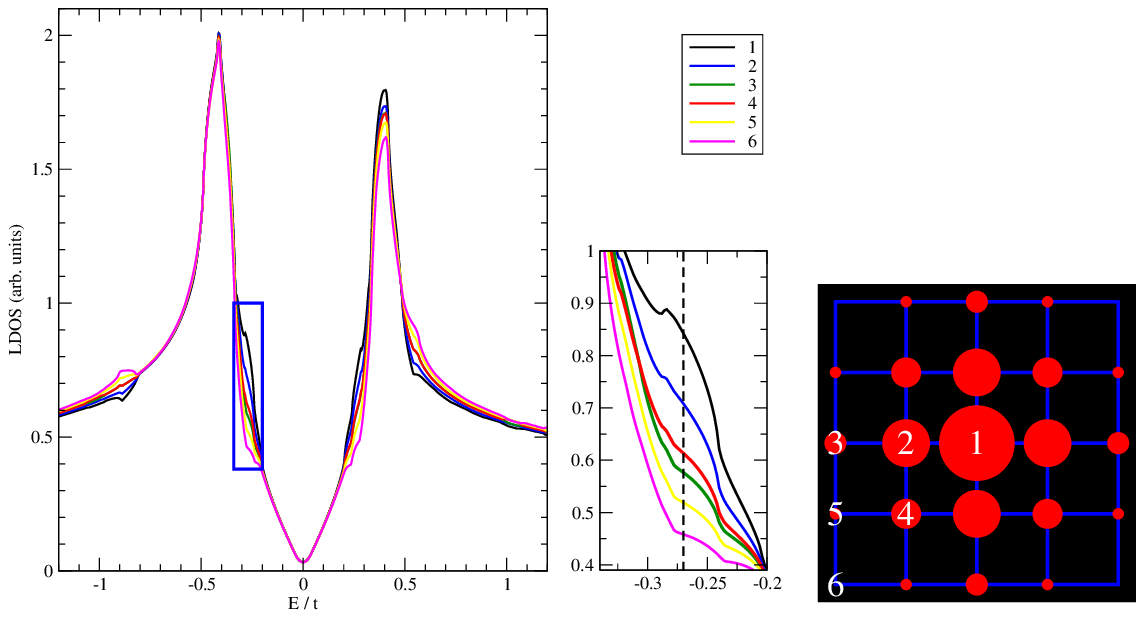}
\caption{\label{ldos}
The local density of states (LDOS) of the lattice $d$-wave superconductor with
the modulated bond gap function $\Delta_{ij}$ 
corresponding to the $4\times 4$ supercell
structure of Fig. \ref{dvortexarray} at doping $x=1/8$.
The calculations are done within the LdSC model of phase fluctuations.
Note that the fluctuations of the gauge fields $v$ and $a$ lead to small broadening of
the peaks but produce no significant changes, as long as one is in the
superconducting state. The parameters are $t^*=1.0$, $\Delta=0.1$ and the
variation in $\Delta$ from the weakest to the strongest bond is $\pm 25\%$. 
The portion enclosed within a rectangle is enlarged in the central panel.
The numbers correspond to the locations of Cu atoms within a $4\times 4$
unit supercell on the CuO$_2$ lattice depicted on the right panel.
Note that Cu atom labeled 1 coincides with the location of the "dual vortex 
hole" in Fig. \ref{dvortexarray} (full green circle) 
where $\Delta$ is the weakest, while the 
one labeled 6 corresponds to the Cu atom in the 
center of Fig. \ref{dvortexarray} which is 
surrounded by four strongest $\Delta$'s. The radii of the red circles indicate 
the magnitude of LDOS at $E = -0.27 t^*$ ( dashed line in the central panel).
This pattern of LDOS is very robust in 
our calculations and is precisely the tight-binding analogue of the 
checkerboard structure observed by Hanaguri et al \cite{davis}
(note that on a tight-binding lattice, the Fourier transforms at
wavevectors $2\pi/4a_0$ and $3\times 2\pi/4a_0$ are not independent since
$3\times 2\pi/4a_0$ and $-2\pi/4a_0$ are equivalent). The symmetry of modulations in
$\Delta_{ij}$ corresponding to Fig. \ref{dvortexarray} imply that there are only six non-equivalent sites 
within the $4\times4 $ unit cell. 
Note that modulation pattern at energies above $\Delta$ is
actually {\em reversed} compared to the pattern at energies below $\Delta$.
Finally,  the nodes remain effectively
intact, in accordance with Ref. \onlinecite{preprint}. Finite LDOS at zero energy is entirely due to artificial
broadening used to emulate finite experimental resolution. 
In the absence of such broadening, the LDOS remains
zero within numerical accuracy (see Fig. \ref{ldos2}).
}
\end{figure*}

\begin{figure*}[tbh]
\includegraphics{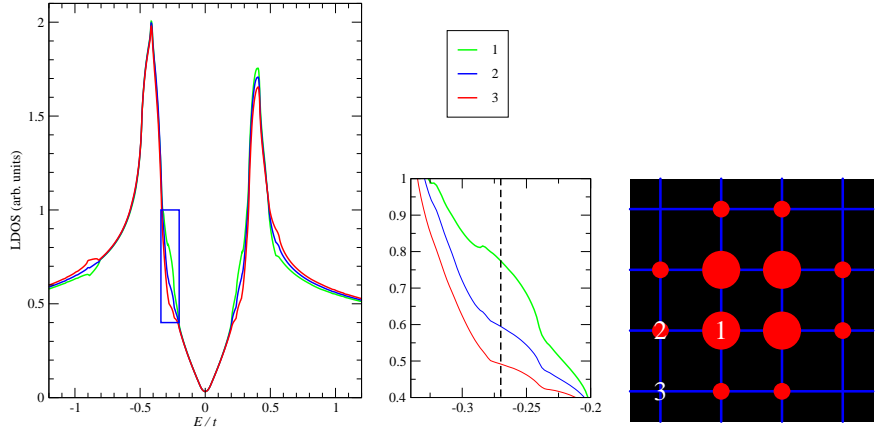}
\caption{\label{ldos1}
Left panel: 
The local density of states (LDOS) of the lattice $d$-wave superconductor with
the modulated bond gap function $\Delta_{ij}$ 
corresponding to the $4\times 4$ supercell
structure of Fig. \ref{dvortexarray1} at doping $x=1/8$.
The parameters used are the same as in Fig. \ref{ldos}. 
Right panel: the local density of states at energy $E = - 0.27 t$ 
(dashed line in the central panel).
The radii of the circles are proportional to the LDOS at a 
given Cu atom. Note that inside the unit cell 
$4\times 4$ there are three (rather than six, as in Fig. \ref{ldos}) 
non-equivalent sites. Even in absence of any underlying
theory, this pattern of
modulation appears to correspond to CPCDW simply by visual inspection.
}
\end{figure*}

\begin{figure*}[tbh]
\includegraphics{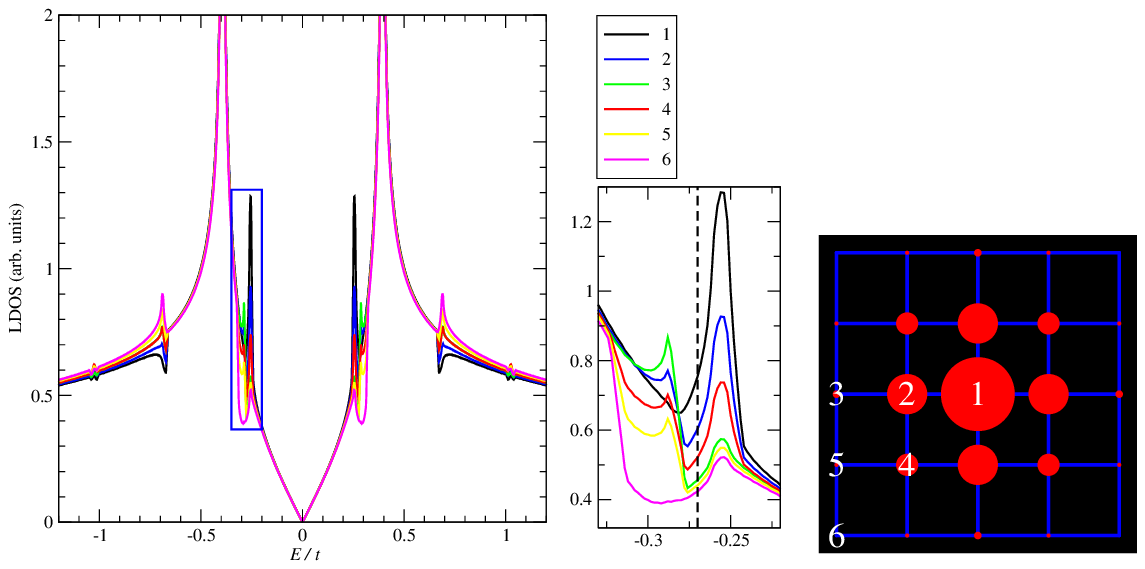}
\caption{\label{ldos2}
The local density of states (LDOS) of the lattice $d$-wave superconductor with
the modulated bond gap function $\Delta_{ij}$ corresponding to the $4\times 4$ supercell structure of Fig. \ref{dvortexarray} at $x=1/8$,
computed under the assumption
of perfect particle-hole symmetry for low-energy fermionic excitations. 
The parameters used are the same as in Fig. \ref{ldos}, 
with the exception of the broadening, which
has been suppressed here in order to demonstrate that
the nodes remain essentially unaffected by CPCDW \cite{preprint}.
}
\end{figure*}

Before we turn to the details of this energetics,
we investigate the signature in the electronic structure
of the above $4\times 4$ elementary block. Such signature could be
detected in the STM experiments of the type performed in
Refs. \onlinecite{yazdani,davis,kapitulnik}. To this
end, we compute the dual induction ${\bf B}_d$ associated with
the pattern of 7 dual vortices in Fig. \ref{dvortexarray} and from
it the spatial modulation of the $d$-wave pseudogap $\delta\Delta_{ij}$
on each bond within the $4\times 4$ supercell. With $\Delta$  thus
fixed, we evaluated the local density of BdG fermion states of our LdSC model. 
The results are shown in Figs. \ref{ldos}-\ref{ldos2}. Note
that we do our computations in the ``supersolid'' region of the
phase diagram in Fig. \ref{dphasediagram}. This enables us to use the
BdG formalism with only moderate smearing in the coherence peaks coming
from the gauge fields $v$ and $a$ and it also allows for rather
direct comparison with experiments. The downside is that we have to
assume that the modulation profile in ${\bf B}_d$ remains the same as
determined from our dual mean-field arguments. Considering the high symmetry
of elementary block in Fig. \ref{dvortexarray} this appears to be a 
rather minor assumption.

\section{Dual Abrikosov-Hofstadter problem}
In this section we present the results of numerical analysis of the H1 model. 
Within this model the values of the matter field 
$\Phi(\br)$ on the black sites are suppressed by very large $m_b^2$, and $\Phi(\br)$ effectively 
lives on the lattice dual to the original copper lattice,
that is on the red sites in Fig. \ref{fig1}. 
Therefore, the dual fluxes  reside inside the red plaquettes, shown in 
Fig. \ref{FigH1model}. In this section we will drop the subscript $d$ in order to make notation more compact
and use  $\bA(\br)$ for the vector potential corresponding to a 
uniform dual magnetic flux equal to $f =  p/q$. Modulation of the field, which will be 
determined numerically, is described by $\delta \bA(\br)$.  Within mean-field approximation, our
problem then  is reduced to minimization  of the following Ginzburg-Landau lattice functional:
\begin{multline}
\label{h1functional2D}
{\cal H}_d= -t\sum_{\br, \bdelta} 
e^{i 
{\cal A}_{\br,\br+\bdelta}+
i\delta{\cal A}_{\br,\br+\bdelta}
}
\Phi^*(\br) \Phi(\br+\bdelta)\\+
\sum_{\br} \left( m^2|\Phi(\br)|^2 + \frac{g}{2} |\Phi(\br)|^4 \right) +
\frac{\kappa^2_d}{2}\sum_{\br} \delta  B^2_d(\br)~~,
\end{multline}
where $m^2=m_r^2$, $\bdelta = \{\pm \hx, \pm \hy\}$,  the link variables ${\cal A}$ are defined as
\begin{equation}
{\cal A}_{\br,\br+\bdelta} = \int_{\br}^{\br+\bdelta}d\br\cdot\bA~~.
\label{definitionA}
\end{equation}
Note that from this point on, we  absorb the factors of $2\pi$ into 
the definition of $\bA$ for numerical convenience 
and to conform with the Abrikosov dimensionless
notation \cite{abrikosov}. The modulated part of the flux  
$\delta B_d(\br)$ is given by the circulation of 
the corresponding $\delta \cA$ around each
plaquette as
\begin{equation}
\delta \cA_{\br, \br+\hx}+ \delta \cA_{\br+\hx, \br+\hx+\hy} + \delta \cA_{\br+\hx+\hy,
\br+\hy}+\delta \cA_{\br+\hy, \br}~~.
\end{equation}
The minimization of 
${\cal H}_d$ with respect to the link variables  $\delta A$ 
is equivalent to the solution of the following set of equations:
\begin{equation}
\begin{aligned}
0 &=
2t
|\Phi(\br)|\,|\Phi(\br+\hx)|  \sin \left( \cA_{\br,\br+\hx} +
\delta \cA_{\br,\br+\hx}+\alpha_{\br+\hx}-\alpha_{\br}\right)\\
&+\kappa^2_d (\delta B_d(\br) - \delta B_d(\br-\hy))\\
0 &=
2t
|\Phi(\br)|\,|\Phi(\br+\hy)|  \sin \left(
\cA_{\br,\br+\hy} +
\delta \cA_{\br,\br+\hy}+\alpha_{\br+\hy}-\alpha_{\br}\right)\\
&+\kappa^2_d (\delta B_d(\br-\hx) - \delta B_d(\br))~~,
\end{aligned}
\label{maxwelllattice}
\end{equation}
where $\alpha_{\br}$ is the phase of the dual matter field $\Phi(\br)$.
Since the last terms in equations (\ref{maxwelllattice}) can be identified as
$x$ and $y$ components of the lattice curl, these are the 
lattice analogs of the (dual) Maxwell's equations 
in two dimensions, providing explicit lattice realization of 
Eq. (\ref{maxwell}).

Before we present the results of the numerical computation, we will
discuss briefly the structure of the solutions that should be
expected on general symmetry grounds.  In the limit of infinite
$\kappa_d$ the gauge field $\delta A$ does not fluctuate, and only
$\Phi(\br)$ should be varied in  minimizing ${\cal H}_d$. To understand
why the solution for a general filling is inhomogeneous consider
first a case of $g=0$. Then the functional  ${\cal H}_d$ simply describes a
particle on a tight-binding lattice  moving in a uniform magnetic
field $\fbar$. The corresponding Hamiltonian is

$$
\hat{H}_{\rm Hof} \Phi(\br)=  - t \sum_{\bdelta=\pm\hx, \pm\hy} e^{i {\cal
A}_{\br,\br+\bdelta}} \Phi(\br+\bdelta)+ m^2\Phi(\br)~~.
$$

The minimization of the functional ${\cal H}_d$ is 
closely related to finding the ground states 
of Hamiltonian $\hat{H}_{\rm Hof}$.
Note that although the dual magnetic field felt by the particles is
perfectly uniform, the gauge-field $\bA(\br)$ is not,
regardless of the gauge used. Indeed, if this were the case 
the circulation of the vector
potential $\bA$ around the primitive plaquette would be zero by periodicity,
which is not possible as the circulation is equal to the flux
of the magnetic field $2\pi p/q$ through the plaquette.
Thus, the Hamiltonian $H_{\rm Hof}$ does not commute with the usual
lattice translation operators. Instead, as noted by Peierls long
time ago,  magnetic translation operators $T_{\bR}$, 
generating lattice translations
complemented by simultaneous gauge transformations, must be
constructed in order to commute with  $\hat{H}_{\rm Hof}$.

Unlike the ordinary translations, operators $T_{\bR}$ do not
commute. Rather,  operators $T_{\bR}$ form  a {\em ray}
representation of the translation group. The theory for irreducible
ray representations of the translation  group 
was constructed by Brown \cite{brown}.
Alternatively, one can use the  magnetic translation group introduced by Zak
\cite{zak}, and use the ordinary representations of that group to
classify the eigenstates of the Hamiltonian. 

\begin{figure*}[tbh]
\includegraphics{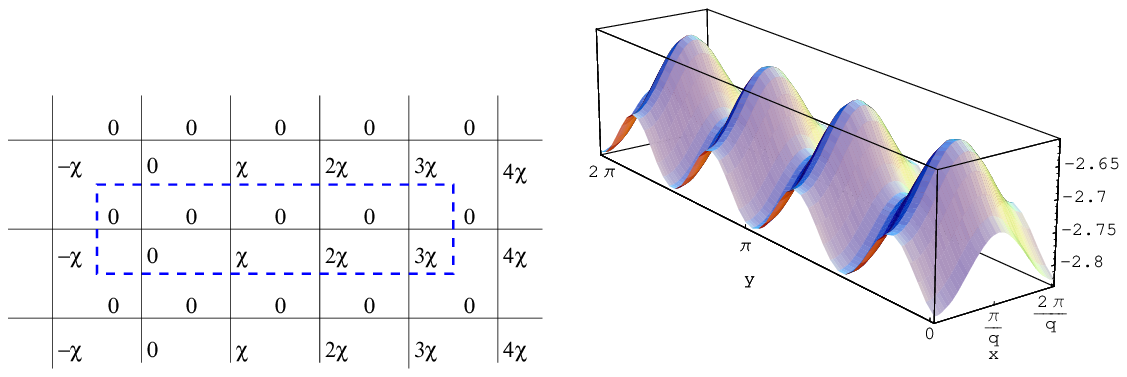}
\caption{
Left panel: the Peierls link variables ${\cal A}_{ij}$ in 
Landau gauge for $q=4$. 
$\chi$ denotes $2\pi p/q$. Both ${\cal A}_{\br,\br+\hx}$ 
and ${\cal A}_{\br,\br+\hy}$
 are periodic in $\hy$-direction. 
 Although ${\cal A}_{\br,\br+\hy}$ increases monotonically with $x$, 
 $\exp(i {\cal A}_{\br,\br+\hx})$ is periodic with the  unit cell shown by the dashed rectangle.
Right panel: the dispersion $E(k_x, k_y)$ of the lowest Hofstadter band for $p/q=1/4$ in 
units of $t^*$. 
There are  $q=4$ ground states at $k_x=0$ and $k_y = 2\pi j/q$, where $j=0,1,{\ldots}, q-1$.
For $q=4$ the energy of the ground states is $E_0=-2\sqrt{2}t$.
}
\label{magcell}
\end{figure*}

On the lattice, the classification of the states is simple
and for completeness, 
we demonstrate  how  the magnetic eigenstates can be constructed.
Consider for example the Landau  gauge $A_x=0$, 
$A_y = 2\pi  xp/q$, in which the unit cell 
spans $q$ elementary plaquettes in
$\hx$-direction. Then the
Peierls factors ${\cal A}_{\br,\br+\bdelta}$, shown in
Fig. \ref{magcell},  are periodic modulo $2\pi$ with enlarged $q\times 1$ unit cell.
The Hamiltonian now can be written as
\begin{multline*}
\hat{H}_{\rm Hof}\Phi(x,y) =
m^2 \Phi(x,y) 
- t\Bigl(\Phi(x+1,y) + \Phi(x-1,y)
\\+\Phi(x,y+1) e^{2\pi i \frac{p}{q} x}
+\Phi(x,y-1)e^{-2\pi i \frac{p}{q} x}
\Bigr)~~.
\end{multline*}
The Hamiltonian, obviously, remains invariant under transformations
$\br\to \br+\hy$ and $\br\to\br+q\hx$. Consequently, 
it can be diagonalized  with the usual Bloch conditions
\begin{align}
\Phi(\br+q\hx) &= e^{iqk_x}\Phi(\br)\\
\Phi(\br+\hy)  &= e^{i k_y}\Phi(\br)~~,
\end{align}
where $(k_x, k_y)$ is the crystal momentum defined in a Brillouin
zone $\frac{2\pi}{q}\times 2\pi$. Using these conditions, we rewrite
the equation for the eigenstates as
\begin{multline}
m^2 g(x)
- t\Bigl(g(x+1) + g(x-1)
+2 g(x)\cos(k_y+2\pi xp/q)
\Bigr) \\= E g(x)~~,
\label{diagonalization}
\end{multline}
where $g(x) = \Phi(x,0)$. Now we have one-dimensional
equation for $g(x)$, where $x = 0,1,2,{\ldots} ,q-1$, 
that has to be solved with Bloch condition 
$$
g(x+q) = e^{iqk_x}g(x)~~.
$$
Thus the problem of diagonalization is reduced to the diagonalization of $q\times q$ matrix for 
each $\bk$. Note, however, that we did not exhaust all the information contained in
the magnetic translation operators, apart from translations by $q$
lattice spacings in $\hx$ direction.  Consider an eigenstate
described by crystal momentum $\bk$ within the Brillouin zone and
characterized with wavefunction $g(x)$. Then function $g_1(x) =g(x+1)$ 
is also an eigenstate of the 
Hamiltonian $\hat{H}_{\rm Hof}$ with the same energy
but with  momentum $(k_x, k_y+2\pi p /q)$. By repeating this operation, one finds
that $q$ states with  crystal momenta described by the same $k_x$ but different $k_y$:
$$
k_y, \;k_y+2\pi p /q,\;k_y+4\pi p /q,\;{\ldots} , \;k_y+ (q-1)2\pi p /q
$$
all have the same energy. Since $p$ and $q$ are mutually prime, this set coincides with
$$
k_y, \;k_y+2\pi /q,\;k_y+4\pi /q,\;{\ldots} , \;k_y+ (q-1)2\pi  /q~~.
$$
Occasionally, in the theory of magnetic translation
groups, this is expressed by using a  reduced Brillouin zone of size
$(2\pi)^2/q^2$, then every band is said to be $q$-fold degenerate. As
a typical example, a dispersion for  $p/q=1/4$ is shown in the
right panel of Fig. \ref{magcell}.

In the gauge we just described, there are $q$ degenerate minima
described by wavefunctions $\Phi_j(x,y)$ that are located
at $k_x =0$ and $k_y = 2\pi mp/q$, where $j=0,1,2,{\ldots} (q-1)$.
Therefore, sufficiently close to the transition,  the minimum of the 
functional ${\cal H}_d$ should be sought as a linear combination of the $q$
degenerate states $\Phi_j(\br)$ and our problem is equivalent 
to minimizing the Abrikosov participation ratio
\begin{equation}
\min \frac{\sum_{\br} |\Phi(\br)|^4}{\left(\sum_{\br} |\Phi(\br)|^2\right)^2}~~,
\qquad {\rm where\;\;} 
\Phi(\br) = \sum_{j=1}^{q} C_j \Phi_j(\br)~~.
\label{participationratio}
\end{equation}
Since  $k_x=0$ for all $\Phi_j(\br)$,
any linear combination of functions $\Phi_j(\br)$ is periodic in
$x$-direction:  $\Phi(\br +q\hx) = \Phi(\br)$.
In addition, for each of the $q$ ground states, $k_y$ is a multiple
of  $2\pi/q$, and consequently $\Phi(\br +q\hy) = \Phi(\br)$. From
these  two properties we find that  any
linear combination  $\sum_{j}C_j\Phi_j(\br)$ must
be periodic in the $q\times q$ unit cell.

The minimization problem can be formulated equivalently in terms of
coefficients $C_j$:
\begin{equation}
{\cal H}_d =
E_0 \sum_{j} |C_j|^2
+ \sum \Gamma^{(4)}_{j_1,j_2,j_3,j_4}
C_{j_1}^*C_{j_2}^*C_{j_3}C_{j_4}~~,
\label{GLC}
\end{equation}
where $E_0$ is the ground state energy of $\hat{H}_{\rm Hof}$ following
from (\ref{diagonalization}) and
\begin{equation}
\Gamma^{(4)}_{j_1,j_2,j_3,j_4} = \sum_{\br}
\Phi_{j_1}^*(\br)\Phi_{j_2}^*(\br)\Phi_{j_3}(\br)\Phi_{j_4}(\br)~~.
\end{equation}
Note that (\ref{GLC}) itself has Ginzburg-Landau form with $q$ order
parameters $C_j$.  In our case, the form of the matrix $\Gamma^{(4)}$ is
dictated by our ``microscopic'' Hamiltonian ${\cal H}_d$ and
corresponds to the Abrikosov participation ratio (\ref{participationratio}).
One could in principle generalize the theory by considering
completely general form of $\Gamma^{(4)}$ compatible with the overall symmetry
requirements. Such procedure is equivalent to introducing
long-ranged quartic interactions with general kernel
$K(\br,\br',\br'',\br''')$:
$$
\sum K(\br,\br',\br'',\br''') \Phi^*(\br)\Phi^*(\br')
\Phi(\br'')\Phi(\br''')~~.
$$
Equation (\ref{GLC}) is the most direct and 
convenient  representation for numerical
minimization of ${\cal H}_d$ in the limit of infinite $\kappa_d$. 
However, in order to  allow for intermediate 
values of $\kappa_d$ and to be able
to analyze the impact of various short-ranged terms describing
interaction of dual fluxes $\delta B_d(\br)$, 
we also have opted to follow a slightly
different route: The numerical
results that are presented below are produced by {\em direct}
numerical minimization of functional 
(\ref{h1functional2D}) with respect to
both the dual  matter field $\Phi$ and the link 
variables $\delta A_{\alpha}$ by imposing 
periodic boundary conditions on $N_x\times N_y$ blocks with varying $N_x$
and $N_y$. At least in the vicinity of the transition, the largest
unit cell one has to consider is $q\times q$. 
It should be stressed that we found
identical results using both approaches whenever direct comparison is
possible (sufficiently large $\kappa_d$ and no additional
short-ranged interactions between the fluxes).

We perform numerical minimization of functional 
(\ref{h1functional2D}) with respect to
both the matter field $\Phi(\br)$ and the link variables 
$\delta A_{\alpha}$ by imposing periodic boundary conditions on
$n\times m$ blocks with variable  $m$ and $n$. The number of 
independent variables grows as $3mn$ and the
largest unit cells we were able to consider are $8\times 8$. 
We used the conjugate gradient minimization
technique with as many as $10^4-10^5$ randomly chosen different  starting points. 

We remind the reader that since the dual matter field variables  $\Phi(\br)$ in the H1 model live on the
on red sites,  the dual fluxes $\delta B_d(\br)$ reside inside the red plaquettes, shown in 
Fig. \ref{FigH1model}. Note that each link of the
original copper lattice is shared by two red plaquettes 
and therefore the enhancement of the $d$-wave gap function $\Delta$
should be interpreted as the average between the fluxes at 
the neighboring dual plaquettes. Thus,
at half-filling $f=\half$ there is no modulation in $\Delta$ and no
charge density modulation emerging from our model H1 even though there is
a checkerboard pattern in the dual flux. 
The checkerboard pattern remains the same for the entire 
range of parameters we were able to check. However, as explained 
in the previous section, since there is a definite
symmetry breaking in the dual sector by the checkerboard array of
$\delta B_d(\br)$ higher order derivatives and other terms not
included in our dual Lagrangian are expected to generate a weak
checkerboard modulation to accompany $\delta B_d(\br)$.

\begin{figure*}[thb]
\includegraphics{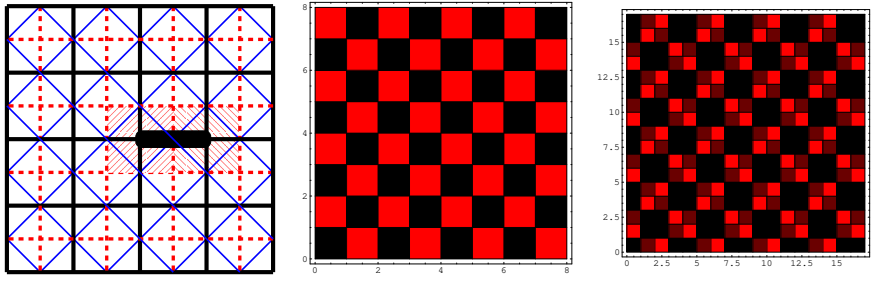}
\caption{Left panel: Any link of the black lattice (thick solid lines), which
corresponds to $\Delta_{ij}$, 
is shared by two dual plaquettes shown in red. Within H1 model, 
therefore, the enhancement or
suppression of $\Delta_{ij}$ is determined by the average of 
the fluxes trough the neighboring plaquettes of
the dual lattice (red dashed lines). 
Central panel: distribution of $\delta B_d$ at 
half filling. In units of $t$, the parameters
used are $\kappa^2_d = 30.0$, $m^2=-1.0$,
and $g =2.0$. The energy per site is $-3.695$.
Right panel: distribution of $\delta B_d$ at
$p/q=1/4$. In units of $t$, the parameters
used are $\kappa^2_d = 30.0$, $m^2=-1.0$,
and $g =2.0$. The energy per site is $-3.564$.
}
\label{FigH1model}
\end{figure*}

\begin{figure*}[thb]
\includegraphics{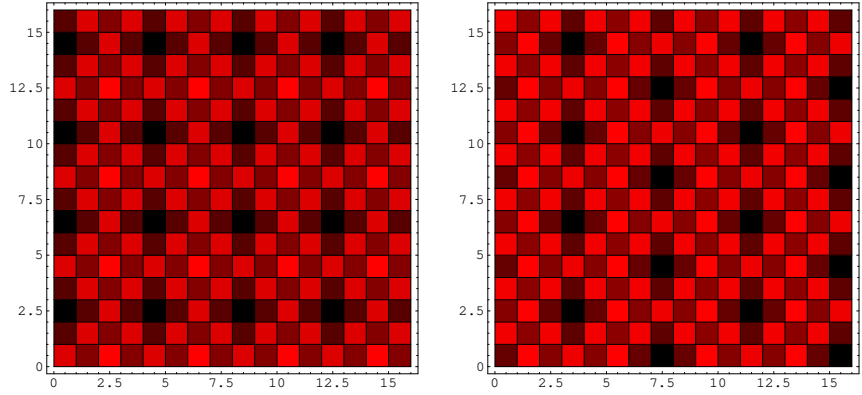}
\caption{Left panel: The pattern of the dual fluxes  $\delta B_d$ for  $p/q = 7/16$ with
periodic boundary conditions in $4\times4$ unit cell. The unit cell $4\times 4$ is repeated four times in
horizontal and vertical directions for presentation purposes. The energy per site is $-3.193$ in the units
of $t$ and the parameters of the model are the same as in the previous figure. The positive(negative) values of
$\delta \bB_d$ are shown in  red (black). Right panel: the same but with $8\times 8$ unit cell. The square
lattice is distorted towards the triangular lattice. The energy
per site is $-3.208t$ and is lower by $0.5\%$ compared to the square arrangement.
}
\label{Figp7q16}
\end{figure*}

At field characterized by $f=p/q =7/16$, which 
within H1 model corresponds to doping  $x=1/8$,
the structure of the configuration is considerably more complex. 
When restricted to a $4\times4$ lattice,
the resulting pattern is the square lattice of ``crosses'' 
separated by four unit cells, shown in Fig.
\ref{Figp7q16}. This configuration, however, is not the 
true global minimum. If a larger unit cell
for modulations of the dual fluxes is allowed, the energy 
can be additionally lowered by $\approx 0.5\%$ by distorting 
the ideal square pattern. The lowest
energy we found corresponds to the quasi-triangular lattice 
of crosses shown in the right panel of Fig.
\ref{Figp7q16}. The lowest energy state that we find has the symmetry of 
this quasi-triangular pattern for all $\kappa_d^2$ from $1.0$ to $10^5$.

The smallness of the energy differences, involving only few
percents of the overall Abrikosov-Hofstadter energy scale,
indicates that the state that emerges
victorious can be changed by the additional
\begin{figure*}[thb]
\includegraphics{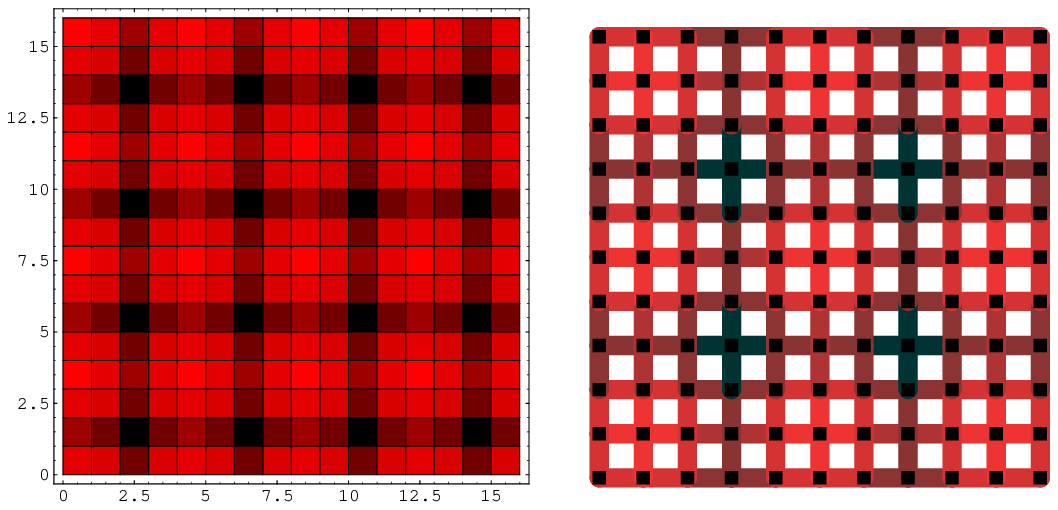}
\caption{
Left panel: The pattern of the dual fluxes $\delta B_d$ for the 
Hofstadter-Abrikosov problem at $f=7/16$ ($x=1/8$) 
obtained for the  $8\times 8$ unit cell. The
parameters of the model are $\kappa^2_d = 30.0$, $\Gamma_0=6.0$, $\Gamma_1 = -12.0$, $m^2=-1.0$, 
and $g=2.0$. The $4\times 4$ checkerboard symmetry
of the pattern is precisely what is shown in Fig. \ref{ldos}.
Right panel: The distribution of modulations in the pairing pseudogap
 $\delta\Delta_{ij}$ corresponding to the pattern of fluxes
shown on the left panel. This ``tartan'' pattern has the symmetry of Fig. \ref{dvortexarray} with
``dual vortex holes'' coinciding with black ``crosses''.
}
\label{Figp7q16Gamma}
\end{figure*}
short range interactions and derivative terms
which we have routinely neglected. Not surprisingly, therefore, the
precise energetics of various low energy Abrikosov-Hofstadter states
is decided by details. We find that inclusion of terms
$|\Phi(\br)|^6$ and dual density-density 
interactions $|\Phi(\br)|^2 |\Phi(\br+\bdelta)|^2$ with moderate coefficients
does not change the patterns we described. On the other hand,
the inclusion of the terms describing short-ranged 
interactions between the dual fluxes produces
significant effects. An example of the typical pattern 
obtained by replacing the self-interaction term
\begin{equation}
\sum_{\br} \frac12\kappa^2_d
(\delta B_d)^2
\label{dualmagneticenergy}
\end{equation}
by 
\begin{equation}
\sum_{\br} \frac12\kappa^2_d (\delta B_d)^2 + 
\Gamma_0  \sum_{nn}\delta B_d(i) B_d(j)
+\Gamma_1  \sum_{nnn} \delta B_d(i) B_d(j)
\label{dualmagneticenergyi}
\end{equation}
is shown in Fig. \ref{Figp7q16Gamma}. The parameters $\Gamma_0$ 
and $\Gamma_1$ in (\ref{dualmagneticenergyi}) are chosen to make the
distribution of the dual flux somewhat smoother than what
is demanded by (\ref{dualmagneticenergy}) only. 
This suffices
to bring the energy of the square pattern in
Fig. \ref{Figp7q16Gamma} from just above to just below that of
the quasi-triangular pattern of Fig. \ref{Figp7q16}. It is
tempting to speculate that this slight additional
``smoothening'' of the dual flux represents the combined effects of nodal
fermions and Coulomb interactions present in real systems.
Note that the symmetry and the qualitative features of
this pattern coincide with our  4$\times$4 ``elementary'' block 
conjectured to be the likely CPCDW ground state in
the previous section (see Figs. \ref{dvortexarray}
and \ref{dvortexarray1}). Once $\delta B_d(i)$ are translated into
$\delta\Delta_{ij}$ the resulting pattern
closely resembles the checkerboard distribution of
local density of states of ``electron crystal'' 
observed in the STM experiments, as illustrated in
Fig. \ref{ldos}.

Let us now consider this observed STM checkerboard pattern in
more detail, in light of our theory. Obviously,  
our analysis being restricted restricted to the
tight-binding lattice, we cannot describe the local
density of states (LDOS) at positions between the sites of Cu lattice -- 
the peaks
in our theoretical LDOS are always located on top of Cu atoms.
In contrast, the STM measurements by Hanaguri {\em et al.} \cite{davis},  have
much better spatial resolution and can image the actual continuous
atomic orbitals. Our tight-binding lattice results 
for LDOS(\br, E) could be viewed as  
``coarse-grained'' representation of the
LDOS $g(\br, E)$ observed experimentally:
$$
{\rm LDOS}(\bR, E) \propto \int g(\br, E) d\br~~,
$$
where the integral extends over a square of size
$a_0\times a_0$ centered around Cu
lattice site $\bR$. Equivalently, the true continuum LDOS signal could be
obtained by broadening our lattice LDOS around each Cu lattice site.

A prominent feature of the checkerboard pattern in STM measurements that has 
received much attention is the presence of a pronounced Fourier signal not
only at wavevectors $\frac{2\pi}{a_0}(\frac14, 0) $ and 
$\frac{2\pi}{a_0}(0,\frac14) $, which correspond to the $4a_0\times 4a_0$
periodicity  just
described, but also at $\frac{2\pi}{a_0}(\frac34, 0) $ and
$\frac{2\pi}{a_0}(0,\frac34) $. 
While it is tempting to associate the $3/4$ peaks
with an entirely independent type of order, we note that for a
periodic lattice of identical $4a_0\times4a_0$ tiles, the Fourier transform is a
{\em discrete series}  with wavevectors  
$Q_{n_x,n_y}=\frac14 \frac{2\pi}{a_0}(n_x , n_y) $, where
$n_x$ and $n_y$ are integers, irrespective of how 
complex is the internal
structure of each tile. 
For a general structure of the tile all of the
harmonics $(n_x, n_y)$ are present, and there is no {\em a priori}
reason for the Fourier
coefficients with $n_x=3$,  $n_y=0$ to be particularly small. 
The presence of a large signal at
$Q_{3,0}$ is only natural, and no more unexpected than the
weakness of Fourier harmonics at $Q_{2,0}$ and $Q_{0,2}$. 

Next, in order to describe and compare
the Fourier transforms of spatially broadened
LDOS patterns in Figs. (\ref{ldos}-\ref{ldos2}) and tunneling LDOS observed
in experiments \cite{davis},
we introduce a simple model which approximates 
each bright spot inside the primitive $4a_0\times 4a_0$ tile as
$$
\eta({\bf r}) = \exp \Bigl[J (\cos \frac{2\pi x}{4a_0}+\cos \frac{2\pi
y}{4a_0} -2)\Bigr]~~.
$$
While the specific functional form of the peak is not important, our choice is
convenient since  $g({\bf r})$ has a period of $4a_0\times 4a_0$ and a gaussian
shape, centered at positions $\br = (4N_x a_0, 4N_y a_0)$, where $N_x$ and $N_y$ are
integers. The  width of the gaussians is $\sim a_0/\sqrt{J}$.

\begin{figure*}
\includegraphics[width=\textwidth]{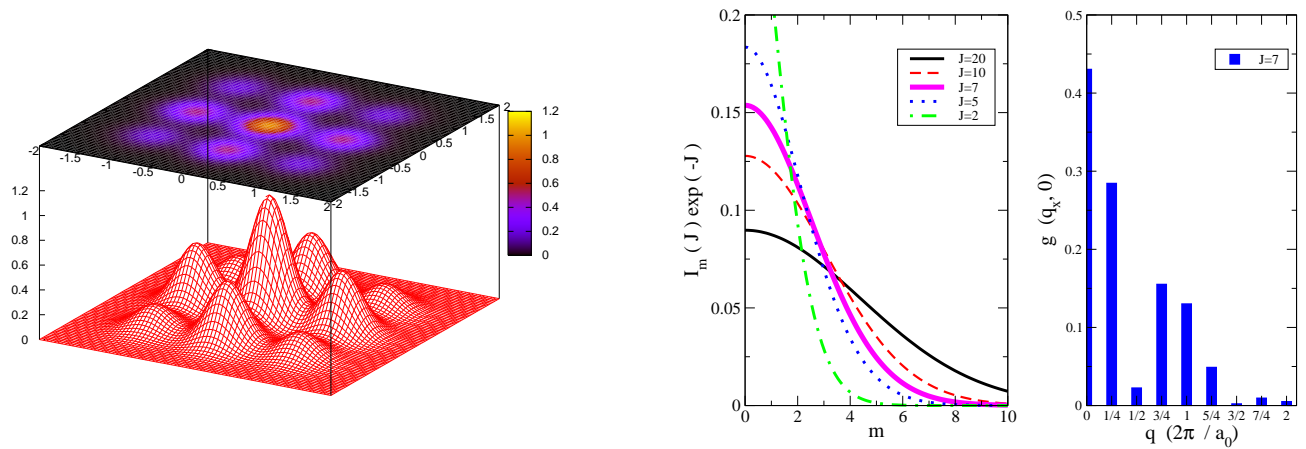}
\caption{\label{ldosmodelfig}
Left panel: Function $g({\bf r})$ used to 
emulate the LDOS signal with parameters
$J=7.0$, $a=1.0$, $b=0.5$ and $c=0.2$. Center panel: 
Regular modified Bessel function $I_m(J)$ shown as a function of its index
$m$ for several values of fixed $J$. Right panel: Fourier coefficients
$g(q_x,0)$ for LDOS pattern shown in the left panel. The reader
should compare this with the similar plot in Fig. 2 of Ref. \onlinecite{davis}.}
\end{figure*}

The real-space tunneling LDOS pattern of Hanaguri {\em et al.}
can now be represented by a function
\begin{equation}
g({\bf r}) = A \eta({\bf r})+ B\sum_{\delta} \eta({\bf r}+\delta) +
C\sum_{\bdelta'} \eta({\bf r}+\delta')~~,
\label{ldosmodel}
\end{equation}
where the first term represents the brightest peak of each tile,
the term proportional to $B$  represents the set of four
second brightest peaks located at $\bdelta=\pm (1+\epsilon)a\hat{x},\; \pm
(1+\epsilon)a\hat{y}$ and the last term represents the
weak peaks at $\bdelta' =\pm a_0(1+\epsilon)\hat{x} \pm a_0(1+\epsilon)\hat{y}$
(see Fig. \ref{ldosmodelfig}). Parameter $\epsilon$ 
equals zero if the real-space  LDOS peaks
are centered at the Cu lattice sites, while $\epsilon=1/3$ if the
maxima of all peaks, except for the central one, are displaced from their
commensurate positions on top of Cu atoms  -- this is 
how the experimental data were
interpreted in Ref. \onlinecite{davis}. We will show that for a wide range of
parameters $J,A,B,C$, and arbitrary $0<\epsilon<1/3$
the $Q_{3,0}$ peaks are much stronger than peaks at $Q_{2,0}$, 
although to explain other features of Fig. 2 in
Ref. \onlinecite{davis}, the ``commensurate''
choice $\epsilon=0$ appears to be more natural.
For the momentum-space direction $n=0$  shown as black 
circles in Fig. 2  of Ref. \onlinecite{davis}, the Fourier coefficients are
$$
g_{m,0} = \frac{1}{16a^2}\int_0^{4a_0} \int_0^{4a_0}   dxdy \;G(x,y)
e^{-i\frac{2\pi}{4a_0}mx}  ~~.
$$
The integral is elementary and the result for $g_{m,0}$ is
\begin{equation}
\left[A+2B+(2B+4C)\cos \frac{2\pi(1+\epsilon) m}{4} \right] I_0(J)I_m(J)e^{-2J},
\label{fm0}
\end{equation}
where $I_m$ is the regular modified Bessel function. 
$I_m(J)$, shown for several values
of $J$ in Fig. \ref{ldosmodelfig}, is a monotonically  
decreasing function of $m$. The non-monotonic  part of
$g_{m,0}$, denoted by $\tilde{g}_{m,0}$, is 
contained in the first factor of (\ref{fm0}). When $\epsilon=0$
(commensurate position of the peaks), this factor is a periodic function of
$m$ with period four, while for $\epsilon=1/3$ the said period is 
equal to three. Table \ref{nonmoni}
 summarizes the  factors $\tilde{g}_{m,0}$ for the two cases.
\begin{table}
\begin{center}
\begin{tabular}{||l|l|l||}
\hline
$m$& $\epsilon=0$&$\epsilon=1/3$\\
\hline
$m\equiv 0\mod 4$ & $A+4B+4C$&$A+4B+4C$\\
$m\equiv 1\mod 4$ & $A+2B$   &$A+B-2C$\\
$m\equiv 2\mod 4$ & $A-4C$   &$A+B-2C$\\
$m\equiv 3\mod 4$ & $A+2B$   &$A+4B+4C$\\
\hline
\end{tabular}
\end{center}
\caption{\label{nonmoni}
Non-monotonic dependence $\tilde{g}_{m,0}$ of the Fourier
transformed LDOS g(\br) defined in (\ref{ldosmodel}).}
\end{table}

We start our analysis with the incommensurate case $\epsilon=1/3$. 
The third column of Table \ref{nonmoni} 
implies that in this case  $\tilde{g}_{1,0}$ and  $\tilde{g}_{2,0}$ 
are  equal and smaller than
the component $\tilde{g}_{3,0}$, as shown in the third column of the table.
Experimentally, however, the Fourier component at $Q_{1,0}$ (the $1/4$
peak) is roughly of the same magnitude as  $Q_{3,0}$, and it is the
component at $Q_{2,0}$ that is the weakest. To account for particularly small
$g_{2,0}\ll g_{1,0}$ one has to select the value of $J$ such that the monotonic
function  $I_m(J)$ decreases rapidly in the region between $m=1$ and $m=2$,
Fig. \ref{ldosmodelfig} indicates that this 
is the case for $3<J$. This choice, 
however, also dramatically suppresses the Fourier transforms  
at $m=3$ and $m=4$, both of which are rather large in experiments.

While the qualitative features of the Fourier 
transformed experimental LDOS could possibly be
reconciled with $\epsilon=1/3$ by fine-tuning parameters $A, B$, and $C$, 
the situation might in fact be better
described by assuming the {\em commensurate} location of the peaks,
in registry with Cu atoms ($\epsilon=0$);
see the second column the Table \ref{nonmoni}. In this case, harmonic $m=2$
is {\em automatically} suppressed compared to the $1/4$ and $3/4$ Fourier
components.
The suppression of Fourier component at $2/4$ can be
qualitatively understood as follows: This Fourier 
component is determined by the overlap of the
LDOS signal  and $\cos(2\frac{2\pi}{4a_0} x)$. Obviously, the 
maxima of the LDOS signal correspond to alternating
maxima and minima of $\cos(2\frac{2\pi}{4a_0} x)$ and  destructive
interference of the two functions occurs. For harmonics $1/4$ and $3/4$ (and of
course for $4/4$) the overlaps are significant and their Fourier coefficients
are larger. Peaks corresponding to larger values of $m$ are strongly
reduced due to the monotonic dependence  $I_m(J)$ of the Fourier transform
on $m$. This suppression can serve as an estimate of parameter $J$: visually,
the last discernible peak in Fig. 2 of Ref. \onlinecite{davis} appears at
$Q_{5,0}$, which places $J$ in the range of $5-10$ (see right panel of
Fig. \ref{ldosmodelfig}). In the right panel of Fig. \ref{ldosmodelfig}
the spatial Fourier transforms of the LDOS (\ref{ldosmodel}) with
parameters $J=7$, $A=1$, $B=0.5$, and $C=0$ and commensurate placement of
the peaks $\epsilon=0$ is shown. This example 
illustrates that the major features
of the Fourier-transformed LDOS obtained by 
Hanaguri {\em et al.} \cite{davis} are
robust properties of an elementary tile of size $4a_0\times 4a_0$ and nine peaks
occupying  {\em commensurate} locations at sites of the Cu lattice
as depicted in our Figs. (\ref{ldos}-\ref{ldos2}): the
large magnitude of $3/4$ peak is simply a higher harmonic 
describing the characteristic intra-tile structure.





\section{Conclusions}

Our main goal in this paper is to devise a more realistic description
of a strongly quantum and thermally fluctuating $d$-wave superconductor,
based on the theory of Ref. \onlinecite{qed}.
Such description applies not only to long distance and low energy properties,
which are the primary
domain of \cite{qed}, but also to intermediate lengthscales,
of order of several lattice spacings, and to energies up to the 
pseudogap scale $\Delta$. This enables us to use the theory to
address the experimental observations of Refs. \onlinecite{yazdani,davis,kapitulnik}.
The charge modulation observed in those experiments is attributed
to the formation of the Cooper pair CDW, the dynamical origin of which
is in strong quantum fluctuations of vortex-antivortex pairs.
These quantum superconducting phase                 
fluctuations reflect enhanced Mott-Hubbard correlations in 
underdoped cuprates as doping approaches zero.
Quantum fluctuating $hc/2e$ (anti)vortices ``see'' physical 
electron as a source of a half-quantum {\em dual} magnetic 
flux and the theory of
the CPCDW can be formulated as the Abrikosov-Hofstadter problem in
a type-II dual superconductor \cite{preprint}.
An XY-type model of such a dual superconductor appropriate for a
lattice $d$-wave superconductor is constructed, both for thermal
and quantum phase fluctuations. The specific translational symmetry
breaking patterns that arise from the dual Abrikosov-Hofstadter
problem are discussed for various dopings $x$, which determines 
the dual flux per unit cell of the CuO$_2$ lattice via
$f=p/q=(1-x)/2$.
In turn, the spatial modulation of the dual magnetic induction ${\bf B}_d$
corresponding to these Abrikosov-Hofstadter patterns is
related to the modulation in the gap function of the
lattice $d$-wave superconductor and is used to compute
LDOS observed in STM experiments. A good agreement is found
for $x=1/8$ ($f=7/16$), which is the dominant fraction
of the Abrikosov-Hofstadter problem in the window of
dopings where our theory applies.

\acknowledgments
The authors thank J.C. Davis, M. Franz, J.E. Hoffman, S. Sachdev, A. Sudb\o , 
O. Vafek, A. Yazdani and S.C. Zhang for useful discussions and correspondence
and to S. Sachdev for generously sharing with 
us unpublished results of Ref. \onlinecite{sachdev}.
A.M. also  thanks P. Hirschfeld, B. Andersen, T. Nunner, and L.-Y. Zhu
for numerous discussions.
The final touches were applied to this paper while one of us (ZT) enjoyed hospitality of the
Aspen Center for Physics. 
This work was supported in part by the NSF grant DMR00-94981.

\appendix*
\section{Two alternative derivations of dual Abrikosov-Hofstadter Hamiltonian}

In this Appendix we present two different and self-contained
derivations of the dual
Abrikosov-Hofstadter Hamiltonian (\ref{ltightbinding},\ref{hofstadter}), 
either one of which can
serve as an alternative to the derivation given in the main text. 
The first approach is
somewhat more detailed and in a sense more ``microscopic'' since it uses
a quantum vortex-antivortex Hamiltonian as a springboard to derive
the effective dual field theory (\ref{dduallagrangian}). In turn, such
vortex-antivortex Hamiltonian in principle can be derived from the
(still unknown) fully 
microscopic theory of cuprates. Incidentally, this derivation
is the (2+1) dimensional analogue of the 3D case presented in
the Appendix of Ref. \onlinecite{zt}.
The second derivation follows the familiar Villain approximation
to the XY model and applies it to our specific situation. The Villain
approximation is less ``realistic'' but provides a transparent
and systematic way of deriving dual representations of XY-like models.

In both derivations the starting point is the effective (2+1)D XY model of a 
quantum fluctuating $d$-wave superconductor:
\begin{multline}
{L}_{XY}^d 
=i\sum_if_i\dot\varphi_i+ \frac{K_0}{2}\sum_i\dot\varphi_i^2
-J\sum_{nn}\cos(\varphi_i - \varphi_j)\\
-J_1\sum_{rnnn}\cos(\varphi_i - \varphi_j)
-J_2\sum_{bnnn}\cos(\varphi_i - \varphi_j)\\
+ {L}_{\rm nodal}[\cos(\varphi_i - \varphi_j)]+{L}_{\rm core}~~,
\label{alxyd} 
\end{multline}
where $\varphi_i(\tau)$ is the fluctuating phase on a site of
the blue lattice in Fig. \ref{fig1}, the first (imaginary) term
is the charge Berry phase corresponding to the overall flux $f$ through
a plaquette the blue lattice, $J$ is the nearest neighbor XY
coupling, $J_{1(2)}$ are the next nearest neighbor XY couplings along
red (black) diagonals, ${L}_{\rm nodal}$ is the contribution
of nodal fermions, and ${L}_{\rm core}$ denotes core
contributions arising from small regions around vortices where the
pairing pseudogap is significantly suppressed -- this, for example, 
generally includes
the mass term (\ref{vortexmass}), the energy cost of core-core overlap, 
the Bardeen-Stephen core dissipation,  etc. The reader should bear in
mind that the last effect is small in cuprates, as explained in the
main text, and will be neglected in the Appendix.
Furthermore, we are neglecting vortex interactions 
with the spin of low energy nodal 
fermions, represented by the Berry gauge field $a$; this is justified
away from the critical point.
The results below are easily adapted to the extended $s$-wave pairing
symmetry. Similarly, both derivations are straightforwardly applied
to a yet simpler case, a fluctuating $s$-wave superconductor:
\begin{equation}
{L}_{XY} =if\sum_i\dot\phi_i+ \frac{K_0}{2}\sum_i\dot\phi_i^2 -
J\sum_{nn}\cos(\phi_i - \phi_j)+ {L}_{\rm core}~~,
\label{alxy} 
\end{equation}
which was used in the main text as a pedagogical example.

Just as the starting points of two derivations coincide,
their final product, the effective dual theory at
long and intermediate lengthscales, will also turn out to be the same.

\subsection{${\cal H}_d$ from a ``microscopic'' vortex-antivortex
Hamiltonian}

The quantum partition function of a phase fluctuating superconductor is:
\begin{equation}
Z_{XY}^d = \int{\cal D}\varphi_i\exp\bigl(-\int_0^{\beta} d\tau
{L}_{XY}^d[\varphi_i(\tau)]\bigr)~~,
\label{azxyd}
\end{equation}
where the functional integral $\int{\cal D}\varphi_i(\tau)$
runs over phase variables $\varphi_i(\tau)$ such that
$\exp(i\varphi_i(\tau))$ is periodic in the interval $\tau\in [0,\beta]$.

The difficulty  in computing (\ref{azxyd}) is twofold:
the fact that $\varphi_i(\tau)$ is a {\em compact} phase variable,
defined on an interval $[0,2\pi)$, rather than an ordinary real field
taking values in $[-\infty,+\infty]$, and the cosine functions in
${L}_{XY}^d$ that couple phases on different sites in a non-linear
fashion. To
deal with the problem one approximates the cosines with quadratic forms.
One popular approximation on a lattice is due to Villain and will be discussed
in the next subsection. In continuum, the approximation amounts to replacing 
$\cos(\varphi_i -\varphi_j)\to 1 -(a^2/2)(\nabla\varphi)^2 +\cdots$,
where $a$ is the lattice spacing and $\varphi(x)$ is now a 
function in continuous 
(2+1) dimensional spacetime. Its compact character is enforced
by writing 
$\partial_\mu\varphi (x)\to \partial_\mu\chi(x) +(\partial_\mu\varphi (x))_v$,
where $\chi$ is an ordinary real field and $(\partial_\mu\varphi (x))_v$
is the part of the phase associated with vortices, defined via 
$\nabla\times(\nabla\varphi (\br,\tau))_v=2\pi\sum_\alpha\delta(\br-\br_\alpha^v(\tau))-2\pi\sum_\alpha\delta (\br-\br_\alpha^a(\tau))$, with 
$\{\br_\alpha^{v(a)}(\tau)\}$ being (anti)vortex positions.
Simultaneously with this decomposition of $\partial_\mu\varphi (x)$,
$\int{\cal D}\varphi_i$ is replaced by the
functional integrations over $\chi(x)$ and (anti)vortex positions
$\{\br_\alpha^{v(a)}(\tau)\}$.
We are assuming here that the (anti)vortices of topological charge $\pm 1$ 
dominate the fluctuation behavior in the regime where the amplitude
of the pseudogap $\Delta$ is large and stiff, allowing us to safely neglect
topological defects corresponding to vorticity 
$\pm 2, \pm 3,\dots$ due to their
higher core energies. This assumption simplifies
the algebra considerably. Furthermore, this assumption is natural within
the theory of Ref. \onlinecite{qed}: The proliferation of defects of high
topological charge is equivalent to strong amplitude
fluctuations and the eventual collapse of the pseudogap -- as
long as we are in the pseudogap regime the $\pm 1$ (anti)vortices
are the only relevant excitations.
With these changes in place, the
partition function (\ref{azxyd}) finally takes the form:
\begin{multline}
Z_{XY}^d \to \sum_{N_v=0}^{\infty}\sum_{N_a=0}^{\infty}\frac{1}{N_v!N_a!}
\int{\cal D}\chi\prod_{\alpha,\gamma=1}^{N_v,N_a}
\int\limits_{\{\br_\alpha^{v}(0)\}=\{\br_\alpha^{v}(\beta)\}}{\cal
D}\br_\alpha^{v}(\tau)\\
\int\limits_{\{\br_\gamma^{a}(0)\}=\{\br_\gamma^{a}(\beta)\}}{\cal D}\br_\gamma^{a}(\tau)
\exp\bigl(-\int d^3x
{L}_{XY}^d[\chi,\{\br_\alpha^{v(a)}(\tau)\}]\bigr)~~,
\label{azxydi}
\end{multline}
where $\int d^3x = \int_0^{\beta} d\tau\int d^2r$ and the
set $\{\br_\alpha^{v(a)}(\tau)\}$ containing $N_{v(a)}$ (anti)vortices
at $\tau=0$ coincides with the one at $\tau =\beta$, to ensure
proper periodicity of $\exp(i\varphi(\br,\tau))$ in imaginary time.
Lastly,
\begin{multline}
{L}_{XY}^d[\chi,\{\br_\alpha^{v(a)}(\tau)\}]=
if(\br)\dot\varphi_v + \frac{K_0}{2}(\dot\chi+\dot\varphi_v)^2\\
+\frac{\tilde J}{2}(\nabla\chi+\nabla\varphi_v)^2 +{L}_{\rm nodal}
+L^d_{\rm core}[\{\br_\alpha^{v(a)}(\tau)\}]~~,
\label{avlxyd}
\end{multline}
where $f(\br) = \sum_if_i\delta(\br - \bR_i)$, $\{\bR_i\}$ are the
sites of the blue lattice, $\tilde J=J+J_1+J_2$, 
$K_0$ has been rescaled by $a^2$, and 
\begin{multline}
\dot\varphi_v (\br,\tau) =
\sum_\alpha \frac{(\br-\br_\alpha^{v}(\tau))\times\hat z}
{|\br-\br_\alpha^{v}(\tau)|^2}\cdot\dot\br_\alpha^{v} -\\
\sum_\gamma \frac{(\br-\br_\gamma^{a}(\tau))\times\hat z}
{|\br-\br_\gamma^{a}(\tau)|^2}\cdot\dot\br_\gamma^{a}
 =\int d^2r' \frac{(\br-\br')\times\hat z}
{|\br-\br')|^2}\cdot {\bf J}(\br,\tau),\\
\nabla\varphi_v (\br,\tau) =
-\sum_\alpha \frac{(\br-\br_\alpha^{v}(\tau))\times\hat z}
{|\br-\br_\alpha^{v}(\tau)|^2} \\ +
\sum_\gamma \frac{(\br-\br_\gamma^{a}(\tau))\times\hat z}
{|\br-\br_\gamma^{a}(\tau)|^2}
=-\int d^2r' \frac{(\br-\br')\times\hat z}
{|\br-\br')|^2}n(\br,\tau)
.
\label{avortexphase}
\end{multline}
In (\ref{avortexphase})  we found it useful to 
introduce vorticity density and current: 
$n(\br,\tau) =\sum_\alpha\delta(\br - \br_\alpha^{v}(\tau))-
\sum_\gamma\delta(\br - \br_\gamma^{a}(\tau))$,
${\bf J}(\br,\tau)=\sum_\alpha\dot\br_\alpha^{v}\delta(\br - \br_\alpha^{v}(\tau))-\sum_\gamma\dot\br_\gamma^{a}\delta(\br - \br_\gamma^{a}(\tau))$,
in terms of which, when combined with vortex particle density and current
$\rho(\br,\tau) =\sum_\alpha\delta(\br - \br_\alpha^{v}(\tau))+
\sum_\gamma\delta(\br - \br_\gamma^{a}(\tau))$,
${\bf j}(\br,\tau)=\sum_\alpha\dot\br_\alpha^{v}\delta(\br - \br_\alpha^{v}(\tau))+\sum_\gamma\dot\br_\gamma^{a}\delta(\br - \br_\gamma^{a}(\tau))$,
we can write 
\begin{multline}
L^d_{\rm core}=
\half\sum_\alpha M\Bigl (\frac{d\br _\alpha^{v}}{d\tau}\Bigr )^2\delta(\br - \br_\alpha^{v})+ 
\\ 
\half\sum_\gamma M\Bigl (\frac{d\br _\gamma^{a}}{d\tau}\Bigr )^2\delta(\br - \br_\gamma^{a}) +
H^d_{\rm core}~~,
\label{alcore}
\end{multline}
where
\begin{multline}
H^d_{\rm core} = V(\br)\rho (\br,\tau) 
+\half \rho (\br,\tau)\int d^2r'V^{(2)}(\br,\br')\rho(\br',\tau) \\
+\half j_k (\br,\tau)\int d^2r'V^{(2)}_{kl}(\br,\br')j_l(\br',\tau) \\ 
+\half n(\br,\tau)\int d^2r'{\cal V}^{(2)}(\br,\br')n(\br',\tau) \\+
\half {\bf J}_k (\br,\tau)\int d^2r'{\cal V}^{(2)}_{kl}(\br,\br'){\bf J}_l(\br',\tau)+(\cdots)~~.
\label{ahcore}
\end{multline}
(Anti)vortex mass terms appearing in (\ref{alcore}) have been 
introduced already in Section IV (see the discussion surrounding
Eq. (\ref{vortexmass})), while $H^d_{\rm core}$ represents a systematic
expansion in vortex core density, including single core, two core terms and
so on. All the short range terms that arise from expanding the cosine
functions (\ref{alxyd}) in continuum limit have been absorbed into 
$H^d_{\rm core}$, as was our habit throughout the text -- in particular,
$V(\br)$ is just the vortex potential on the blue lattice, containing
crucial information on $d$-wave pairing, which is extensively
discussed in Section III.

We can now integrate out $\chi$, the regular (XY ``spin-wave'') part of the
phase. The quadratic phase stiffness terms in (\ref{avlxyd}) are
decoupled as:
\begin{multline}
\frac{K_0}{2}(\dot\chi+\dot\varphi_v)^2
+\frac{\tilde J}{2}(\nabla\chi+\nabla\varphi_v)^2 \to
i W_0(\dot\chi +\dot\varphi_v) \\ + i{\bf W}\cdot (\nabla\chi+\nabla\varphi_v)
+\frac{1}{2K_0}W_0^2 + \frac{1}{2\tilde J}{\bf W}^2
\label{ahubbard}
\end{multline}
via the Hubbard-Stratonovich vector field $W=(W_0,{\bf W})$ (note
that we have set the ``dual speed of light'' $\sqrt{\tilde J/K_0}$ to
unity). Integration by parts gives
$iW\cdot\partial\chi\to -i(\partial\cdot W)\chi$ and is followed by
functional integration over $\chi(x)$, resulting in the local
$\delta$-function constraint $\delta (\partial\cdot W)$. 
The constraint is solved by introducing
a non-compact gauge field $A_d$ such that $W=\partial\times A_d$, ensuring
$\partial\cdot W=\partial\cdot (\partial\times A_d)=0$. What remains
of (\ref{ahubbard}) is further transformed by another partial integration:
\begin{multline}
i (\partial\times A_d)\cdot(\partial\varphi)_v
+\frac{1}{2K_0}(\partial\times A_d)_0^2+ \frac{1}{2\tilde J}(\partial\times A_d)_{\perp}^2 \\ \to -iA_d\cdot\partial\times(\partial\varphi)_v
+\frac{1}{2K_0}(\partial\times A_d)_0^2+ \frac{1}{2\tilde J}(\partial\times A_d)_{\perp}^2~~,
\label{ahubbardi}
\end{multline}
where $(\partial\times A_d)_{0,\perp}$ denotes temporal and  spatial components
of $\partial\times A_d$, respectively. Now observe that (\ref{avortexphase}) 
implies $\partial\times(\partial\varphi)_v =(2\pi n,2\pi {\bf J})$.
This allows us to finally write the partition function
of the quantum vortex-antivortex system as:
\begin{multline}
Z_v^d=\sum_{N_v=0}^{\infty}\sum_{N_a=0}^{\infty}\frac{1}{N_v!N_a!}
\prod_{\alpha,\gamma=1}^{N_v,N_a}
\int_{\{\br_\alpha^{v}(0)\}=\{\br_\alpha^{v}(\beta)\}}{\cal
D}\br_\alpha^{v}(\tau)\\
\times
\int_{\{\br_\gamma^{a}(0)\}=\{\br_\gamma^{a}(\beta)\}}{\cal D}\br_\gamma^{a}(\tau)
\exp\bigl(-\int d^3x
{L}_v^d\bigr)~~,
\label{azv}
\end{multline}
where $L^d_v$ equals
\begin{multline}
\half\sum_\alpha M\Bigl (\frac{d\br _\alpha^{v}}{d\tau}\Bigr )^2\delta(\br - \br_\alpha^{v})+ 
\half\sum_\gamma M\Bigl (\frac{d\br _\gamma^{a}}{d\tau}\Bigr
)^2\delta(\br - \br_\gamma^{a})\\+
{L}_{\rm nodal}+ H^d_{\rm core}
-2\pi i A_{d0}n -i2\pi ({\bf A}_d^{(0)} +{\bf A}_d)\cdot {\bf J}
\\+\frac{1}{2K_0}(\partial\times A_d)_0^2+ \frac{1}{2\tilde J}(\partial\times A_d)_{\perp}^2~~,
\label{alv}
\end{multline}
and $\nabla\times{\bf A}_d^{(0)} = {\bf B}_d^{(0)}=f(\br)\hat z$.

Eqs. (\ref{azv},\ref{alv}) are an important result of this Appendix.
We recognize $Z_v^d$ as equivalent to a partition function 
of two species of non-relativistic quantum bosons expressed 
in the Feynman path integral representation over particle
worldline trajectories. These vortex and antivortex bosons have
identical mass $M$ and carry dual charges $+2\pi$ and $-2\pi$,
respectively, through which they couple to a dynamical gauge field $A_d$.
The dual photons of $A_d$ mediate long range ``electrodynamic'' interactions
between the bosons, which are just the familiar Biot-Savart 
interactions between (anti)vortices. In addition, the particles
interact through an assortment of short range interactions contained
in $H^d_{\rm core}$ (\ref{ahcore}). Furthermore, ${L}_{\rm nodal}$ describes
the interactions generated by nodal Dirac-like fermions which
will be included explicitly once we arrive at the dual representation
of (\ref{alv}), as detailed in Section IV. Finally,
the vortex potential $V(\br)$ contains important information about
the underlying lattice structure and the symmetry of the order
parameter, as emphasized throughout the text.

We have derived  $Z_v^d$ as a continuum limit approximation to
the partition function  $Z_{XY}^d$ (\ref{azxyd}) of the quantum XY-type
model (\ref{alxyd}). Actually, in real cuprates and in all other
physical systems, the opposite is true: It is the quantum XY-type
representation that is an approximation to $Z_v^d$. $Z_v^d$ captures
the general description of the quantum vortex-antivortex system,
applicable to all superconductors and superfluids whose order
parameter is a complex scalar. To appreciate this, imagine
that for each given configuration of the phase,
with (anti)vortex positions fixed in Euclidean spacetime, the 
microscopic action of a physical system is minimized with respect
to the amplitude, after all other degrees of freedom have been
integrated out. The subsequent summation over all distinct
(anti)vortex positions leads to precisely $Z_v^d$ as the final result
(again, we remind the reader that 
the core dissipative terms will also generically appear
in $Z_v^d$ but, being small in underdoped
cuprates, are neglected here as explained in the main text). 
In practice, this procedure is difficult to carry out explicitly
and the actual values of various terms that enter $Z_v^d$ are hard
to determine from ``first principles''. This is particularly
true for core-core interaction terms appearing in $H^d_{\rm core}$.
For our purposes it will suffice to approximate 
$V^{(2)}(\br,\br')\to g\delta(\br-\br')$, where $g>0$, and drop the rest.

There is one crucial feature which distinguishes $Z_v^d$ from the
standard Feynman partition function: The vortex and
antivortex quantum bosons are {\em not conserved}. As particles
make their way through imaginary time, vortices and antivortices can
{\em annihilate} each other; similarly, they can also be {\em created}
at an instant in time; this is depicted in 
Fig. \ref{vortexloops}. All such processes of creation
and annihilation proceed in {\em pairs} of vortices and antivortices.
Consequently, while the individual number of vortices and antivortices
is not conserved, the {\em vorticity}, measured by dual charge
$e_d=\pm 2\pi$, {\em is conserved} and the gauge symmetry associated
with $A_d$ is always maintained (unless, of course, it is spontaneously
broken by a dual Higgs mechanism in dual superfluid). In other words,
these nonrelativistic (anti)vortex bosons propagate through spacetime
permeated by a vortex-antivortex condensate, of strength $\Delta_v$.

Feynman path integrals are beautiful but difficult to calculate with.
Following the standard mapping \cite{negele} we can express $Z_v^d$
as a functional integral over complex fields
$\Psi_v(\br,\tau)$ and $\Psi_a(\br,\tau)$, which are the eigenvalues
of vortex and antivortex annihilation operators, respectively, in
the basis of coherent states:
\begin{eqnarray}
&Z_v^d&\to \int{\cal D}\Psi_v{\cal D}\Psi_a\int{\cal D}A_d 
\exp\bigl[-\int d^3x \bigl(\Psi_v^*(\partial_\tau +ie_dA_{d0})\Psi_v 
\nonumber \\
&+&
\Psi_a^*(\partial_\tau -ie_dA_{d0})\Psi_a +\nonumber \\
&+&\frac{1}{2M}|(\nabla +ie_d{\bf A}_d^{(0)} +ie_d{\bf A}_d)\Psi_v|^2 \nonumber \\
&+&
\frac{1}{2M}|(\nabla -ie_d{\bf A}_d^{(0)} -ie_d{\bf A}_d)\Psi_a|^2 
-\mu_v(|\Psi_v|^2+|\Psi_a|^2)\nonumber \\
&+& V(\br)(|\Psi_v|^2+|\Psi_a|^2) 
+\Delta_v \Psi_v^*\Psi_a^* + \Delta_v^* \Psi_a\Psi_v \nonumber \\
&+& \frac{g}{2}(|\Psi_v|^2+|\Psi_a|^2)^2 +
{L}_{\rm nodal}+ \frac{1}{2K_\mu}(\partial\times A_d)_\mu^2\bigr)\bigr]~~,
\label{azquantum}
\end{eqnarray}
where the meaning of various terms is straightforward in light of
our earlier discussion; we have consolidated the notation so that
$e_d=2\pi$, $K_\mu=(K_0,\tilde J,\tilde J)$, the chemical potential
for vortices is $\mu_v=\mu_a$, and $g$ describes the short range 
core-core repulsions. The ``vortex-antivortex pairing'' function $\Delta_v$ is
crucial since it regulates the frequency of vortex-antivortex pair
creation and annihilation processes.

The form of $Z_v^d$ can be further simplified by exploiting the
vorticity conservation law. We observe that the action in
(\ref{azquantum}) is invariant under gauge transformations:
$\Psi_v\to\exp(i\zeta)\Psi_v$, $\Psi_a\to\exp(-i\zeta)\Psi_a$,
$A_d\to A_d -(\partial\zeta)/e_d$. This prompts us to
introduce bosonic ``spinors'' 
$\bar\Psi = (\Psi_v^*,\Psi_a)$ which carry a conserved
dual charge $e_d$ and couple minimally to $A_d$:
\begin{multline}
Z_v^d\to \int{\cal D}\Psi{\cal D}A_d {\cal D}\sigma
\exp\bigl[
 -\int d^3x \bigl(\bar\Psi {\cal L}\Psi \\+ {L}_{\rm nodal} +\frac{1}{2g}\sigma^2 +
 \frac{1}{2K_\mu}(\partial\times A_d)_\mu^2\bigr)
 \bigr]
~~,
\label{azspinor}
\end{multline}
where
\begin{multline}
{\cal L} = \begin{bmatrix}
(\partial_\tau +ie_dA_{d0})
&
\Delta_v \\
\Delta_v^*
&
-(\partial_\tau +ie_dA_{d0}) \\
\end{bmatrix} \\
+ \bigl(-\frac{1}{2M}(\nabla +ie_d{\bf A}_d^{(0)} +ie_d{\bf A}_d)^2 -\mu_v
+V(\br) + i\sigma\bigr)\openone~~,
\label{ahspinor}
\end{multline}
and a Hubbard-Stratonovich scalar field $\sigma$ was deployed to decouple
short range repulsion. Now we set $A_d\to 0$ and 
ignore ${L}_{\rm nodal}$-- they will be easily
restored later --  and note that the integration over vortex matter
fields $\Psi$ gives:
\begin{multline}
Z_v^d\to \int{\cal D}\sigma\exp\bigl(-\int d^3x\frac{1}{2g}\sigma^2\bigr)
\times
\\
{\rm det}\begin{bmatrix}
\partial_\tau -\frac{1}{2M}\nabla^2 -\mu_v
+V(\br) + i\sigma
&
\Delta_v \\
\Delta_v^*
&
-\partial_\tau -\frac{1}{2M}\nabla^2 -\mu_v
+V(\br) + i\sigma\\
\end{bmatrix}~~.
\label{adet}
\end{multline}
The above partition function has a transition at $\mu_v=-|\Delta_v|$ 
(we are assuming that the minimum of $V(\br)$ occurs at zero, with
$E_c^r$, or $E_c^b$ as the case may be, having been absorbed into
$\mu_v$). For $\mu_v<-|\Delta_v|$ the system of (anti)vortex bosons
is in its ``normal'' state, with $\langle\Psi\rangle =0$. For
$\mu_v>-|\Delta_v|$ (anti)vortex bosons condense and $\langle\Psi\rangle$
becomes finite. This is nothing but the dual description  of the
superconducting transition discussed in the main text. In the general
vicinity of the transition it is useful to introduce 
$m^2=\mu_v^2-|\Delta_v|^2$, where now $m^2>0$ and $m^2< 0$ indicate
dual normal and superfluid states, respectively.
Focusing on distances longer than $\sqrt{\Delta_vM}$ and energies
lower than $\Delta_v$, the determinant in (\ref{adet}) can be
further reduced to:
\begin{multline}
{\rm det}\Bigl[
-\partial_\tau^2-\frac{|\Delta_v|+V(\br)+i\sigma}{2M}\nabla^2+m^2-
\frac{1}{M}[(\nabla^2V)+i(\nabla^2\sigma)]\\ +
2|\Delta_v|(V(\br)+i\sigma)+(V(\br)+i\sigma)^2\Bigr]+(\cdots)
~~,
\label{adeti}
\end{multline} 
with additional terms $(\cdots)$ contributing unimportant derivatives.

By setting  $V(\br)$ and $\sigma(x)$ to zero, we 
observe that the above expression
assumes the form of the partition function determinant for a system of 
{\em relativistic} quantum bosons
of mass $m$ :
${\rm det}\Bigl[-\partial_\tau^2-c^2\nabla^2+m^2c^4\Bigr]$, with
the speed of ``light'' $c=\sqrt{|\Delta_v|/2M}$ set to unity henceforth.
The terms involving $V(\br)$ and $\sigma(x)$
describe the underlying
potential and various short range interactions of these relativistic bosons.
We therefore can reexpress the determinant (\ref{adeti}) as a functional
integral over the relativistic boson field $\Phi(x)$; this is a faithful
representation of the original (anti)vortex partition function at
distances longer than $\sqrt{\Delta_vM}$ and energies lower than $\Delta_v$:
\begin{eqnarray}
&Z_v^d \to {\cal Z}_d=\int{\cal D}\Phi\int{\cal D}A_d\int {\cal D}\sigma
\nonumber \\
&\exp\bigl[-\int d^3x \bigl(
|(\partial +i2\pi A_d)\Phi |^2 + m^2(\br)|\Phi |^2 \nonumber \\
&+2|\Phi|^2[|\Delta_v|+V(\br)-\frac{\nabla^2}{2M}](i\sigma)+
\frac{|{\bf D}\Phi|^2}{M}(i\sigma) \nonumber \\
&+
[\frac{1}{2g}-|\Phi|^2]\sigma^2+ {\cal L}_{\rm nodal}
+\frac{1}{2K_\mu}(\partial\times A_d)_\mu^2
\bigr],
\label{azdual}
\end{eqnarray}
where $m^2(\br) = m^2 +2|\Delta_v|V(\br)+V(\br)^2- (\nabla^2V(\br)/M)$
and we have restored ${\cal L}_{\rm nodal}$ and dual gauge field $A_d$, 
through covariant derivatives 
$\partial\to D=(D_0,{\bf D})=(\partial_0+i2e_dA_{d0},\nabla+ie_d{\bf A}_d^{(0)} +ie_d{\bf A}_d)$.
The minimal coupling of $A_d$ is mandated by dual charge conservation:
$\Phi\to\exp(i\zeta)\Phi$, $A_d\to A_d -(\partial\zeta)/e_d$.

The dual partition function ${\cal Z}_d$ 
(\ref{azdual}) is the final result of this
subsection. It describes the system of relativistic
quantum bosons of mass $m$ and charge $e_d=2\pi$ in a magnetic field
${\bf B}_d=\nabla\times{\bf A}_d^{(0)}$. 
The virtual particle-antiparticle creation
and annihilation processes in the vacuum of this theory are nothing
but quantum vortex-antivortex pair excitations evolving in imaginary
time (see Fig. \ref{vortexloops}). In the ``normal'' 
vacuum, $m^2>0$, the average size of such pairs is
$\sim m^{-1}$. This is just the superconducting ground state of
physical underdoped cuprates.
For $m^2<0$, this ``normal'' vacuum is unstable to a
Higgs phase, with a finite dual 
condensate $\langle\Phi\rangle$. The vortex-antivortex pairs unbind
as infinite loops of virtual
particle-antiparticle excitations of $\Phi$
permeate this dual Higgs vacuum -- this is
the pseudogap state of cuprates.
The integration over the Hubbard-Stratonovich field $\sigma$ produces
a short range repulsion $\half (4|\Delta_v|^2)g|\Phi|^4$ followed 
by an assortment of other short range interactions including $V(\br)$,
powers of $|\Phi|$ higher than quartic and various derivatives.
All these additional interactions are irrelevant in the sense of
long distance behavior but might play some quantitative role at intermediate
lengthscales. For simplicity, we shall mostly ignore them in this paper.
Finally, with the change of notation $4|\Delta_v|^2g\to g$ and
${\cal L}_{\rm nodal}$ incorporated into the self-action for $A_d$
as detailed in Section IV,
the dual Lagrangian in (\ref{azdual}) reduces to ${\cal L}_d$ 
(\ref{dduallagrangian}). The arguments in the text can then be
followed to arrive at ${\cal H}_d$ (\ref{ltightbinding},\ref{hofstadter}).

\subsection{${\cal H}_d$ in the Villain approximation}

A useful approximation to the ordinary XY model is due to Villain. In this
approximation the exponential of the 
cosine function is replaced by an infinite sum of the exponentials
of parabolas. We will illustrate this approximation first for the 2D case and 
follow up with the (2+1)D quantum XY-like model.


\subsubsection{Classical (thermal) phase fluctuations}

Here we apply the Villain approximation to our classical XY model of a 
phase fluctuating two-dimensional $d$-wave
superconductor (\ref{hxy}). As we will see presently, the final
Coulomb gas representation coincides with 
(\ref{latticecoulomb}) -- however, the advantage of the Villain
approximation is that it will allow us to obtain explicit expressions
for the core energies $E_c^{r(b)}$ in terms of the coupling constants
of the original Hamiltonian (\ref{hxy}).

Denoting the sites of the blue lattice by $\brho  = (x, y)$,
the partition function of the model can be written as
\begin{multline}
Z= \int  \left(\prod_{\brho} d\phi_{\brho}\right)
\exp
  \sum_{\brho}  \Bigl[
                   J\left(
         		   \cos(\nabla_{\hx}\phi_{\brho})
			   + \cos(\nabla_{\hy}\phi_{\brho}) 
		   \right)\\
		   + J_{12}(\brho)
		   \left(  \cos(\phi_{\brho+\hx+\hy}-\phi_{\brho})
 		           +\cos(\phi_{\brho+\hy}-\phi_{\brho+\hx})
		   \right)       
	       \Bigr]
\label{Z0}
\end{multline}
where the lattice operator $\nabla$ is defined according to $\nabla_{\bdelta}
f(\br) = f_{\br+\bdelta}-f_{\br}$,  and  coefficients $J_{12}(\brho)$ denote $J_1$ (or $J_2$) if $\brho$ is in 
the lower left corner of a red (or black) plaquette.

In Villain approximation the exponent of a cosine is 
replaced by a sum of Gaussian exponents that has the
same periodicity $2\pi$:
\begin{equation}\label{villain}
\exp [\beta\cos\gamma]
\approx
R_V(\beta)
\sum_{n=-\infty}^{+\infty}\exp [-\frac{\beta_V(\beta)}{2}(\gamma-2\pi n)^2]~~.
\end{equation}
The fitting functions $\beta_V(\beta)$ and $R_V(\beta)$ are determined by the
requirement that the lowest Fourier coefficients of 
the two functions coincide. In particular, the 
function $\beta_V(\beta)$ has the following asymptotic behavior for
low and high temperatures:
\begin{equation}
\beta_V(\beta)\approx  \left\{
\begin{array}{ll}
\beta & {\text for } \beta\gg 1\\
\left(2\ln\frac{\beta}{2}\right)^{-1} & {\text for } \beta\ll 1
\end{array}
\right.
\label{betaV}
\end{equation}
The sum over $n$ in (\ref{villain}) can be transformed 
by  decoupling the quadratic term in the exponent via the
Hubbard-Stratonovich transformation:
\begin{equation}
\label{villain1}
\exp [\beta\cos\gamma]\propto
\sum_{n=-\infty}^{\infty}\exp [-\frac1{2\beta_V(\beta)} n^2+i \gamma n]~~.
\end{equation}
By introducing integer fields $u_x(\brho)$, $u_y(\brho)$
and $w_{\pm}(\brho)$ to apply (\ref{villain1}) to the
cosine terms in (\ref{Z0}) we obtain:
\begin{multline}
Z =\sum_{u_{\alpha}} \sum_{w_{\pm}}\int \prod_{\brho} d\phi_{\brho} 
e^{-\sum_{\brho}
  \left[
   \frac{u_{\alpha}^2(\brho)}{2J'}
   +\frac{w_{+}^2(\brho)+w_{-}^2(\brho)}{2J_{12}'(\brho)}
  \right]}
\times\\
e^{i \sum_{\brho}
  \left[
     u_{\alpha}(\brho)\nabla_{\alpha} \phi_{\brho}+
     w_+(\brho)(\phi_{\brho+\hx+\hy}-\phi_{\rho})+
w_-(\brho)(\phi_{\brho+\hy}-\phi_{\brho+\hx})
  \right]
}~~,
\label{Z1}
\end{multline}
where $\alpha=x,y$, and  coefficients $J_{12}(\brho)$ and $J'$ are defined as
\begin{align}
J'&=\beta_V( J)\\
J_{12}(\brho)&=\beta_V(J_{12}(\brho))~~.
\end{align}
The integration over the angles $\phi_{\brho}$ can  be performed after
applying the discrete analog of the integration by parts to 
the sums in the last exponent in (\ref{Z1}):
\begin{multline}
\sum_{x} f(\brho) \nabla_x g(\brho) =
\sum_{x} f(\brho) (g(\brho+\hx)-g(\brho))\\
=
\sum_{x} (f(\brho-\hx)-f(\brho)) g(\brho) = - \sum_x\overline{\nabla}f(\brho) \; g(\brho)~~.
\label{sumbyparts}
\end{multline}
Note that we distinguish between the ``right difference'' operator $\nabla$
and ``left difference'' operator $\overline{\nabla}$. Integration over the
phases $\phi_{\brho}$  yields
the following expression for the partition function:
\begin{equation}
\sum_{u_x,u_y. w_{\pm}}
e^{-\sum_{\brho}
  \left[
    \frac{u_x^2(\brho)+u_y^2(\brho)}{2J'}+
    \frac{w_+^2(\brho)+w_-^2(\brho)}{2J_{12}'(\brho)}
  \right]
   }
   \delta(\overline{\mathbf\nabla}\cdot\bv(\brho)+{\ldots})
~~.
\end{equation}
where $\overline{\mathbf \nabla}\cdot\bv(\brho)$ denotes
two-dimensional lattice divergence of $\bv(\brho)$ and dots denote
$$
(w_+(\brho)-w_+(\brho-\hx-\hy))+(w_-(\brho-\hx)-w_-(\brho-\hy)).
$$
We rewrite the constraint appearing as the Kronecker 
delta function in the sum as
$$
\lnabla\cdot (\bv+{\mathbf w}[w_+,w_-]) = 0~~,
$$
where ${\mathbf w}=(w_x,w_y)$ denote the following  
linear combinations of integer fields  $w_{\pm} (\brho)$:
\begin{align}
\label{w_xy}
w_x &= \frac{w_+(\brho)+w_+(\brho-\hy)}{2}-\frac{w_-(\brho)+w_-(\brho-\hy)}{2}\\
w_y &= \frac{w_+(\brho)+w_+(\brho-\hx)}{2}+\frac{w_-(\brho)+w_-(\brho-\hx)}{2}~~.
\end{align}
The constraint can then be resolved as
$$
\bv = \bmB -{\mathbf w}[w_+,w_-]~~,
$$
where
\begin{equation}
\bmB = (\lnabla_y \Lambda, -\lnabla_x \Lambda)
\label{magField}
\end{equation}
and  $\Lambda(\brho)$ has the meaning of the time-like 
component of a vector potential.
At this point it is useful to
pause and  establish a simple geometrical interpretation of the 
various fields we have introduced.
Variables $u_x(\brho)$ and $u_y(\brho)$ are coupled to the phase 
differences $\phi_{\brho+\hx}-\phi_{\brho}$
and $\phi_{\brho+\hy}-\phi_{\brho}$
and therefore they reside on the links emanating from $\brho$ 
in positive $x$ and $y$ directions respectively.
Integers $\Lambda(\brho)$, on the other hand are related to link variable 
$u_x(\brho)$ through the difference $\Lambda(\brho)-\Lambda(\brho-\hy)$. Consequently we must associate $\Lambda(\brho)$ with the
centers of the blue plaquettes, which coincide 
with either red or black sites. Nevertheless, we will
continue to use notation $\Lambda(\brho)$ tacitly 
implying that $\brho$ refers to
the lower left corner of the (red or black) 
plaquette associated with $\Lambda$.
Having resolved the constraint,  we find that  
the partition function  $Z$ can be written in terms of integer-valued fields
$w_{\pm}(\brho)$ and $\Lambda(\brho)$ only:
\begin{equation}
\sum_{\Lambda, w_{\pm}}
\exp \sum_{\brho}
\left[
  -\frac{(\bmB[\Lambda]-{\mathbf w}[w{_{\pm}}])^2_{\perp}}{2J'}
  -\frac{w_+^2(\brho)+w_-^2(\brho)}{2J_{12}'(\brho)}
\right].
\end{equation}
To obtain the description in terms of continuous rather than integer-valued
fields $\Lambda(\brho)$,  we use the Poisson summation formula:
\begin{equation}
\label{poissonsummation}
\sum_{\Lambda=-\infty}^{\infty} f(\Lambda) = \int_{-\infty}^{\infty}
d\Lambda \sum_{l=-\infty}^{\infty} e^{2\pi i
l \Lambda} f(\Lambda)~.
\end{equation}
The partition function $Z$ assumes the following form:
\begin{equation}
\int\limits_{-\infty}^{\infty}  \prod_{\brho} d\Lambda(\brho)
\sum_{\bl(\brho)}
\exp\sum_{\brho} 2\pi i l(\brho) \Lambda(\brho)\}
\exp\{F[\bmB[\Lambda(\brho)]]~,
\label{Z4}
\end{equation}
where we have defined a functional 
$\exp\{F[\bmB(\brho)]\}$
according to 
\begin{equation}
 \sum_{w_{\pm}} 
 \exp
 \left[
   \sum_{\brho}
      \left(
      -\frac{(\bmB-{\mathbf w}[w{_{\pm}}])^2_{\perp}}{2J'}
      -\frac{w_+^2(\brho)+w_-^2(\brho)}{2J_{12}'(\brho)}
      \right)
\right]~.
\end{equation}
%

In the limit when constants $J_{1}$ and $J_2$ are 
infinitesimally small, only the configurations 
$w_{\pm}(\brho) =0 $ contribute to the $F[\bmB]$. This limit, 
which corresponds to  
the usual  2D XY model, is described by partition function $Z_0$ given by
\begin{equation}
\int\limits_{-\infty}^{\infty}  \prod_{\brho} d\Lambda(\brho)
\sum_{\bl(\brho)}
\exp\left[
 \sum_{\brho}\left(
                2\pi i l(\brho) \Lambda(\brho)
                 -\frac{(\lnabla_{\alpha} \Lambda)^2}{2J'}
           \right)
\right].
\label{Zsimple}
\end{equation}

For finite $J_{12}$ we must resort to approximate 
evaluation of the functional $F(\mB(\brho))$:
\begin{multline}
\exp\{F[\mB_x(\brho),\mB_y(\brho)]\} =
\sum_{w_{\pm}}
\exp
\Bigl[-\sum_{\brho}
  \Bigl(
    \frac{(\mB_{\alpha}-w_{\alpha})^2}{2J'}\\
+\frac{w_+^2(\brho)+w_-^2(\brho)}{2J_{12}'(\brho)}\Bigr)\Bigr]~.
\end{multline}
The quadratic terms containing $\bmB$ can be decoupled using the
Hubbard-Stratonovich transformation:
\begin{multline}
e^{F[\bmB(\brho)]} = \int \prod_{\brho}dZ_x(\brho)dZ_y(\brho) 
\times\\
\sum_{w_{\pm}}
\exp
\Bigl[
  \sum_{\brho}
  \Bigl(
     -\frac{J'Z_{\alpha}^2(\brho)}{2}\\
     +iZ_{\alpha}(\brho)\left(b_{\alpha}(\brho)-w_{\alpha}(\brho)\right)
     -\frac{w_+^2(\brho)+w_-^2(\brho)}{2J'_{12}(\brho)}
  \Bigr)
\Bigr]
,
\end{multline}
where $\alpha = x,y$. Using explicit expressions for $w_{\alpha}$ we obtain
\begin{multline}
e^{F[\bmB]} =
\int \prod_{\brho}dZ_x(\brho)dZ_y(\brho)\times\\ 
\sum_{w_{\pm}} \exp 
\Bigl[
\sum_{\brho}
 \Bigl(
 -\frac{J' Z_{\alpha}^2(\brho)}{2} 
 -\frac{w_+^2(\brho)+w_-^2(\brho)}{2J'_{12}(\brho)}
 +iZ_{\alpha}(\brho)\mB_{\alpha}(\brho)\\
 - \frac{i}{2} w_{+}(\brho)(Z_x(\brho)+Z_x(\brho+\hy)+Z_y(\brho)+Z_y(\brho+\hx))\\
 - \frac{i}{2} w_{-}(\brho)(-Z_x(\brho)-Z_x(\brho+\hy)+Z_y(\brho)+Z_y(\brho+\hx))
\Bigr)
\Bigr].
\end{multline}
The sums over $w_{\pm}(\brho)$  can be performed by employing
the Villain approximation
(\ref{villain}) backward. Note that the coupling constants $J_{\pm}'$ are
restored to the original values of coupling constants $J_{\pm}$:
\begin{multline}
e^{F[\bmB(\brho)]} =  \int \prod_{\brho}dZ_x(\brho)dZ_y(\brho)\times\\
\exp
\Bigl[
  \sum_{\brho}
  \Bigl(
     -\frac{J' Z_{\alpha}^2(\brho)}{2}
     +iZ_{\alpha}(\brho)\mB_{\alpha}(\brho)\\
     +J_{12}(\brho)\Bigl(\cos\frac{Z_x(\brho)+Z_x(\brho+\hy)+Z_y(\brho)+Z_y(\brho+\hx)}{2}\\
+\cos\frac{-Z_x(\brho)-Z_x(\brho+\hy)+Z_y(\brho)+Z_y(\brho+\hx)}{2}\Bigr)
  \Bigr)
\Bigr]~.
\end{multline}
To quadratic order, the expression in the exponent is 
\begin{multline}
\sum_{\brho}
\Bigl[
-\frac{J' Z_{\alpha}^2(\brho)}{2}
+iZ_{\alpha}(\brho)\mB_{\alpha}(\brho)
-J_{12}(\brho)\times\\
\frac{
  (Z_x(\brho)+Z_x(\brho+\hy))^2+(Z_y(\brho)+Z_y(\brho+\hx))^2
   }{4}
\Bigr].
\label{Gquad}
\end{multline}
Note that $Z_x$ and $Z_y$ components are completely decoupled 
at the quadratic level. To proceed, one can
double the unit cell, in which case the expression in the 
exponent becomes diagonal in the momentum space. Alternatively, one can use an
equivalent, but technically simpler procedure of keeping the 
original unit cell. In this latter case, the
momentum space problem reduces to the diagonalization 
of a $2\times 2$ matrix connecting modes at
wavevectors $\bq$ and $\bq-\bg$, where ${\mathbf g} = \pi(\hx+\hy)$.

It is convenient to represent $J_{12}(\brho)$ as
$$
J_{12}(\brho) = \frac{J_1+J_2}{2}+ e^{i \bg \brho} \frac{J_1-J_2}{2} \equiv
\Jbar + \delta J  e^{i {\mathbf g} \brho}~~.
$$
After Fourier transformation, the exponent 
in  (\ref{Gquad}) becomes
\begin{multline}
iZ_x(\bq)\mB_x(-\bq) -Z_x(\bq)Z_x(-\bq) \frac{J'+\Jbar(1+\cos q_y)}{2}
\\
-\frac{i \delta J }{2}\sin q_y  Z_x(\bg-\bq)Z_x(\bq) + (x\leftrightarrow y)
~~.
\end{multline}
Since $Z_{\alpha}(\brho)$ and $\mB_{\alpha}(\brho)$ 
are real, their Fourier components satisfy 
\begin{align}
Z_{\alpha}(-\bq)&=Z_{\alpha}^*(\bq)\\
\mB_{\alpha}(-\bq)&=\mB_{\alpha}^*(\bq)~~.
\end{align}
In the last expression for the partition function 
we found that the terms with $Z_x$ 
and $Z_y$  decouple and we thus can integrate 
over $Z_x(\brho)$ and $Z_y(\brho)$ separately. 

The expressions in this and especially the next subsection can be
significantly economized by using a check mark to denote
two-component vectors:
$$
\spb(\bq) = 
\begin{pmatrix}
b(\bq)\\
b(\bq-\bg)
\end{pmatrix}
$$
Using this notation, the contribution due to $Z_x(\brho)$ can be
written as
%
\begin{equation}
\frac{i}{2}
\spb_x^T(-\bq) \spZ_x(\bq)
-\frac14
\spZ_x^T(-\bq)({\ldots})
\spZ_x(\bq).
\end{equation}
where 
$({\ldots}) =J'+\Jbar+\Jbar\cos q_y\sigma_3 + \delta J \sin q_y\sigma_2$ and
superscript $T$ denotes the transpose of a matrix.
Now $Z_{\alpha}$ can be integrated out. Apart from the overall
normalization constant, $F[\bmB]$ is given by:
\begin{equation}
F[\bmB(\brho)] =
-\frac14\int \frac{d\bq}{(2\pi)^2} \left[
\spb_x^T(-\bq) G(q_y) \spb_x(\bq)
+(x\leftrightarrow y)
\right]~,
\label{F_of_B}
\end{equation}
where matrix $G$ is defined as
%
\begin{equation}
G=\frac1{\Delta(q_y)}
(J'+\Jbar -\Jbar \cos q_y \sigma_z-\delta J \sin q_y \sigma_2)
\label{matrixG}
\end{equation}
and the determinant $\Delta$ equals
\begin{equation}
\Delta(q_y) =
J'(J'+2\Jbar)  +(\Jbar^2+(\delta J)^2) \sin^2q_y~.
\end{equation}
When $\Jbar=\delta J = 0$ we find
$$
F[\bmB(\bq)] \to -\frac1{2J'} \mB_{\alpha}(\bq) \mB_{\alpha}(-\bq)~~,
$$
which restores the limit of an ordinary 2D XY model (\ref{Zsimple}).


Returning to the partition function (\ref{Z4}) and
using the expression (\ref{F_of_B}) for $F[\bmB]$ we just found, we are now
in position to integrate out the gauge 
field $\Lambda(\brho)$ and obtain the analogue of the Coulomb gas
representation for our model.  Note that (\ref{magField}) implies
\begin{align}
\mB_x(\bq) &= \left(1-e^{-iq_y}\right) \Lambda(\bq)\\
\mB_y(\bq) &= - \left(1-e^{-iq_x}\right) \Lambda(\bq)~~.
\end{align}
The partition function now becomes
\begin{multline}
Z =
\sum_{\bl(\brho)}\int_{-\infty}^{\infty}  \prod_{\brho} d \Lambda(\brho) 
\times\\
\exp 
\left[\int  \frac{d\bq}{(2\pi)^2}
 \left(
   i \pi 
     \spl^T(-\bq) \spLambda(\bq)
   -
     \spLambda(-\bq)    
     \tilde{M}
     \spLambda(\bq)
 \right)
\right]~~,   
\label{ZZ}
\end{multline}
where $2\times 2$ matrix $\tilde{M}$ is given by
\begin{multline}
\tilde{M}=
\frac1{2\Delta(q_y)}
\Bigl(
   (1-\sigma_3\cos q_y)\bigl[J'+\Jbar(1-\sigma_3\cos q_y)\bigr] \\
   +\sigma_1 \delta J \sin^2 q_y)
\Bigr)
+(x\leftrightarrow y)~~.
\label{Mtilde}
\end{multline}
After integration over $\Lambda(\brho)$ we obtain
\begin{equation}
Z =
\sum_{\bl(\brho)}
\exp
  \Bigl[
     -\frac{\pi^2}{4}\int \frac{d\bq}{(2\pi)^2}
     \spl^T(-\bq) M(\bq)  \spl(\bq)
   \Bigr]
~~,
\label{ZZ1}
\end{equation}
where matrix $M$ is the inverse of $\tilde{M}$. 
The elements of matrix $M$ satisfy the following 
simple identities:
\begin{align}
M_{11}(\bq) &= M_{22}(\bq-\bg)\\
M_{12}(\bq) &= M_{21}(\bq)~~.
\end{align}
Consequently, the integrand in the exponent of partition function (\ref{ZZ1}) can be written as
\begin{equation}
2   \Bigl(
     l(-\bq)M_{11}(\bq)l(\bq)+
     l(\bg-\bq)M_{12}(\bq) l(\bq)
     \Bigr)
~~.
\label{M11M12}  
\end{equation}
The explicit form of $M=\tilde{M}^{-1}$ is rather cumbersome. 
Fortunately, we will only need the leading
and subleading order terms in the long wavelength ($q\to 0$) expansion:
\begin{eqnarray}
M_{11}(\bq)&=&
   \frac{4(J'+2\Jbar)}{q_x^2+q_y^2}\nonumber \\
   &+& \frac{(J'-4\Jbar+18\dJ)(q_x^4+q_y^4)+12\dJ q_x^2q_y^2}
   {3J'(q_x^2+q_y^2)^2}\nonumber \\
&+&O(q^2)\\
M_{12}(\bq)&=& -\dJ+O(q^2)~~.
\end{eqnarray}
The terms of order $O(q^2)$  correspond to $M(\brho-\brho')$ 
decreasing at least as fast as $|\brho-\brho'|^{-4}$.
Returning to the real space representation, we obtain
\begin{multline}
\sum_{\bl(\brho)}
\exp\{-\frac{\pi^2}{2}\sum_{\brho, \brho'}
   \bigl(
     l(\brho) M_{11}(\brho-\brho') l(\brho')+\\
     e^{i\bg\cdot\brho}l(\brho)M_{12}(\brho-\brho') l(\brho')
     \bigr)
  \}~~,
\label{ZZ2}
\end{multline}
where 
\begin{equation}
M_{\alpha\beta}(\brho-\brho') = 
\int_{-\pi}^{\pi} \frac{dq_x}{2\pi}
\int_{-\pi}^{\pi} \frac{dq_y}{2\pi} 
e^{i\bq \cdot (\brho-\brho')} M_{\alpha\beta}(\bq)~~.
\label{Mft}
\end{equation}
The two terms in the exponent of (\ref{ZZ2}) are easy to interpret. 
Recall that integers $l(\brho)$ coupled to $\Lambda(\brho)$
effectively reside at the centers of the plaquettes of the blue lattice
corresponding to either black or red sites in Fig. \ref{fig1}. 
The terms containing $M_{11}$ clearly describe the average 
interaction between two plaquettes {\em irrespective} of their ``color'', 
while the terms with $M_{12}$ reflect
the difference between the red and black sites. For example, 
the strength of interaction between two black
sites separated by two lattice spacings with $\brho=0$ 
and $\brho=2\hx$ will be different from interaction
between two red sites at $\brho=\hx$, $\brho'=3\hx$  
due to the factor $\exp(i\bg\cdot\brho)$ that multiplies
$M_{12}$.

At large distances the Fourier transform can be evaluated 
by comparison to the standard lattice 
Green's function in two dimensions. The difference
$$
M_{11}(\bq) - 4(J'+2\Jbar) \frac1{4-2\cos q_x-2\cos q_y}
$$
is finite  at $\bq=0$, and therefore the Fourier
transform of this difference vanishes at large distances. Thus
\begin{multline}
M_{11}(\brho) = 4 (J'+2\Jbar)\times\\
\int_{-\pi}^{\pi} \frac{dq_x}{2\pi}\int_{-\pi}^{\pi} 
\frac{dq_y}{2\pi} \frac{e^{i\brho \bq}}{4-2\cos q_x-2\cos q_y}+{\ldots}~~~.
\end{multline}
Using the well known asymptotic behavior of the last integral \cite{kleinert}, we find
\begin{multline}
M_{11}(\brho-\brho') = M_{11}(0)- \\
-\frac{4 (J'+2\Jbar)}{2\pi} 
\left[\ln |\brho-\brho'| +()+{\ldots} 
\right]~~,
\label{M11}
\end{multline}
where  $C_1$ can be related to  the
Euler-Mascheroni constant $\gamma\approx 0.5772$  as $C_1=\gamma +\ln \left(2\sqrt{2}\right)$.
Note that $M_{11}(\brho)$ formally logarithmically  
diverges because $M_{11}(\bq)$ is
proportional to $q^{-2}$
at small momenta. The difference, $M_{11}(\brho)-M_{11}(0)$, 
however, is finite. The overall infinite
additive constant has a simple physical 
interpretation, just like for an ordinary two-dimensional XY model,
as will become clear in a moment.

The real space expression for $M_{12}$ can be easily calculated directly:
\begin{equation}
M_{12}(\brho-\brho') = -\delta J \; \delta_{\brho\brho'}+{\ldots},
\label{M12}
\end{equation}
where {\ldots} denotes terms that decrease at least as fast
as $|\brho-\brho'|^{-4}$. Combining (\ref{Mft}),
(\ref{M11}), and (\ref{M12}), we obtain  the following expression, after
separating off the terms with $\brho=\brho'$:
\begin{multline}
Z = \sum_{l(\brho)} 
\exp
\Bigl[
   -\frac{\pi^2}{2}
   \Bigl(   
   M_{11}(0)\sum_{\brho} l^2(\brho)+\\
   +\sum_{\brho\neq\brho'} M_{11}(\brho-\brho')l(\brho)l(\brho')
   -\delta J \sum_{\brho} e^{i\bg\cdot\brho} l^2(\brho)
   \Bigr)
\Bigr].
\end{multline}
By applying the long distance expansion of 
$M_{11}(\brho)$ in the sum containing terms with 
$\brho\neq\brho'$ we find that the partition function $Z$ becomes
\begin{multline}
\sum_{l(\brho)} 
\exp
\Bigl[
   -\frac{\pi^2}{2}
   \Bigl(   
   M_{11}(0)\Bigl(\sum_{\brho} l(\brho)\Bigr)^2  -\delta J \sum_{\brho} e^{i\bg\cdot \brho} l^2(\brho)\\
   -\frac{4(J'+2\Jbar)}{2\pi}\sum_{\brho\neq\brho'}
l(\brho)l(\brho')\Bigl( \ln |\brho-\brho'|+ C_1\Bigr) 
   \Bigr)
\Bigr]~.
\end{multline}

We now return to the discussion of the 
formally divergent constant $M_{11}(0)$. This divergence is a
reflection of the logarithmic dependence of a single vortex energy on the 
system size. If the number of the sites $N$ were finite, we would 
have obtained a constant of order $\ln N$ for $M_{11}(0)$ 
instead of an outright divergence. Although finite, this constant becomes
large in the thermodynamic limit $N\to \infty$, with the effect of suppressing
all configurations of the integer-valued field
$l(\brho)$ except those that satisfy 
\begin{equation}
\sum_{\brho} l(\brho)= 0~~.
\label{neutrality}
\end{equation}
This is nothing but the charge neutrality condition in
the partition function of a 2D Coulomb plasma. 
Restricting ourselves only to such configurations, 
we obtain:
\begin{multline}
Z = \sum_{l(\brho)} 
\exp
\Bigl[
   \frac{\pi^2}{2}
   \Bigl(   
   \frac{4(J'+2\Jbar)}{2\pi}\sum_{\brho\neq\brho'}
   l(\brho)l(\brho')\bigl( \ln |\brho-\brho'|\\
+ C_1\bigr)
   +\delta J \sum_{\brho} e^{i\bg\cdot \brho}l^2(\brho)
   \Bigr)
\Bigr]~.
\end{multline}
A further simplification is achieved by noticing that 
$$
\sum_{\brho\neq\brho'}l(\brho)l(\brho') =
\left(\sum_{\brho} l(\brho)\right)^2-\sum_{\brho} l^2(\brho)~~.
$$
Since the first term on the right hand 
side vanishes by  virtue of (\ref{neutrality}), the partition
function equals
\begin{multline}
Z = \sum_{l(\brho)} 
\exp
\Bigl[
   \frac{\pi^2}{2}
   \Bigl(   
   \frac{4(J'+2\Jbar)}{2\pi}\sum_{\brho\neq\brho'}
   l(\brho)l(\brho')\ln |\brho-\brho'|\\
   - \frac{4C_1(J'+2\Jbar)}{2\pi} \sum_{\brho} l^2(\brho)
   + \delta J \sum_{\brho} e^{i\bg\cdot \brho}l^2(\brho)
   \Bigr)
\Bigr]~.
\end{multline}
This is the  Coulomb gas representation of our model
describing charges $l(\brho)$ residing on black and red plaquettes 
and interacting with long-ranged forces.  Condition 
(\ref{neutrality}) therefore is simply an expression
of the overall neutrality of the system -- only 
the configurations with the same number of vortices and
antivortices contribute to the partition function.

The Hamiltonian of the system can be finally recast as
\begin{multline}
{\cal H}^d_{v} = -\pi (J'+J_1+J_2)\sum_{\brho\neq\brho'}
    l(\brho)l(\brho')\ln |\brho-\brho'| \\
   +E_c^r \sum_{\brho\in {\cal R}} l^2(\brho)
   +E_c^b \sum_{\brho\in {\cal B}} l^2(\brho)~~,
\label{ashothamiltonian}
\end{multline}
where the core energies of the vortices on 
red ($\cal R$) and black (${\cal B}$) plaquettes are expressed
through the original parameters of the model as
\begin{align}
E_c^r &= \pi C_1(J'+J_1+J_2) - \frac{\pi^2}{4} (J_1-J_2)\\
E_c^b &= \pi C_1(J'+J_1+J_2) + \frac{\pi^2}{4} (J_1-J_2)~.
\end{align}
The Hamiltonian (\ref{ashothamiltonian})
is of the form equivalent to Eq. (\ref{latticecoulomb}) 
derived in the main text from the continuum
formulation. Note that in the ``low temperature'' 
limit $J\gg1$, coefficients  $J$ and $J'=\beta_V(J)$
coincide (see Eq. (\ref{betaV})), and the agreement with 
the continuum formulation is complete: the effective
strength of the long range interaction between vortices is 
$$
\tilde{J} = J+J_1+J_2~~.
$$

\subsubsection{Quantum phase fluctuations}
The derivation in 2+1 dimensions follows closely the steps of 
the two-dimensional case considered in the
previous subsection. Denoting the imaginary time by $\tau$, 
the partition function of the model is
$$
Z=\int \prod_{\brho}{\cal D}\phi_{\br}(\tau) \exp\left[-\int_0^{\beta}
d\tau \sum_{\brho} L(\brho,\tau) \right]~~,
$$
where $\brho$ is defined precisely like in the 2D case
and $\br=(\brho,\tau)$. The Lagrangian of our quantum model is defined as
\begin{multline}
-L(\brho,\tau)=-\frac{K_0}{2}\phidot_{\br}^2+i\fbar \dot{\phi}_{\br}+J\bigl(
                \cos(\nabla_x\phi_{\br}) + \cos(\nabla_y\phi_{\br})
		\bigr) \\
+J_{12}(\brho)
\bigl(  \cos(\phi_{\br+\hx+\hy}-\phi_{\br})
       +\cos(\phi_{\br+\hy}-\phi_{\br+\hx})
\bigr)       
.
\end{multline}
The sign of the Berry  phase is chosen to be positive for 
later convenience; obviously the partition function is
not affected by the change. 
It is convenient to replace the integrals over continuous variable $\tau$ by
sums over discrete $\tau_n$ (abbreviated often as $\tau$ below) separated by
intervals of ``length'' $\epsilon$. 
For brevity, we will  use $\phi_{\br+\htau}$
to denote $\phi(\brho, \tau+\epsilon)$.

The terms containing time derivatives can be transformed as follows:
\begin{multline}
\exp\Bigl[\int d\tau \sum_{\brho}
\Bigl(-\frac{K_0}{2}\phidot^2_{\brho,\tau}+i\fbar\phidot_{\brho,
\tau}\Bigr)\Bigr]\\
= 
\exp\Bigl[ \sum_{\br}
-\frac{K_0\epsilon}{2}\bigl(\frac{\phi_{\br+\tau}-\phi_{\br}}{\epsilon}\bigr)^2+ \sum_{\br}i\fbar\epsilon
\frac{\phi_{\br+\tau}-\phi_{\br}}{\epsilon}
\Bigr]~~.
\end{multline}
After completing the square we have
$$
\exp
\sum_{\br}
\left[
-\frac{K_0}{2\epsilon}\left(\phi_{\br+\tau}-\phi_{\br}-i\frac{\fbar}{K_0}\epsilon\right)^2-\frac{\epsilon\fbar^2}{2K_0}
\right]~~. 
$$
Note that this expression can be formally replaced by a sum
$$
\sum_{m(\br)}\exp
\sum_{\br}
\left[
-\frac{K_0}{2\epsilon}(\nabla_{\tau}\phi_{\br}-i\frac{\fbar}{K_0}\epsilon-2\pi
m(\br) )^2-\frac{\epsilon\fbar^2}{2K_0}
\right]~~,
$$
since, clearly, only the term $m=0$ survives in the limit of
small $\epsilon$. The latter form is convenient because 
now the Poisson identity
\begin{multline}
\sum_{n=-\infty}^{\infty} \exp\left[-\frac{a}2 n^2 + i n \phi\right] \\=
\sqrt{\frac{2\pi}{a}}\sum_{m=-\infty}^{\infty} \exp\left[-\frac{(\phi-2\pi
m)^2}{2a}\right]
\label{poisonidentity}
\end{multline}
can be applied. The result is
\begin{multline}
\exp
\Bigl[\int d\tau \sum_{\brho}
\Bigl(-\frac{K_0}{2}\phidot^2_{\brho,\tau}+i\fbar\phidot_{\brho,
\tau}\Bigr)\Bigr]\\ 
\propto
\sum_{u_{\tau}(\br)}
\exp
\Bigl[
   \sum_{\br}
   \Bigr(
        -\frac{\epsilon}{2K_0}u_{\tau}(\br)^2\\
	+iu_{\tau}(\br)
	\bigl(
              \phi_{\br+\htau}-\phi_{\br}-\frac{i\fbar\epsilon}{K_0}
	\bigr)
	-\epsilon\frac{\fbar^2}{2K_0}
   \Bigr)
\Bigr],
\end{multline}
where $u_{\tau}(\br)$ is an integer-valued field. 
Using the identity the partition function $Z$ can now be rewritten as
\begin{multline}
\sum_{\bu} \sum_{w_{\pm}}\int
\prod_{\br} d\phi_{\br} 
\exp\Bigl[i\sum_{\br}\Bigl(
u_{i}(\br)\nabla_{i}\phi_{\br}\\+
w_+(\br)(\phi_{\br+\hx+\hy}-\phi_{\br})
+w_-(\br)(\phi_{\br+\hy}-\phi_{\br+\hx})
\Bigr)
\Bigr]\times
\\
\exp
\Bigl[
    -\sum_{\br}
    \Bigl(
        \frac{u_{\alpha}^2(\br)}{2J'}
	+\frac{w_{+}^2(\br)+w_{-}^2(\br)}{2J_{12}'(\brho)}
	+\frac{\epsilon (u_{\tau}-\fbar)^2}{2 K_0}
    \Bigr)
\Bigr]
~~.
\label{Z0_3D}
\end{multline}
The reader should bear in mind that throughout the appendix 
the Greek indices exclusively denote the space-like components $x,y$ 
of a three-vector, 
while latin indices denote both space-like and time-like components, as
the case may be.
The coefficients $J'$ and $J'_{12}(\brho)$ are defined as 
\begin{align}
J'&=\beta_V(\eps J)\\
J_{12}(\brho)&=\beta_V(J_{12}(\eps\brho))~~.
\label{Jprimes_3D}
\end{align}
We now proceed to
transform the above expression by shifting the 
differences of the phases $\phi_r$ onto the 
difference of the fields $\bu(\br)$ and 
$w_{\pm}(\br)$ by using discrete integration by parts (\ref{sumbyparts}),
just as it was done in the two-dimensional case:
\begin{multline}
Z =\sum_{\bu} \sum_{w_{\pm}}\int
\prod_{\br}
\Bigl[
    d\phi_{\br} \delta\Bigl( 
    \overline{\mathbf\nabla}\cdot\bu(\br)+
\bigl(w_+(\br)\\-w_+(\br-\hx-\hy)\bigr)+\bigl(w_-(\br-\hx)
-w_-(\br-\hy)\bigr)  \Bigr)
\Bigr]\\
\exp
\Bigl[
   -\sum_{\br}\bigl(
      \frac{u_{\alpha}^2(\br)}{2J'}
      +\frac{\epsilon (u_{\tau}-\fbar)^2}{2 K_0}
      +\frac{w_{+}^2(\br)+w_{-}^2(\br)}{2J_{12}'(\brho)}
   \bigr)
\Bigr]~,
\label{Z1_3D}
\end{multline}
where bold letters stand for {\em three-dimensional} vectors and
$\lnabla \cdot \bu$ denotes {\em three-dimensional} divergence.

In the absence of the next-nearest coupling terms represented by $w_{\pm}$,
the $\delta$-function constraint in (\ref{Z1_3D}) is resolved by
$\bu=\overline{\nabla}\times  \ba$, where the lattice curl is defined as
$$
u_i = \epsilon_{ijk} \overline{\nabla}_j \Lambda_k (\br-\be_k)~~, \quad \be =(\hx,\hy,\htau)~~.
$$
In our case of a $d$-wave superconductor and finite
$w_{\pm}$, we rewrite the constraint as
\begin{equation}
\overline{\bf
\nabla}\cdot(\bu(\br)+\bw[w_+,w_-] =0~~,
\end{equation}
where $\bw=(w_x,w_y,0)$ and $w_{x(y)}$ are defined in the 
previous subsection section (\ref{w_xy}).
The solution is clearly
$$
\bu = \lnabla\times \bLambda -{\bf w}[w_+,w_-]~~.
$$

One can easily check that the choice of $\bLambda$ is not unique: 
for arbitrary scalar function $\xi(\br)$
\begin{multline}
\left[\lnabla\times(\bLambda+\nabla\xi(\br))\right]_i -
\left[\lnabla\times\bLambda\right]_i=\\ =
\epsilon_{ijk}\lnabla_j \nabla_k \xi(\br-\bdelta_k) = 
\epsilon_{ijk}\lnabla_j \lnabla_k \xi(\br) =0
\end{multline}
This gauge invariance implies that a gauge-fixing term 
must be introduced when replacing the sums over integer field  $\bu$ 
by summation over $\bLambda$ in order to avoid multiple-counting. 
The next step is most easily derived in the temporal gauge:
$$
\delta(\Lambda_3)\equiv \prod_{\br} \delta_{\Lambda_3(\br),0}~~.
$$
Afterwards, the results will be generalized to an arbitrary 
gauge-fixing condition. We proceed by rewriting the partition function as
\begin{multline}
Z=\sum_{\bLambda}
\sum_{w_{\pm}}
\exp
\Bigl[-
   \sum_{\br}
   \Bigl(
      \frac{(\lnabla\times\bLambda-
      {\bf F}[w{_{\pm}}])^2_{\perp}}{2J'}\\
      +\frac{\epsilon}{2K_0}
      \left(
        (\lnabla\times\bLambda)_0-\fbar
      \right)^2
      +\frac{w_+^2(\br)+w_-^2(\br)}{2J_{12}'(\brho)}
   \Bigr)
\Bigr]
\delta_{\Lambda_3(\br),0}
\end{multline}
and apply the Poisson formula in order to obtain a theory
depending on continuous rather than integer valued gauge field $\bLambda$:
\begin{multline}
\!\sum_{\bLambda(\br)} \delta_{\Lambda_3,0}
f(\Lambda_1(\br),\Lambda_2(\br),\Lambda_3(\br)) 
=\prod_{\br} \!\int\!  d\Lambda_1(\br) d\Lambda_2(\br)\times \\
\sum_{l_1(\br), l_2(\br)}e^{2\pi i
\sum_{\br} l_{\alpha}(\br)\Lambda_{\alpha}(\br)}
 f(\Lambda_1(\br),\Lambda_2(\br),0)\\
=\prod_{\br} \int_{-\infty}^{\infty}  d\Lambda_1(\br) d\Lambda_2(\br) d\Lambda_3(\br)
\delta(\Lambda_3)\times \\
\sum_{\bl(\br)} \delta_{\lnabla\cdot l} e^{2\pi i\sum_{\br}
l_j(\br)\Lambda_j(\br)}
 f(\Lambda_1(\br),\Lambda_2(\br),\Lambda_3(\br))~.
\end{multline}
In performing the last step, we formally introduced 
 $l_3\Lambda_3$ and an additional sum over
$l_3(\br)$. The delta function $\delta(\Lambda_3(\br))$ 
ensures that the exponent
is not affected. All terms in the  sum over $l_3$ are therefore equal, and in
order to avoid multiple-counting we need to impose a 
constraint, chosen as $\lnabla\cdot l=0$,  by assigning
$$
l_3 = -(\lnabla_3)^{-1} (\lnabla_1 l_1+\lnabla_2 l_2)~~.
$$
Note that the result of applying operator  $(\lnabla_3)^{-1}$ 
to an integer field is another integer.
The integer-valued field $\bl(\br)$ with zero divergence 
describes non-backtracking closed loops on the 2+1 space-time lattice.
The field $\bLambda(\br)$ is now continuous and the temporal gauge condition
$$
\Theta_{\tau}[\bLambda]=\prod_{\br}\delta(\Lambda_3(\br))
$$
can be replaced \cite{kleinert} by an arbitrary 
gauge-fixing condition $\Theta[\bLambda]$; examples are  
$\lnabla \cdot \bLambda =0$ (Landau gauge) or 
$\lnabla_{\perp} \cdot\bLambda_{\perp} =0$ (radiation gauge):
\begin{multline}
\sum_{\bLambda(\br)} \delta_{\Lambda_3,0}
f(\Lambda_1(\br),\Lambda_2(\br),\Lambda_3(\br))=\\ 
\int_{-\infty}^{\infty}  \prod_{\br} d\bLambda(\br) \Theta[\bLambda(\br)]
\sum_{\bl(\br)} e^{2\pi i\sum_{\br}
l(\br)\cdot \bLambda(\br)}
f(\bLambda(\br))\delta_{\lnabla\cdot l}~~.
\end{multline}
This final identity allows us to rewrite our partition function as
\begin{multline}
Z =
\int_{-\infty}^{\infty}  \prod_{\br} d\bLambda(\br) \Theta[\bLambda(\br)]
e^{ \sum_{\tau}  F[(\lnabla\times\bLambda)_{\perp}]}\times\\
\sum_{\bl(\br)}\delta_{\lnabla\cdot l} e^{\sum_{\br}\left[2\pi i l(\br)\cdot \bLambda(\br)
-(\epsilon/2K_0)\left((\lnabla\times\bLambda)_0-\fbar\right)^2\right]}
\label{Z4_3D}
\end{multline}
where $F[(\lnabla\times\bA)_{\perp}]$ has already been calculated in
(\ref{F_of_B}) and can be used as is,
provided that proper definitions 
(\ref{Jprimes_3D}) of $J'$ and $J'_{12}(\brho)$ are replaced.

The remaining steps leading to the ``Coulomb'' 
gas representation of 3D vortex loops are conceptually 
similar to the 2D case from the previous subsection.
The algebra, however, is considerably more involved. 
We therefore will go slowly and first wade
through the derivation for the simple case $J_1=J_2=0$. This
is just the ordinary (2+1)D XY model, appropriate for our $s$-wave
pedagogical exercise from the main text and the beginning
of this Appendix (\ref{alxy}). 
Only the configurations $w_{\pm}(\br) =0 $ 
contribute to the functional $F[(\lnabla\times\bA)_{\perp}]$ (\ref{F_of_B}) and
we recover the usual \cite{fisherlee} anisotropic 3D XY model in a
uniform  magnetic field ${\bf H} = \fbar \htau$: 
\begin{multline}
Z_0 =
\int_{-\infty}^{\infty}  \prod_{\br} d\bLambda(\br) \Theta[\bLambda(\br)]
\sum_{\bl(\br)}\delta_{\lnabla\cdot l}\times\\
e^{\sum_{\br}
\Bigl[
   2\pi i l(\br)\cdot \bLambda(\br)
   -\frac{(\lnabla\times \bLambda)_{\perp}^2}{2J'}-\frac{\epsilon}{2K_0}
   \Bigl(
       (\lnabla\times\bLambda)_0-\fbar
   \Bigr)^2
   \Bigr]
}~~.
\label{Zsimple_3D}
\end{multline}

To obtain the lattice loop gas representation, we need to 
integrate out the gauge field $\bLambda$.
The most transparent  connection with the results 
for the 2D model is obtained 
by using the radiation gauge $\lnabla_{\alpha}\Lambda_{\alpha}=0$:
\begin{equation}
\sum_{\br} (\lnabla\times \Lambda)_0^2 =
\sum_{\br}  [\lnabla_x \Lambda_y(\br-\hy) - \lnabla_y \Lambda_x(\br-\hx)]^2
\end{equation}
Expanding the square and 
shifting the difference operators via (\ref{sumbyparts}) we have 
\begin{multline}
\sum_{\br} (\lnabla\times \Lambda)_0^2 =
\sum_{\br} 
-\Lambda_y(\br-\hy)\nabla_x\lnabla_x \Lambda_y(\br-\hy)\\ 
-\Lambda_x(\br-\hx)\nabla_y \lnabla_y \Lambda_x(\br-\hx) -
2 \nabla_y \Lambda_y(\br-\hy) \nabla_x \Lambda_x(\br-\hx)
\\
=\sum_{\br} 
-\Lambda_{\mu}(\br)
     \nabla_{\nu}\lnabla_{\nu}
\Lambda_{\mu}(\br)
-
(\lnabla_{\mu} \Lambda_{\mu}(\br))^2~~.
\label{curlLambda0}
\end{multline}

We introduce the following  notation for the 
lattice analogues of wavevectors $q_j$ appearing from discrete left-sided
or right-sided derivatives after Fourier transformation:
\begin{align}
Q_j(\bq) &= \frac{e^{iq_j}-1}{i}\\
\Qbar_j(\bq) &= \frac{1-e^{-iq_j}}{i}\\
Q_j^{\bg}(\bq) &= Q_j(\bq-\bg)=\frac{-e^{-iq_j}-1}{i}\\
\Qbar_j^{\bg}(\bq) &= \Qbar_j(\bq-\bg)=\frac{1+e^{-iq_j}}{i}~~.\\
\end{align}
The arguments of $Q_j$ and $\Qbar_j$ is assumed to be $\bq$ unless specified otherwise.

We define the  Fourier  transformation as
$$
f(\br)=\frac1{\beta}\sum_{q_0}\int\frac{d\bq_{\perp}}{(2\pi)^2}
e^{i\bq\cdot\br} f(\bq)\\
$$
where the sum over frequencies $q_0$ runs through
$q_0 =  0, \frac{2\pi}{\beta}, {\ldots} ,\frac{2\pi}{\eps}$ and the
integrals over $q_x, q_y$ extend from $-\pi$ to $\pi$.  Using
the definition and properties that follow from it
\begin{align}
f(\bq)&= \eps\sum_{q_0}\int\frac{d\bq_{\perp}}{(2\pi)^2}e^{-i\bq\cdot\br} f(\bq)\\
f^2(\br) &= \frac1{\beta\eps}
\sum_{q_0}\int\frac{d\bq_{\perp}}{(2\pi)^2}f(\bq) f(-\bq)~~,
\end{align}
we obtain in the radiation gauge:
\begin{multline}
\sum_{\br} (\lnabla\times \Lambda)_0^2 =- \sum_{\br}
\Lambda_{\mu}(\br)\nabla_{\nu}\lnabla_{\nu}
\Lambda_{\mu}(\br)\\ = 
\sum_{q_0}\int \frac{d\bq_{\perp}}{\beta\eps(2\pi)^2}
\Lambda_{\mu}(-\bq)Q_{\nu}\Qbar_{\nu}\Lambda_{\mu}(\bq)~~,
\end{multline}
Similarly, 
\begin{multline}
\sum_{\br} (\lnabla\times \Lambda)_{\perp}^2  =  
\sum_{\br}\Bigl(
-\Lambda_0(\br) \lnabla_{\mu}\nabla_{\mu} \Lambda_0(\br)\\
-\Lambda_{\mu}(\br) \lnabla_0 \nabla_0 \Lambda_{\mu}(\br)
-2\nabla_0 \Lambda_0(\br-\htau)
\nabla_{\alpha}\Lambda_{\alpha}(\br-\be_{\alpha})\Bigr)~~,
\end{multline}
where the last term vanishes due to our choice of the gauge. In 
the momentum space we have:
\begin{multline}
\sum_{\br} (\lnabla\times \Lambda)_{\perp}^2  =  
\sum_{q_0}\int \frac{d\bq_{\perp}}{\beta\eps(2\pi)^2}
\Bigl(
   \Lambda_0(-\bq)\Lambda_0(\bq) Q_{\mu}\Qbar_{\mu}\\ 
   + \Lambda_{\mu}(-\bq)\Lambda_{\mu}(\bq) Q_0\Qbar_0
\Bigr)~~.   
\end{multline}

The above definitions are generally valid but now we focus
again on the simple case of $J_1=J_2=0$. 
The partition function $Z_0$, given by
(\ref{Zsimple_3D}), can be written as
\begin{multline}
Z_0 =
\sum_{\bl(\br)}\delta_{\lnabla\cdot l}
 e^{\sum_{\br}   2\pi i l(\br)\cdot \bLambda_f(\br)}
\int_{-\infty}^{\infty}  \prod_{\br} d\bLambda(\br)\times\\
\delta[\lnabla_{\alpha}\Lambda_{\alpha}]
 e^{\sum_{\br}
\left[
   2\pi i l(\br)\cdot \bLambda(\br)
   -\frac{(\lnabla\times \bLambda)_{\perp}^2}{2J'}-\frac{\epsilon}{2K_0}
   \left(
       \lnabla\times\bLambda
   \right)_0^2
   \right]
}~~.
\label{Zsimple_3D_1}
\end{multline}
In arriving at the expression above we performed a
shift of variable $\bLambda\to \bLambda + \bLambda_f$, 
where $\Lambda_f$ is a time {\em independent} vector potential
corresponding to a constant and uniform 
magnetic field $\fbar \htau$. After Fourier transformation the
expression in the exponent can be written as
\begin{multline}
\sum_{q_0}\int \frac{d\bq_{\perp}}{\beta\eps(2\pi)^2}
\Bigl[
   2\pi i l(\bq)\cdot \bLambda(-\bq)
   -\frac{\Lambda_0(-\bq)\Lambda_0(\bq) Q_{\mu}\Qbar_{\mu}}{2J'}\\ 
   -\Lambda_{\mu}(-\bq)\Lambda_{\mu}(\bq) 
   \Bigl(
       \frac1{2J'}Q_{0}\Qbar_{0}+\frac1{2K'}Q_{\nu}\Qbar_{\nu}
   \Bigr)
\Bigr]~~.
\label{Zsimple_3D_2}
\end{multline}
Note that temporal and space-like components are
independent and can be integrated out separately.
Integration over $\Lambda_0$ is trivial and yields
\begin{multline}
\exp
\Bigl[
  -2\pi^2 J' \sum_{q_0}\int\frac{dq_x dq_y}{\beta\eps(2\pi)^2}\frac{l_0(-\bq)l_0(\bq)}{Q_{\mu}\Qbar_{\mu}}
\Bigr]\\
=
\exp
\Bigl[
  -2\pi^2 J' \sum_{q_0}\int\frac{dq_x dq_y}{\beta\eps(2\pi)^2}\frac{l_0(-\bq)l_0(\bq)}{4-2\cos q_x - 2\cos q_y}
\Bigr]~.
\end{multline}
The remaining integral
\begin{multline}
\int_{-\infty}^{\infty}  \prod_{\bq} d\bLambda(\bq) \delta[\Qbar_{\alpha}\Lambda_{\alpha}]
\exp\Bigl[\sum_{\br}
\Bigl(
   2\pi i l_{\nu}(-\bq) \Lambda_{\nu}(\bq)\\
   -\bigl(\frac1{2J'}Q_0\Qbar_0+\frac1{2K'}Q_{\nu}\Qbar_{\nu}\bigr)
   \Lambda_{\mu}(-\bq)\Lambda_{\mu}(\bq) 
\Bigr)
\Bigr]
\label{Z_simple_3D_2}
\end{multline}
can be computed by switching to 2D 
transverse and longitudinal components of $\Lambda$, which we define on
the lattice as
\begin{align}
\Lambda_L(\bq) &= i\frac{\Qbar_x \Lambda_x(\bq) + \Qbar_y \Lambda_y(\bq)}{Q_{\perp}}\\
\Lambda_T(\bq) &= i\frac{-Q_y \Lambda_x(\bq) + Q_x \Lambda_y(\bq)}{Q_{\perp}}~~,
\end{align}
where
$$
Q_{\perp}= \sqrt{\Qbar_{\alpha}Q_{\alpha}} = \sqrt{\Qbar_x Q_x+\Qbar_y Q_y}~~.
$$
The  $\Lambda_x(\br)$ and $\Lambda_y(\br)$ can be expressed through
the transverse and longitudinal components of the gauge filed
$\Lambda$ as
\begin{align}
\Lambda_x(\bq) &= -i \frac{Q_x \Lambda_L(\bq) - \Qbar_y \Lambda_T(\bq)}{Q_{\perp}}\\
\Lambda_y(\bq) &= -i \frac{Q_y \Lambda_L(\bq) + \Qbar_x \Lambda_T(\bq)}{Q_{\perp}}~~.
\end{align}

Now observe that the Jacobian of the 
transformation $(\Lambda_x, \Lambda_y)\to(\Lambda_L, \Lambda_T)$ is unity,
2D divergence of $\bLambda$ is proportional to $\Lambda_L$ as expected:
$$
\Qbar_{\alpha}\Lambda_{\alpha} =  -i \Lambda_L Q_{\perp}~~,
$$
and
$$
l_{\alpha}(-\bq) \Lambda_{\alpha}(\bq) = l_T(-\bq) \Lambda_T(\bq)+ l_L(\bq) \Lambda_L(\bq)~~.
$$
The integral (\ref{Z_simple_3D_2}) can be written as
\begin{multline}
\int_{-\infty}^{\infty}  \prod_{\bq} d\Lambda_T(\bq) 
\exp\Bigl[
\sum_{q_0}\int\frac{dq_x dq_y}{\beta\eps(2\pi)^2}
\Bigl(
   2\pi i \,l_T(-\bq) \Lambda_T(\bq)\\
   -\left(\frac1{2J'}Q_0\Qbar_0+\frac1{2K'}Q_{\alpha}\Qbar_{\alpha}\right)
   \Lambda_T(-\bq)\Lambda_T(\bq)    
\Bigr)   
\Bigr]~~,
\label{Z_simple_3D_3}
\end{multline}
which finally gives
\begin{multline}
Z_0 \propto
\sum_{\bl(\br)}\delta_{\lnabla\cdot l}
\exp
\Bigl[
  - \pi^2\sum_{q_0}\int\frac{dq_x dq_y}{\eps\beta(2\pi)^2}
  \Bigl(
       2J'\frac{l_0(-\bq)l_0(\bq)}{\Qbar_{\alpha}Q_{\alpha}}\\
      +
      \frac{l_T(-\bq)l_T(\bq)}{\frac1{2J'}\Qbar_0 Q_0+\frac1{2K'}\Qbar_{\alpha}Q_{\alpha}}
  \Bigr)
\Bigr]~~.
\label{Z0finalresult}
\end{multline}
This is the desired vortex loop gas representation of our model.
Rather than integrating out the gauge field $\bLambda$ in (\ref{Zsimple_3D}), one can
partially perform the sum over over the 
integers $\bl(\br)$ and arrive at yet another (dual) representation
of partition function $Z$. The constraint $\lnabla\cdot\bl(\br) =0$ 
in the partition
function $Z_0$ (\ref{Zsimple_3D}) is rewritten using
auxiliary variables $\alpha(\br)$ as
\begin{multline}
\sum_{\bl(\br)} \delta_{\lnabla \cdot \bl,0}
\exp\left[-\frac{\sum_{\br}\bl^2(\br)}{2\beta_V(\beta')}+2\pi i
\sum_{\br}\bl(\br)\cdot\bLambda(\br)\right] =\\
\sum_{\bl(\br)} \prod_{\br}\int_0^{2\pi} d\alpha(\br)
\exp\Bigl[-\frac1{2\beta_V(\beta')}\sum_{\br}\bl^2(\br)\\
+2\pi i
\sum_{\br}\bl(\br)\cdot\bLambda(\br)+ i \sum \alpha(\br) \lnabla\cdot \bl(\br)\Bigr]=\\
\sum_{\bl(\br)} \prod_{\br}\int_0^{2\pi} d\alpha(\br)
\exp\Bigl[-\frac1{2\beta_V(\beta')}\sum_{\br}\bl^2(\br)\\
+ i \sum_{\br}\bl_i (2\pi\bLambda_i(\br)-\nabla_i\alpha(\br)) \Bigr]\approx\\
\prod_{\br}\int_0^{2\pi} d\alpha(\br)
\exp\left[\sum_{\br,
i}\beta'\cos(\nabla_i\alpha(\br)-2\pi \Lambda_i) \right]~~.
\label{xysup}
\end{multline}
The last equation describes a lattice superconductor. Note that the partition
function (\ref{Zsimple_3D}) does not contain quadratic terms
$\bl^2(\br)$. Instead, one introduces a vortex core energy term
$$
-\frac1{2\beta_V(\beta')}\sum_{\br}\bl^2(\br)
$$
by hand, and then, in the final lattice superconductor
representation, a limit $\beta\to \infty$ is taken. Such a system is
called a frozen superconductor. Alternatively, terms 
proportional to $\bl^2$ can be kept finite. Such ``unfrozen
lattice superconductor'' is equivalent to an ensemble of vortex loops,
namely an XY-model augmented with an
additional core energy that makes formation of vortices more
difficult. Thus, applying (\ref{xysup}) to (\ref{Zsimple_3D}) we
obtain  the partition function $Z_0$ describing lattice superconductor in
a field $\fbar$ coupled to fluctuating gauge field $\bLambda$:
\begin{multline}
\prod_{\br}\int_0^{2\pi} d\alpha(\br)
\int_{-\infty}^{\infty} d\bLambda(\br) \Theta[\bLambda(\br)]\times\\
\exp
\Bigl[
    \sum_{\br,i} \beta'\cos(\nabla_i\alpha(\br)-2\pi \Lambda_i)\\
    -\sum_{\br}
    \left(
        \frac1{2J'} (\lnabla\times\bLambda)_{\perp}^2
	+\frac{1}{2K'}\left((\lnabla\times\bLambda)_0-\fbar\right)^2
    \right)
\Bigr]~~.
\label{Z5_simple}
\end{multline}
The last step of our derivation is the standard\cite{kleinert}  Ginzburg-Landau expansion
of the action and for completeness we reproduce here the derivation
following Kleinert \cite{kleinert}.

First, we introduce  a complex   field
$U_{\br} = \exp (i\alpha(\br))$ and define covariant derivative operators $D_i$,
$\Dbar_i$ according to
\begin{align}
 D_x \Phi(\br) &= \Phi(\br+\hx)e^{-2\pi i \bLambda_x}-\Phi(\br)\\
\Dbar_x \Phi(\br) &=\Phi(\br)- \Phi(\br-\hx)e^{2\pi i
\bLambda_x(\br-\hx)}~~.
\end{align}
The following identity, which  expresses the cosine in (\ref{Z5_simple})
through $U_{\br}$,  can be proved easily:
$$
\sum_{\br} \cos(\nabla_x\alpha(\br)-2\pi \Lambda_Ax) = \sum_{\br} U^*_{\br}
(1+\frac12\Dbar_x D_x)U_{\br}~~.
$$
Thus, for a given fixed configuration of dual gauge
field $\Lambda$ in (\ref{Z5_simple}), the sum over all configurations of angular variables
$\alpha(\br)$  is
$$
Z_{XY}[\bLambda] = \int {\cal D}\alpha(\br)
e^{\sum_{\br,i} \beta'\cos(\nabla_i\alpha(\br)-2\pi \Lambda_i)}
$$
and can be  transformed into
\begin{equation}
Z_{XY}[\bLambda]= \int {\cal D} \alpha \exp[3 \beta'\sum_{\br}
U^*_{\br} \clD U_{\br}]~~.
\label{ZxyUDU}
\end{equation}
where $\clD = \left(1+\frac16 \Dbar_i D_i \right)$. Operator $\clD$ is
Hermitian, and therefore allows decomposition  $\clD = \clK^2$. Let us show that  $Z_{XY}$
is proportional to
\begin{equation}
\int\!{\cal D} [\alpha, \Phi, \Phi^*] e^{-\frac1{12 \beta'}\sum_{\br}
|\Phi_{\br}|^2+\frac12\sum_{\br,\br'} (\Phi_{\br}^*\clK_{\br\br'}
U_{\br'}+U^*_{\br}\clK_{\br\br'} \Phi_{\br'})}~.
\label{ZxyUDU1}
\end{equation}
where notation
$$
\int {\cal D}[\Phi, \Phi^*] {\ldots} = \prod_{\br} \int_{-\infty}^{\infty} d {\rm Re}\Phi_{\br}
\; d  {\rm Im}\Phi_{\br}\; {\ldots}
$$
is used. To establish the equivalence, we group the terms in the exponent as 
\begin{multline}
-\frac1{12\beta'} (\Phi^*_{\br}-6\beta'\sum _{\br'}U^*_{\br'}
\clK_{\br'\br})(\Phi_{\br}-6\beta' \sum_{\br'} \clK_{\br'\br} U_{\br'})\\+
3\beta' \sum_{\br}\sum_{\br'\br''} U^*_{\br''}\clK_{\br''\br} \clK_{\br\br'}
U_{\br'}~~.
\end{multline}
After a shift of variables and integrating out the auxiliary fields
$\Phi_{\br}$ the result coincides with (\ref{ZxyUDU}) up to an unimportant
proportionality  factor. To obtain the equivalent description in terms of
field $\Phi$, we now integrate over the angular variables  $\alpha_{\br}$.
To simplify notation, we define $\chi_1 = \clK \Phi/2$ and $\chi_2 =
\clK^{T}\Phi^*/2$, or more explicitly,

$$
\left\{
\begin{aligned}
\chi_1(\br) =\frac12\sum_{\br'} \clK_{\br\br'} \Phi_{\br'}\\
\chi_2(\br) = \frac12 \sum_{\br'}  \clK_{\br'\br} \Phi^*_{\br'}
\end{aligned}
\right.
~~.
$$
Integration over the phases in  (\ref{ZxyUDU}) now amounts to calculation of
disentangled integrals at separate $\br$:
$$
\int d \alpha(\br) e^{e^{i\alpha(\br)} \chi_2(\br)+e^{-i\alpha(\br)}
\chi_1(\br)} =  2\pi  I_0(\sqrt{4\chi_1(\br)\chi_2(\br)})~~,
$$
where $I_0$ denotes the modified Bessel function. Thus, omitting
non-essential overall prefactors the expression for $Z_{XY}[\bLambda]$ assumes the
following form:
$$
\int {\cal D}[\Phi,\Phi^*]
e^{-\frac1{12\beta'}\sum_{\br}|\Phi_{\br}|^2+
\sum_{\br} \ln I_0(\sqrt{4\chi_1(\br)\chi_2(\br)})}~~.
$$
Finally, after applying the Taylor expansion
$$
\ln I_0(x)  = \frac{x^2}{4} -\frac{x^4}{64}+{\ldots}~~,
$$
and retaining only the leading terms, we obtain
\begin{multline}
Z_0 =
\int \prod_{\br} d\Phi(\br) d\Phi^*(\br) d\bLambda(\br) \Theta[\bLambda(\br)]\times
\\ 
\exp
\Bigl[-
    \sum_{\br}
    \Bigl(
    \frac1{24}(\overline{D_i}\Phi)^*( D_i\Phi)
    +\frac14\left(\frac1{3\beta'}-1\right)|\Phi(\br)|^2\\
    +\frac{|\Phi(\br)|^4}{64}
    +\frac1{2J'} (\lnabla\times\bLambda)_{\perp}^2    
    +\frac{1}{2K'}\left((\lnabla\times\bLambda)_0-\fbar\right)^2
\Bigr)
\Bigr]~~.
\label{Z6_simple}
\end{multline}

The partition function (\ref{Z6_simple}) 
is the desired dual representation of our
initial anisotropic $XY$ model with the Berry phase, in the simple
case $J_1=J_2=0$.

Armed with the experience from the above derivation
we now return to the  
partition function $Z$ of the full-fledged  model (\ref{Z4_3D})
containing $J_1$ and $J_2$. As we will demonstrate, the effect of the  next nearest
neighbor interactions will be rather modest: to the leading order
only the term proportional to
$|\Phi(\br)|^2$ will be modified. The prefactors of this term will
be modulated, having different values on the red and the black
plaquettes.

To arrive at the dual representation of $Z$, we seek a gauge that will ensure
the decoupling of the temporal and
spatial components of $\Lambda(\bq)$, similarly to the radiation 
gauge in the simple example above. The
bilinear terms in $\Lambda_{\mu}$ appearing 
in the exponent can be classified as following:
first, there is a contribution from the 
term $(\lnabla\times A)_0^2$ which in an arbitrary (yet unknown)
gauge has been already calculated in (\ref{curlLambda0})
\begin{equation}
-\frac1{2K'} 
\Bigl(
   \Lambda_{\mu}(-\bq)\Lambda_{\mu}(\bq)\Qbar_{\nu}Q_{\nu}
- Q_{\nu}  \Qbar_{\mu} \Lambda_{\mu}(-\bq) \Lambda_{\nu}(\bq)
\Bigr)~~.
\label{curlA0squared}
\end{equation}
To facilitate the
bookkeeping of various terms resulting from 
$F[(\lnabla \times \bLambda)_{\perp}]$
we use the following set of identities:
\begin{equation}
\spb_x^T(-\bq)
=
-i e^{iq_0}
\spLambda_0^T(-\bq)
\begin{pmatrix}
Q_y &0\\
0&Q_y^{\bg} 
\end{pmatrix}
+
ie^{iq_y}Q_0
\spLambda_x^T(-\bq)
\sigma_3
\label{auxbLambda1}
\end{equation}

\begin{equation}
\spb_x(\bq)
=
-i e^{iq_0}
\begin{pmatrix}
\Qbar_y&0\\
0&\Qbar_y^{\bg}
\end{pmatrix}
 \spLambda_0(\bq)
-
ie^{-iq_y}\Qbar_0
\sigma_3
\spLambda_x(\bq)
~~.
\label{auxbLambda2}
\end{equation}
The expressions for $\spb_y(\bq)$ can be obtained 
by replacing $x\leftrightarrow y$ and the overall
change of sign. Using these identities the bilinear form
$$
\spb_x^T(-\bq)
\begin{pmatrix}
G_{11}(q_y) & G_{12}(q_y)\\
G_{21}(q_y) & G_{22}(q_y)
\end{pmatrix}
\spb_x(\bq)\\
+(x\leftrightarrow y)
$$
can be written as a sum of two groups: the diagonal terms are
\begin{multline}
\spLambda_0^T(-\bq)
\begin{pmatrix}
\Qbar_y Q_y G_{11}(q_y) & \Qbar_y^{\bg} Q_y G_{12}(q_y)\\
\Qbar_y^{\bg} Q_y^{\bg} G_{21}(q_y) & \Qbar_y^{\bg} Q_y^{\bg} G_{22}(q_y)
\end{pmatrix}
\spLambda_0(\bq)\\
+
\spLambda_y^T(-\bq)
\begin{pmatrix}
\Qbar_0 Q_0  G_{11}(q_y) & - \Qbar_0 Q_0 G_{12}(q_y)\\
-\Qbar_0 Q_0 G_{21}(q_y) &  \Qbar_0 Q_0G_{22}(q_y)
\end{pmatrix}
\spLambda_y(\bq)\\
+(x\leftrightarrow y)~~.
\end{multline}
In addition, we obtain  cross-terms that couple the spatial 
and temporal components of $\spLambda_{i}$
\begin{equation}
 Q_0 
\spLambda_0^T(-\bq)\Bigl(P(q_y)\spLambda_y(\bq)+P(q_x)\spLambda_x(\bq)\Bigr)+c.c.~~,
\end{equation}
where $c.c.$ denotes complex conjugation and matrices
$P(q_{\alpha})$ are defined as
$$
P(q_{\alpha})=\begin{pmatrix}
-\Qbar_{\alpha}      G_{11}(q_{\alpha}) &  \Qbar_{\alpha}       G_{12}(q_{\alpha})\\
\Qbar_{\alpha}^{\bg} G_{21}(q_{\alpha}) & -\Qbar_{\alpha}^{\bg} G_{22}(q_{\alpha})
\end{pmatrix}~~.
$$
Note that the terms diagonal in $\Lambda_0$ are exactly 
what we encountered in  (\ref{Mtilde}) when we
considered the 2D example. The off-diagonal terms 
can be eliminated altogether by choosing a gauge defined 
by the following relation
between $\Lambda_x(\bq)$ and $\Lambda_y(\bq)$:
\begin{equation}
P(q_y)
\spLambda_y(\bq)
+
P(q_x)
\spLambda_x(\bq)
=0~~.
\end{equation}
The  matrix equation can be resolved by
$\spLambda_y(\bq)=\Gamma(\bq)\spLambda_x(\bq)$ where matrix  $\Gamma$
is defined as $\Gamma(\bq) = -P^{-1}(q_y) P(q_x)$.
The spatial part of the action in momentum space becomes
\begin{multline}
\sum_{q_0}\int\frac{dq_x dq_y}{\beta\eps(2\pi)^2}
\Bigl[
  i\pi
  \Bigl(\spl_x^T(-\bq)+\spl_y^T(-\bq)\Gamma(\bq)\Bigr)
  \spLambda_x(\bq)\\
  -
  \spLambda_x(-\bq)
  \tilde{X} (\bq)
  \spLambda_x(\bq)
\Bigr]~~,
\end{multline}
where $2\times2$ matrix $\tilde{X}$ is defined as
\begin{multline}
\frac1{4K'}
\left[
  \begin{pmatrix}
  Q_{\perp}^2 & 0\\
  0& (Q^{\bg}_{\perp})^2
  \end{pmatrix}+
  \Gamma^T(-\bq)
  \begin{pmatrix}
  Q_{\perp}^2 & 0\\
  0& (Q^{\bg}_{\perp})^2
  \end{pmatrix}
  \Gamma(\bq)
\right]\\
+
\frac1{4}
\Bigl[
   \begin{pmatrix}
   G_{11}(q_x)& -G_{12}(q_x)\\
   -G_{21}(q_x)& G_{22}(q_x)
   \end{pmatrix}\\
+
   \Gamma^T(-\bq)
   \begin{pmatrix}
   G_{11}(q_x)& -G_{12}(q_x)\\
   -G_{21}(q_x)& G_{22}(q_x)
   \end{pmatrix}
   \Gamma(\bq)
\Bigr]~~.
\end{multline}
Note that we omitted the second term in (\ref{curlA0squared})
as it is only of the order $Q_{\perp}^6$ and can be safely neglected
for extracting the long distance behavior. After integrating out $\Lambda_x$, we obtain
\begin{multline}
\exp
\Bigl[
-
\frac{\pi^2}{4}
\sum_{q_0}\int\frac{dq_x dq_y}{\beta\eps(2\pi)^2}
\times \\\Bigl(\spl_x^T(-\bq)+\spl_y^T(-\bq)\Gamma(\bq)\Bigr) X \bigl(\spl_x(\bq)
+\Gamma^T(-\bq)\spl_y(\bq)\bigr)
\Bigr]~~,
\end{multline}
where $X=\tilde{X}^{-1}$. 
Expanding this expression in the 
region of small momenta as in (\ref{M11M12}) and
retaining only the leading order terms, as in 
the two-dimensional example, we find
\begin{multline}
-\pi^2
\sum_{q_0}\int\frac{dq_x dq_y}{\beta\eps(2\pi)^2}
\frac{1}{\frac1{2\tilde{J}} q_0^2+\frac1{2K'}q_{\perp^2}}\times \\
\frac{\Bigl(q_y l_x(-\bq)-q_x l_y(-\bq)\Bigr)
\Bigl(q_y l_x(\bq)-q_x l_y(\bq)\Bigr)}{q^2_{\perp}}
\label{spaceZ}~~.
\end{multline}
Observe that there are no cross-terms that couple
modes at wavevectors $\bq$ and $\bq-\bg$ to the order of $q^0$ and $q^{-2}$.
The spatial part of the action after integrating out the gauge fields
is equivalent to the result (\ref{Z0finalresult}) 
obtained in the framework of simple model
$Z_0$ where  $J'$ is replaced by the effective 
coupling constant $\tilde{J} = J'+J_1+J_2$.

Combining (\ref{spaceZ}) and (\ref{ashothamiltonian}), the final
form of the  action $Z$ in terms of closed vortex loops
$\bl(\br)$ can be written as
\begin{multline}
Z=\sum_{\bl(\br)} \delta_{\lnabla \cdot \bl} 
\exp\Bigl[
   \sum_{\br}
   \Bigl(
       2\pi i \,\bl(\br) \cdot \bA_{f}(\br)   - E_c'(\brho)\,l_0^2(\br)
   \Bigr)\\
   -\pi^2
   \sum_{q_0}\int \frac{dq_x \,dq_y}{\beta\eps(2\pi)^2}
   \bigl(
   2 \tilde{J}  \frac{l_0(-\bq) l_0(\bq)}{q_{\perp}^2}\\   
   +\frac{1}{\frac1{2\tilde{J}} q_0^2+\frac1{2K'}q_{\perp^2}} l_T(\bq) l_T(-\bq)
   \bigr)
\Bigr]~~,
\end{multline}
where $E'_c(\brho) = \pm \pi^2(J_1-J_2)/4$ depending on whether
$\brho$ corresponds to a black or a red plaquette.

We had intentionally used momentum representation for the last two terms
in the exponent. It is important to recognize
that these terms are precisely what one would
have obtained for the usual 3D XY model with no next nearest 
neighbors interaction and the effective nearest neighbors coupling
constant equal to $\tilde{J} = J'+J_1+J_2$. Thus, we may introduce 
a dual gauge field $\bA(\br)$  and present the partition function as
\begin{multline}
Z=
\sum_{\bl(\br)} 
\int_{-\infty}^{\infty}  \prod_{\br} d\bA(\br) \Theta[\bA(\br)]
\delta_{\lnabla\cdot \bl}\times \\
\exp
\Bigl[
   \sum_{\br}
   \Bigl(
       2\pi i \;\bl(\br)\cdot (\bA(\br)+\bA_{f}(\br))\\
       -
       E_c'(\brho)l^2(\br)
       -\frac{(\lnabla\times\bA)_{\perp}^2}{2\tilde{J}}
       -\frac{(\lnabla\times\bA)_0^2}{2K'}       
   \Bigr)
\Bigr]~~.
\end{multline}
By shifting $\bA\to \bA-\bA_{f}$ back we obtain
\begin{multline}
Z=
\int_{-\infty}^{\infty}  \prod_{\br} d\bA(\br) \Theta[\bA(\br)]
\exp
\Bigl[
   \sum_{\br}
   \Bigl(
       2\pi i \;\bl(\br)\cdot \bA(\br)\\
       -
       E_c'(\brho)l^2(\br)
       -\frac{(\lnabla\times\bA)_{\perp}^2}{2\tilde{J}}
       -\frac{[(\lnabla\times\bA)_0-\fbar]^2}{2K'}       
   \Bigr)
\Bigr]~~.
\label{ZZZ}
\end{multline}
Note that in the absence of the next nearest neighbors interactions,
a similar expression  (\ref{Zsimple_3D}) contained 
no quadratic terms $\bl^2(\br)$ and 
the core energy  $-\frac1{\beta_V(\beta')} \bl^2(\br)$ was introduces 
by hand. In the present case,  the
difference $\Delta E_c'$ of the core energies on the black and red
sites is {\em finite} due to the anisotropic next 
nearest neighbors interactions.
However, the average magnitude is still zero within our model, and
here we also need to introduce a constant average core
energy term $ \frac1{\beta_V(\beta'_0)}\bl^2(\br)$. Thereby we replace
$E_c'(\brho)$ in (\ref{ZZZ}) by
$$
\frac1{\beta_V(\beta'(\brho))} \to\frac1{\beta_V(\beta'_0)}+E_c'(\brho)~~,
$$
where the function $\beta'(\brho)$ is implicitly defined by this
equation. 

The remaining steps repeat the derivation  leading from (\ref{xysup}) to
(\ref{Z6_simple}) with the replacement $\beta'\to\beta'(\brho)$ and
result in the  Ginzburg-Landau expansion of our dual theory:
\begin{multline}
Z =
\int \prod_{\br} d\Phi(\br) d\Phi^*(\br) d\bLambda(\br) \Theta[\bLambda(\br)]\times \\
\exp
\Bigl[-
    \sum_{\br}
    \Bigl\{
    \frac1{24}(\overline{D_i}\Phi)^*( D_i\Phi)\\
    +\frac14\left(\frac1{3\beta'(\brho)}-1\right)|\Phi(\br)|^2
    +\frac1{64}|\Phi(\br)|^4\\
    +\frac1{2J'} (\lnabla\times\bLambda)_{\perp}^2    
    +\frac{1}{2K'}\left((\lnabla\times\bLambda)_0-\fbar\right)^2
\Bigr\}
\Bigr]~~.
\label{Z6}
\end{multline}

This is our final result -- the partition function
(\ref{Z6})  represents the Ginzburg-Landau functional
of a dual type-II superconductor appropriate for our
model and subjected to a constant dual magnetic field $\fbar$.

\vfill
\eject





\begin{thebibliography}{999}

\bibitem{corson} 
J. Corson, R. Mallozzi, J. Orenstein, J. N. Eckstein, and I. Bozovic, Nature {\bf 398}, 221 (1999).
\bibitem{ong} Z. A. Xu, N. P. Ong, Y. Wang, T. Kakeshita, and S. Uchida, Nature {\bf 406}, 486 (2000);
Y. Wang, S. Ono, Y. Onose, G. Gu, Y. Ando, Y. Tokura, S. Uchida, and N. P. Ong, Science {\bf 299}, 86 (2003).
\bibitem{campuzano} J. C. Campuzano {\em et al.}, unpublished.
\bibitem{emerykivelson} V. J. Emery and S. A. Kivelson, Nature {\bf 374}, 434 (1995).
\bibitem{qed} M. Franz and Z. Te\v sanovi\' c, \prl {\bf 87}, 257003
(2001); Z. Te\v sanovi\' c, O. Vafek, and M. Franz, \prb {\bf 65}, 180511 (2002);
M. Franz, Z. Te\v sanovi\' c, and O. Vafek, {\it ibid.} {\bf 66}, 054535 (2002).
\bibitem{zt} Z. Te\v sanovi\' c, \prb {\bf 59}, 6449 (1999).
\bibitem{herbut} I. F. Herbut, \prl {\bf 88}, 047006 (2002);
B.H. Seradjeh and I. F. Herbut, \prb {\bf 66}, 184507 (2002).
\bibitem{yazdani} M. Vershinin, S. Misra, S. Ono, Y. Abe, Y. Ando, and A. Yazdani,
Science {\bf 303}, 1995 (2004).
\bibitem{davis} T. Hanaguri {\it et al.}, to appear in Nature;
K. McElroy, D.-H. Lee, J. E. Hoffman, K. M. Lang, E. W. Hudson, H. Eisaki, S. Uchida, J. Lee, J. C. Davis
, cond-mat/0404005;
J. E. Hoffman, E. W. Hudson, K. M. Lang, V. Madhavan, H. Eisaki, S. Uchida, and J. C. Davis,
Science {\bf 295}, 466 (2002).
\bibitem{kapitulnik} C. Howald, H. Eisaki, N. Kaneko,
M. Greven, and A. Kapitulnik, \prb {\bf 67}, 014533 (2003);
M. A. Steiner and A. Kapitulnik, cond-mat/0406227.
\bibitem{zhang0} H. D. Chen, O. Vafek, A. Yazdani, and S. C. Zhang, \prl
{\bf 93}, 187002 (2004).
\bibitem{preprint} Z. Te\v sanovi\' c, \prl {\bf 93}, 217004 (2004).
\bibitem{fisherlee} M. P. A. Fisher and D. H. Lee, \prb {\bf 39}, 2756 (1989).
\bibitem{zhang}  H. D. Chen, S. Capponi, F. Alet, and S. C. Zhang,
cond-mat/0312660; H. D. Chen, C. Wu, and S. C. Zhang, 
\prl {\bf 92}, 107002 (2004).
\bibitem{auerbach} E. Altman and A. Auerbach, 
\prb {\bf 65}, 104508 (2002).
\bibitem{footothers} While the Cooper and the real-space pairs correspond to
two distinct limiting behaviors there is still a sense in which they
are two sides of the same coin: in both cases charge modulation 
arises primarily from the particle-particle
channel. Similarly, a recent preprint, P. W. Anderson,
cond-mat/0406038, also examines the
effect on the LDOS of the modulations in the particle-particle
channel. This puts these works in the category different from theories which
focus on non-uniformities in the particle-hole channel, for example
H. C. Fu, J. C. Davis, and D. H. Lee, cond-mat/0403001.
Other prominent examples of such theories are
J. Zaanen and O. Gunnarson, \prb {\bf 40}, 7391 (1989);
K. Machida, Physica C {\bf 158}, 192 (1989); S. A. Kivelson, I. P. Bindloss, E. Fradkin,
V. Oganesyan, J. M. Tranquada, A. Kapitulnik, and C. Howald, \rmp {\bf 75}, 1201 (2003);
S. Chakravarty, R. B. Laughlin, D. K. Morr, and C. Nayak,
\prb {\bf 63}, 094503 (2001); Y. Zhang, E. Demler, and 
S. Sachdev, \prb {\bf 66}, 094501 (2002). 
A useful general study of a non-uniform superconductor is found
in D. Podolsky, E. Demler, K. Damle, and B.I. Halperin,
\prb {\bf 67}, 094514 (2003). 
\bibitem{sachdev} A Hofstadter-type problem
is also at the root of various inhomogeneous phases of 
quantum spin-dimer models; 
see L. Balents, L. Bartosch, A. Burkov, S. Sachdev, and
K. Sengupta, cond-mat/0408329 and references therein.
For early discussions of valence-bond-solids in such
quantum spin-dimer models see S. Sachdev and N. Read,
Int. J. Mod. Phys. B {\bf 5}, 219 (1991) and
M. Vojta and S. Sachdev, \prl {\bf 83}, 3916 (1999).
\bibitem{vafek} O. Vafek, A. Melikyan, M. Franz, and Z. Te\v sanovi\' c, 
\prb {\bf 63}, 134509 (2001) and references therein.
\bibitem{taillefer}
M. Sutherland, D. G. Hawthorn, R. W. Hill, F. Ronning, S. Wakimoto, 
H. Zhang, C. Proust, E. Boaknin, C. Lupien, L. Taillefer, R. Liang, D. A. Bonn, 
W. N. Hardy, R. Gagnon, N. E. Hussey, T. Kimura, M. Nohara, H. Takagi, cond-mat/0301105.
\bibitem{footqed} There is a large and growing field-theory literature
on non-compact, parity-preserving QED$_3$, which is the low-energy
limit of (\ref{lagrangianqed3}). Those interested will find 
T. Appelquist and L. C. R. Wijewardhana, hep-ph/0403250,
C. S. Fischer, R. Alkofer, T. Dahm, and P. Maris, hep-ph/0407104,
S. J. Hands, J. B. Kogut, L. Scorzato, and C. G. Strouthos, hep-lat/0404013,
and J. Alexandre, K. Farakos, and N. E. Mavromatos, hep-ph/0407265 to
provide good overview of frontier issues and useful source of additional
references.
\bibitem{footupdown} Representing $a_\mu$ as a U(1) gauge
field constitutes the ``natural'' gauge choice for this problem 
due to the following fundamental feature of a spin-singlet superconductor:
consider a system made up of two {\em distinct} species of fermions,
``up'' ($u$)  and ``down'' ($d$). 
The normal part has the form consisting entirely of bilinears
$u^\dag u$ and $d^\dag d$ so that $u$ and $d$ flavors are {\em separately}
conserved. Note that we are not at all concerned here with the symmetry 
with respect to rotations between $u$ and $d$ -- such symmetry 
may or may not be present and the relevant symmetry of the
normal part is just the global U$_u$(1)$\times$U$_d$(1). The inclusion
of pairing terms of the form $ud$ and $d^\dag u^\dag$ 
(but not $uu$ or $dd$ and their complex conjugates!) breaks this symmetry
by violating the conservation law for the {\em total} fermion number,
the sum of ``up'' and ``down'' flavors. There remains, however, an
intact {\em continuous} U$_{u-d}$(1) 
symmetry associated with the {\em relative}
fermion number, the difference between ``up'' and ``down'' flavors. This
continuous symmetry signals the remaining conservation law
(spin conservation in spin-singlet superconductors).
Actually, our lattice $d$-wave superconductor model 
(\ref{ldsc}) is a simple illustration of this general feature:
consider screening the bond phase factor of $\Delta_{ij}$ by site
phase factors arising from the gauge transformed electron
fields $c_{i\sigma}$: there are $2N$ bond
phase factors $\exp(i\theta_{ij})$ 
versus only $N$ site phases $\exp(-i\varphi_i)$.
The most natural solution is to introduce $2N$ site phase factors
$\exp(-i\varphi_{i\uparrow})$ and $\exp(-i\varphi_{i\downarrow})$ and
attach them via gauge transformation
to $c_{i\uparrow}$ and $c_{i\downarrow}$. In this way
one can completely eliminate 
the {\em center-of-mass} $\exp(i\theta_{ij})$ \cite{vafek} from (\ref{ldsc}) 
by a judicious choice of $\exp(i\varphi_{i\sigma})$:
$\exp(i\theta_{ij})\exp(-i\varphi_{i\uparrow}-i\varphi_{j\downarrow})\to
\exp(-ia_{ij})$;
$\exp(i\theta_{ij})\exp(-i\varphi_{i\downarrow}-i\varphi_{j\uparrow})\to
\exp(ia_{ij})$, where $a_{ij}$ is a bond phase antisymmetric under
$\uparrow\leftrightarrow\downarrow$ exchange.
This is nothing but the tight-binding 
lattice version of the FT transformation and leads directly to
the U(1) representation of the Berry gauge field $a\leftrightarrow a_{ij}$.
Here $\exp(2ia_{ij})$ is determined by
$\exp(i\varphi_{i\uparrow}-i\varphi_{j\uparrow}-i\varphi_{i\downarrow}+i\varphi_{j\downarrow})$, where $\exp(i\varphi_{i\sigma})$ are found in terms
of {\em center-of-mass} $\theta_{ij}$'s from
$\exp(i\varphi_{i\uparrow}+i\varphi_{j\uparrow}+i\varphi_{i\downarrow}+i\varphi_{j\downarrow})\leftrightarrow \exp(2i\theta_{ij})$. Note also that the
hopping term in the Hamiltonian acquires a gauge field factor
$\exp(i\varphi_{i\uparrow(\downarrow)}-i\varphi_{j\uparrow(\downarrow)})$ for
spin $\uparrow(\downarrow)$ fermions, with $\varphi_{i\uparrow(\downarrow)}-\varphi_{j\uparrow(\downarrow)}$ being the lattice equivalents of the
gauge fields $v_{A(B)}$ featured in the continuum FT transformation.
Since $a_{ij}$ is ultimately given by the (half of) phase differences
$(\varphi_{i\uparrow} - \varphi_{j\uparrow}) -
(\varphi_{i\downarrow}-\varphi_{j\downarrow})$ -- expressed
in terms of $\theta_{ij}$'s -- 
its configurations are non-compact, {\em i.e.} monopole free, by construction.
Note that these arguments do not generally apply to
superconductors which are not spin-singlet. For example, if we have a single
flavor of spinless fermions $f$, the normal part made up of $f^\dag f$ 
bilinears has only a single U(1) symmetry -- pairing terms of the form 
$f^\dag f^\dag$ and $ff$ (in the odd angular momentum
channel) break this U(1) symmetry down to {\em discrete} Z$_2$. 
\bibitem{footcompactqed} In short, the gauge theory 
(\ref{lagrangiani}, \ref{lagrangianqed3}) is non-compact by
construction. Recently, there has been much interest in quantum spin 
systems where the underlying effective gauge theory
is compact but its monopole (instanton) configurations are 
dynamically irrelevant at a critical point or in a critical phase. 
In such cases, one is again back to a non-compact QED$_3$ with massless
bosons or fermions; see
M. Hermele, T. Senthil, M. P. A. Fisher, P. A. Lee, N. Nagaosa, and X.-G. Wen, 
cond-mat/0404751 and references therein.
\bibitem{thermalmetal} This chirally symmetric ``algebraic Fermi liquid'' 
phase has unusual thermodynamic and transport properties; see
O. Vafek and Z. Te\v sanovi\' c, \prl {\bf 91}, 237001 (2003). For
example, its specific heat $\sim T^2$, just as in a 
neighboring superconducting state (see Fig. \ref{figqed}).
This translates into heat transport similar to that of a 
nodal $d$-wave superconductor, i.e. the critical pseudogap 
state is a ``thermal metal''. On the
other hand, since vortex-antivortex pairs are unbound, the same
state is also a ``charge insulator'', as emphasized in Ref. \cite{qed}
and detailed in this manuscript. This implies breakdown of
the Wiedemann-Franz law in the pseudogap state. The $T^2$ specific heat and
heat transport are due to spin excitations carried by nodal BdG fermions.
Remarkably, the pseudogap is also a ``spin dielectric'' in the sense
that Pauli spin susceptibility vanishes as $\chi\sim q^2$,
in contrast to $\chi\sim q$ in a superconductor. Consequently,
the Wilson ratio also vanishes as temperature goes to zero.
\bibitem{footcompact} In general, the Hamiltonian (\ref{hxy}) will
also contain terms which do not involve the standard XY phase
differences of bond phases but are instead due to $\Delta_{ij}$
itself being a ``hopping term'' in a $d$-wave superconductor (unlike
the case of a simple $s$-wave superconductor). The leading
such term is $K\cos(\theta_{12}-\theta_{23}+\theta_{34}-\theta_{41})$
around a plaquette of the CuO$_2$ lattice. Such terms are down
by a factor $\sim\Delta^2/t^{*2}$ relative to the XY terms kept
in (\ref{hxy}) and one can neglect them in cuprates, where
$\Delta$ remains significantly smaller than
$t^*$ for most of the underdoped regime, judged by the
ratio of $v_\Delta$ to $v_F$ \cite{taillefer}. However, such
terms might become important, along with many other longer range terms
excluded from (\ref{hxy}) and (\ref{dduallagrangian}), 
in the calculation of vortex
core energies later in the text since $K$ is not necessarily
smaller than $J_1-J_2$.
\bibitem{nelson} D. R. Nelson, in Phase Transitions and Critical Phenomena, edited by
C. Domb and J. L. Lebowitz, (Academic Press, London, 1983), 
Vol. 7, p. 1.
\bibitem{jose} J. V. Jose, L. P. Kadanoff, S. Kirkpatrick and D. R. Nelson, \prb {\bf 16}, 1217 (1977).
\bibitem{mizel} The vortex core mass $M$ can be computed in a specific
microscopic model, say a $d$-wave BCS model of vortex quantum tunneling
(A. Mizel, unpublished); also, see 
J. H. Han, J. S. Kim, M. J. Kim, and  P. Ao, cond-mat/0407156 and references therein.
Similarly, one can also envision computing the mass with a Gutzwiller correlated BCS wavefunction.
\bibitem{core} Ch. Renner, B. Revaz, K. Kadowaki, I. Maggio-Aprile, and \O.
Fischer, \prl {\bf 80}, 3606 (1998);
E. W. Hudson, S. H. Pan, A. K. Gupta, K.-W. Ng, and J. C. Davis,
Science, {\bf 285}, 88 (1999); 
S. H. Pan, E. W. Hudson, A. K. Gupta, K.-W. Ng, H. Eisaki, S.
Uchida, and J. C. Davis, \prl {\bf 85}, 1536 (2000).
\bibitem{mottvortex} M. Franz and Z. Te\v sanovi\' c, \prb {\bf 63},
064516 (2001).
\bibitem{ogata} H. Tsuchiura, M. Ogata, Y. Tanaka, and S. Kashiwaya,
\prb {\bf 68}, 012509 (2003) and references therein.
\bibitem{dhlee} Q.-H. Wang, J. H. Han and D.-H. Lee, Phys. Rev. Lett.
{\bf 87}, 167004 (2001).
\bibitem{palee} P. A. Lee and X.-G. Wen, \prb {\bf 63}, 224517 (2001).
\bibitem{millis} L. B. Ioffe and A. J. Millis, \prb {\bf 66}, 094513 (2002). 
\bibitem{galilean} Viewing cuprates as entirely analogous to superfluid 
$^4$He, however, should be studiously avoided as emphasized elsewhere in
this paper -- the essential role of gapless
fermions, the Galilean invariance being broken by the CuO$_2$ lattice,
the conversion of the Goldstone mode to gapped plasmon in
superconductors are but a few
examples of the calamities that will be visited upon those who take
the similarities too far. 
\bibitem{footviscosity} In transport calculations a small empirical
Bardeen-Stephen core dissipation should be included for better quantitative
accuracy.
\bibitem{kleinert} H. Kleinert, {\it Gauge Fields in Condensed Matter}
 (World Scientific, Singapore, 1989).
\bibitem{abrikosov} A. A. Abrikosov, Zh. Eksp. Teor. Fiz. {\bf 32}, 1442 (1957); Sov. Phys. JETP {\bf 5}, 1174 (1957).
\bibitem{footswave} Since we are discussing an $s$-wave case here,
the pseudogap $\Delta$ in the fermionic action would have to be
changed from a $d_{x^2-y^2}$- to an $s$-wave form.
\bibitem{singer} J. M. Singer, M. H. Pedersen, T. Schneider, H. Beck,
and H.-G. Matuttis, \prb {\bf 54}, 1286 (1996) and references therein.
\bibitem{footdeltadelta} Although 
$\delta\Delta_{ij}\leftrightarrow\delta n_i +\delta n_j$ 
remains a useful approximation and will be used 
occasionally in this paper.
\bibitem{footdeltabdeltadelta} A more explicit  argumentation for
the $\delta {\bf B}_d(\br)\leftrightarrow \delta\Delta_{ij}$ correspondence
goes as follows: Note that $\delta\Delta_{ij}$ as defined in 
(\ref{dberryi}) is nothing but the Hubbard-Stratonovich field decoupling
the $\dot\theta_{ij}^2$ term in the action. By the reasoning of the
previous ``pedagogical''
subsection this immediately makes $\delta\Delta_{ij}$ equivalent to
the modulation of the dual induction ${\bf B}_d$. But why should this
modulation be translated into the modulation of the pairing gap amplitude
$\Delta_{ij}$ as our notation seems to imply? 
Consider the spatial region where ${\bf B}_d$ is larger (smaller)
than its average. This region attracts (repels) dual vortices, i.e. vortices in
the dual field $\Phi$. Consequently, the amplitude of $\Phi$ is reduced
(enhanced) in this region which translates into 
the (anti)vortices in $\theta_{ij}$ itself
staying away from (being drawn to)
the same region. This, by the analiticity of the
complex gap function $\Delta_{ij}$ near a (anti)vortex position, finally 
implies that its amplitude must be larger (smaller) than its average 
in the regions where ${\bf B}_d$ is larger (smaller) than its average. 
Consequently, 
$\delta {\bf B}_d(\br)\leftrightarrow \delta\Delta_{ij}$ to the leading
order, where $\delta\Delta_{ij}$ indeed assumes the meaning of 
the modulation in
the $d$-wave pairing amplitude, up to an overall factor which can only
be determined from a fully microscopic theory. In the present paper, this
overall factor should be treated as an adjustable parameter.
Finally, note that the above reasoning does not imply that 
$|\Delta_{ij}|$ is the operator canonically conjugate to $\theta_{ij}$.
Rather, it simply relates the modulation in the ground state expectation
value of such (unknown) operator to $\delta\Delta_{ij}$.
\bibitem{read} N. R. Cooper, S. Komineas, and N. Read, cond-mat/0404112.
\bibitem{franzaffleck} M. Franz, I. Affleck, and M. H. S. Amin,
\prl {\bf 79}, 1555 (1997).
\bibitem{footnospin} Again, we should remind the reader that
this dual Lagrangian contains the contribution from the charge
channel but ignores the spin. The coupling to the spin channel,
associated with nodal fermions, enters through the Berry gauge
field $a_\mu$ and its contribution to ${\cal L}_d$ is
important near a critical point but not
otherwise. In the parlance of effective field theory ${\cal L}_d$
describes the ``high energy'' physics of charge sector relative
to the ``low energy'' physics of spin. Of course, the 
effect of $a$ remains {\em essential} for
the low energy fermiology throughout the pseudogap state. 
\bibitem{brown} E. Brown, Phys. Rev {\bf 133}, A1038 (1964).
\bibitem{zak} J. Zak, Phys. Rev. {\bf 134}, A1602 (1964); {\bf 134}, A1607 (1964).
\bibitem{negele} J. W. Negele and H. Orland, {\it 
Quantum Many-Particle Systems} (Addison-Wesley, New York, 1988).
\bibitem{fiveindependent} Only five of the six parameters are
independent: $\delta\Delta_{ij}$ are related to dual fluxes $\delta B_d$, whose average over a 
unit cell is zero.


\end{thebibliography}
\end{document}